\documentclass[final,3p,times]{elsarticle}

\usepackage[utf8]{inputenc}
\usepackage[english]{babel}
\usepackage{natbib}
\biboptions{sort&compress}
\usepackage{amsmath, amssymb, amsfonts}
\usepackage{eqnarray}
\usepackage{graphicx}
\usepackage{subfigure}
\usepackage{tabularx}
\usepackage{caption}
\usepackage[hidelinks]{hyperref}
\usepackage{url}
\usepackage{cleveref}
\usepackage{midpage}
\usepackage{setspace}
\usepackage{todonotes}
\usepackage{soul}
\usepackage{xcolor}
\usepackage{lipsum}
\usepackage{xargs}
\usepackage[normalem]{ulem} 

\definecolor{amber(sae/ece)}{rgb}{1.0, 0.49, 0.0}
\definecolor{applegreen}{rgb}{0.55, 0.71, 0.0}
\definecolor{ao(english)}{rgb}{0.0, 0.5, 0.0}
\definecolor{blue(munsell)}{rgb}{0.0, 0.5, 0.69}
\definecolor{forestgreen(web)}{rgb}{0.13, 0.55, 0.13}

\newcommand{\rev}[1]{\textcolor{black}{#1}}

\newcommand{\ee}{\mathbf{e}}
\newcommand{\ii}{\mathbf{i}}
\newcommand{\jj}{\mathbf{j}}
\newcommand{\vv}{\mathbf{v}}
\newcommand{\xx}{\mathbf{x}}

\newcommand{\qoi}{\mathbf{f}}
\newcommand{\qoiscal}{f}
\newcommand{\mcI}{\mathcal{I}}
\newcommand{\mcS}{\mathcal{S}}
\newcommand{\mcU}{\mathcal{U}}
\newcommand{\s}{\, \, \, \,\,}
\DeclareMathOperator*{\argmax}{arg\,max}
\DeclareMathOperator*{\argmin}{arg\,min}

\begin{document}
\begin{frontmatter}
		
\title{\rev{Sparse-grids uncertainty quantification of part-scale additive manufacturing processes}}
\author[UniPV]{Mihaela Chiappetta}
\author[IMATI,TUM]{Chiara Piazzola}
\author[UniPV]{Massimo Carraturo}
\author[IMATI]{Lorenzo Tamellini}
\author[UniPV,IMATI]{Alessandro Reali}
\author[UniPV,IMATI,IRCCS]{Ferdinando Auricchio}
\address[UniPV]{Department of Civil Engineering and Architecture - University of Pavia, Via Ferrata 3, 27100, Pavia, Italy}
\address[IMATI]{Istituto di Matematica Applicata e Tecnologie Informatiche “E. Magenes” - Consiglio Nazionale delle Ricerche, Via Ferrata, 5/A, 27100, Pavia, Italy}  
\address[TUM]{Department of Mathematics - Technical University of Munich, Boltzmannstra\ss e, 3, 85748, Garching bei M\"unchen, Germany}  
\address[IRCCS]{Fondazione IRCCS Policlinico San Matteo, 27100 Pavia, Italy}

\begin{abstract} 
The present paper aims at applying uncertainty quantification methodologies to process simulations of powder bed fusion of metal. In particular, for a part-scale thermomechanical model of an Inconel 625 super-alloy beam, we study the uncertainties of three process parameters, namely the activation temperature, the powder convection coefficient and the gas convection coefficient. First, we perform a variance-based global sensitivity analysis to study how each uncertain parameter contributes to the variability of the beam displacements. The results allow us to conclude that the gas convection coefficient has little impact and can therefore be fixed to a constant value for subsequent studies. Then, we conduct an inverse uncertainty quantification analysis, based on a Bayesian approach on synthetic displacements data, to quantify the uncertainties of the two remaining parameters, namely the activation temperature and the powder convection coefficient. Finally, we use the results of the inverse uncertainty quantification analysis to perform a data-informed forward uncertainty quantification analysis of the residual strains. \rev{Crucially, we make use of surrogate models based on sparse grids to keep to a minimum the computational burden of every step of the uncertainty quantification analysis. The proposed uncertainty quantification workflow allows us to substantially ease the typical trial-and-error approach used to calibrate power bed fusion part-scale models, and to greatly reduce uncertainties on the numerical prediction of the residual strains. In particular, we demonstrate the possibility of using displacement measurements to obtain a data-informed probability density function of the residual strains, a quantity much more complex to measure than displacements}.
\end{abstract}
		
\end{frontmatter}
\section{Introduction}
\label{sec:introduction}
Additive Manufacturing (AM) \cite{gibson2021additive,wong2012review} is a technology that is able to produce components with complex geometries and high mechanical performances by means of a layer-by-layer strategy. Among the different AM technologies available for the production of metal components {\cite{frazier2014metal}}, Powder Bed Fusion  (PBF) of metals is nowadays the most widespread \cite{king2015laser,chowdhury2022laser}. PBF processes start by distributing a layer of metal powder particles on a build plate within a closed chamber where an inert gas (argon or nitrogen) is inflated. The metal powder is then selectively melted by means of a high-energy-density laser source following a predefined scan path. Once the layer scan is completed, a new layer of metal powder is deposited on top of the previous one by means of a roller and the process is repeated until the final component is terminated. 
	
Due to the multi-scale and multi-physics nature of the process, complex process-structure-property relationships occurring in PBF processes are not fully understood yet \cite{smith2016linking}; thus, proper calibration of a large set of parameters still requires long and expensive trial-and-error experimental approaches. To this end, computer-aided simulations of the PBF process can play a crucial role in the AM production of metal components \cite{king2015overview,keller2017application}. In the literature, we can find a large set of numerical models suitable for AM simulations, which can be classified into three main classes based on their reference length scale, namely: micro-, meso-, and macro-scale \cite{bayat2021review}. Micro-scale models, also called powder-scale models, provide information on the effect of the laser beam on the microstructure evolution and grain structure; they are typically solved using, e.g., Phase-Field methods \cite{keller2017application, karayagiz2020finite,ghosh2018simulation,lu2018phase} or Cellular Automata \cite{zinoviev2016evolution}. Meso-scale models investigate the effect of the scanning strategy and the laser parameters on melt-pool dynamics and are typically solved using e.g., the Lattice-Boltzmann method \cite{attar2011lattice,klassen2014evaporation} and the Discrete Element method \cite{hooper2018melt,steuben2016discrete}. Finally, macro-scale models, also called part-scale models, allow the prediction of the mechanical response of the component, including the study of quantities such as residual stresses and thermal distortions at part-scale. The most popular method in part-scale AM numerical models is the Finite Element (FE) method \cite{chiumenti2017numerical,denlinger2017thermomechanical,carraturo2020modeling,li2018modeling,tan2019thermo,ghosh2018simulation,carraturo2022two,viguerie2022spatiotemporal}.
	
PBF processes include several sources of uncertainty, due to the inherent variability of the process parameters, e.g., powder particle radius, mechanical properties of powder particles, physical-chemical properties of the material (see e.g. \cite{king2015laser} for a general survey). Uncertainty Quantification (UQ) methodologies \cite{hu2017uncertainty, ahu2017uncertainty} are a suitable tool to quantify and reduce the influence of such uncertainties on both the process and product quality. In particular, forward UQ analysis studies the propagation of uncertainties from the parameters -- modeled as random variables with uncertainty described by the associated probability density function (PDF) -- to the outputs of the model, also called Quantities of Interest (QoIs) of the problem. On the contrary, inverse UQ analysis aims at reducing the uncertainty on the model parameters, estimating the plausibility of the different possible values that the parameters can take given a set of experimental data, using, e.g., Bayesian inversion techniques \cite{berger2013statistical,stuart:acta.bayesian,thanh-bui.gattas:MCMC}. Therefore, inverse UQ analysis can be seen as closely related to parameter calibration, but -- instead of providing a specific value for the calibrated parameters -- it returns a \lq\lq data-informed\rq\rq~PDF.
	
The effectiveness of UQ techniques for AM processes is widely recognized for applications in production-level experiments \cite{kamath2016data,ahu2017uncertainty,garg2014state}, whereas their application to AM numerical models (e.g., for validation of computational models) still remains an open challenge. \rev{Some discussion concerning forward UQ analysis can be found in \cite{lopez2016identifying,grasso2017process,korshunova2021uncertainty,nath2017mutli}. In particular, \citet{lopez2016identifying} presented a discussion on the origin and propagation of uncertainty in PBF models;  \citet{grasso2017process} developed a method based on principal component analysis for spatial identification of defects during the PBF process; \citet{korshunova2021uncertainty} introduced a random field model in combination with the Finite Cell Method to efficiently evaluate the influence of microstructure on the variability of the mechanical behavior of AM products; finally, \citet{nath2017mutli} proposed a framework for modeling and quantifying the uncertainty of material properties using a multi-level approach}.

	
\rev{Coming to inverse UQ analysis, on the one hand the capability of this kind of analysis to provide appropriate PDFs for input parameters would dramatically increase the reliability of simulated results; on the other hand, their application to AM models remains critical, mainly due to the large number of numerical simulations required, whose computational cost could be vary large.}
To reduce such computational burden, a viable approach is to replace the results of the full AM model with cheaper evaluations of so-called surrogate models. Surrogate models are obtained by first running a \emph{limited number} of full model simulations; these full model results are then interpolated or approximated by, e.g., least-squares to create a response surface to compute inexpensive problem solutions avoiding full model, computationally demanding runs. Therefore, surrogate modeling approaches can be considered the key to overcome the computational burden affecting UQ technology. In the literature, most of the surrogate models suitable for UQ of AM processes are based on Gaussian processes and focus on experimental melt-pool parameters. \citet{xie2022bayesian} quantified model and data uncertainties in a melt-pool model using measurement data on melt-pool geometry (length and depth). \rev{\citet{wang2019uncertainty} proposed a data-driven UQ framework for efficient investigation of uncertainty propagation from process parameters to material micro-structures, then to macro-level mechanical properties through a combination of advanced AM multi-physics simulations and data-driven surrogate modeling.} \citet{wang2020uncertainty} proposed a sequential Bayesian calibration method to perform calibration of experimental parameters and model correction to significantly improve the validity of the melt-pool surrogate model. \citet{nath2017mutli} presented an inverse UQ framework predicting the microstructure evolution during solidification by coupling a surrogate meso-scale melt-pool model with a micro-scale cellular automata model. \rev{\citet{ghosh2019uncertainty} used surrogate models to quantify the contribution of different sources of uncertainty to micro-segregation variability during PBF processes.}
	
In the present paper, we propose and apply a UQ approach to quantify the uncertainties involved in the simulation of a PBF process using a part-scale thermomechanical model of an Inconel 625 beam (i.e., at a different scale compared to the above-mentioned contributions \cite{lopez2016identifying,grasso2017process,nath2018modeling,korshunova2021uncertainty,xie2022bayesian,wang2019uncertainty, wang2020uncertainty,nath2017mutli,ghosh2019uncertainty}). In particular, we want to study the influence of the activation temperature, the powder convection coefficient and the gas convection coefficient in order to obtain reliable residual strains in a PBF produced part.
We remark that the part-scale thermomechanical model considered in the present work highly simplifies the physics of the process and therefore the parameters that we consider are in a way only conceptual (i.e, they cannot be directly measured in reality): for this reason an inverse UQ analysis is mandatory to suitably define their ranges and PDFs. Conversely, the UQ analysis in \cite{lopez2016identifying,grasso2017process,nath2018modeling,korshunova2021uncertainty,xie2022bayesian,wang2019uncertainty, wang2020uncertainty,nath2017mutli,ghosh2019uncertainty} focuses on melt-pool model parameters, that can actually be measured (either directly or indirectly) to obtain a prior PDF that is at least partially consistent with reality. Moreover, our approach is not based on mechanical responses obtained by experiments but rather by part-scale thermomechanical numerical simulations: this allows us to remove any errors due to the inherent mismatch between reality and our part-scale thermomechanical model; consequently, any error/sub-optimal result (either in the calibration or in the subsequent forward uncertainty propagation) can be attributed solely to the adopted methodology.

A further technical difference between \cite{xie2022bayesian,wang2019uncertainty,wang2020uncertainty,nath2017mutli,ghosh2019uncertainty} and our work is that we use a different surrogate modeling methodology, namely sparse grids \cite{bungartz2004sparse,babuska.nobile.eal:stochastic2,xiu.hesthaven:high}. \rev{Sparse grids are an effective method due to their simplicity of use, since they require only a limited number of hyperparameters to be tuned, and moreover it is straightforward to construct a sparse-grid surrogate model for problems that depend on uncertain parameters following a non-uniform probability distribution, a feature that will be extensively exploited in the present contribution. Furthermore, sparse grids and UQ functionalities are available off-the-shelf in the Sparse-Grids Matlab Kit software \cite{piazzola2022sparse}, which considerably simplifies the coding effort. In the context of AM, sparse grids are used also, e.g., in \citet{KnappUQ}, where the parameters of a melt-pool model are calibrated by Bayesian inversion accelerated by a sparse-grids surrogate model, as well as in \citet{tamellini2020parametric}, where a sparse-grid surrogate model is used to replace the full simulation of a PBF process for shape optimization purposes. A comparison between sparse grids and other surrogate modeling techniques (e.g., Gaussian regression processes) is however not provided herein since it would exceed the scope of the current work. The interested reader is referred to \cite{ghanem2017handbook, sudret2017surrogate,back.nobile.eal:comparison} for a thorough survey on this topic.}

\rev{In light of the above discussion, the main contribution of the present work is to provide a structured methodological procedure  that needs little to no user intervention to calibrate the parameters of a PBF model at part-scale, and at the same time to quantify its uncertainties. In details, this procedure (which we call UQ workflow in the following) consists of three steps: i) a Global Sensitivity Analysis (GSA) \cite{sobol2001global,saltelli2004sensitivity,saltelli2008global,sudret:sobol,iooss2015review} to identify the most influential parameters; ii) an inverse UQ analysis based on displacements data to compute data-informed PDFs of such parameters; iii) a forward UQ analysis based on such data-informed PDFs to predict residual strains. The present results show that the proposed methodology allows to substantially ease the typical trial-and-error approach used to calibrate models describing PBF processes. Moreover, employing sparse grids-based surrogate models, we are able to substantially reduce the overall number of AM numerical simulations required in the procedure. We highlight also that this procedure allows us not only to predict residual strains but also to quantify the uncertainty in the prediction, which is a highly desirable feature.}
        
The present work is structured as follows. In \Cref{sec:Metal-AM model}, we present the part-scale thermomechanical model to describe PBF process simulation. In \Cref{sec:UQ analysis}, we describe the developed UQ approaches, both forward and inverse. In \Cref{sec:Results and Discussion}, UQ results are reported and discussed. Finally, in \Cref{sec:Conclusion}, we draw the main conclusions and possible further perspectives of the present work. We also report some technical background information in \ref{Surrogate Model} and \ref{sec:Leja_points}. 
	
\section{Part-scale thermomechanical model}
\label{sec:Metal-AM model}	
\begin{figure}[!h]
	\begin{center}
	\includegraphics[width=0.8\textwidth]{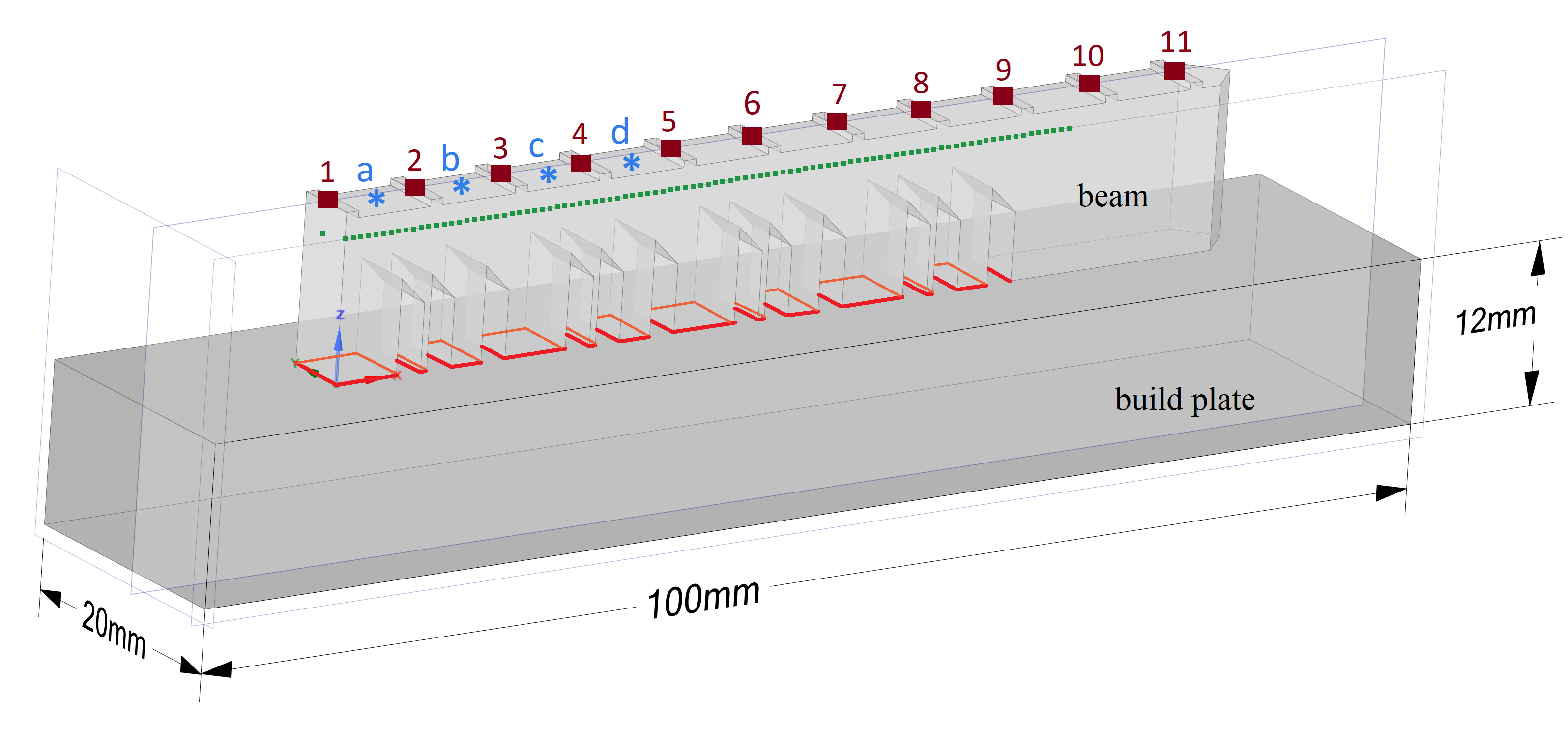}
	\caption{\rev{AM model of a beam 75 mm long, 12 mm high and 5 mm wide with a build plate measuring 85 mm long, 12 mm high and 20 mm wide. Points marked by (purple) numbers and (blue) letters will be used during the UQ analysis; the (green) dotted line marks the location where we will compute residual strains (i.e., at $z=11$ mm); the (red) continuous line contours indicate the removal area ending at $x=56$ mm. A color version of every figure in this manuscript is available online.}}
	\label{fig:stl}
	\end{center}
\end{figure}

In the present work, we employ a part-scale thermomechanical model \cite{chiumenti2017numerical,denlinger2017thermomechanical,carraturo2020modeling,li2018modeling,tan2019thermo,ghosh2018simulation,carraturo2022two,viguerie2022spatiotemporal} to simulate the PBF process of an Inconel 625 beam (\Cref{fig:stl}) according to the design experiment proposed by the National Institute of Standards and Technology \rev{(hereafter, we will use the terms \lq\lq beam\rq\rq~and \lq\lq component\rq\rq~interchangeably)}. In \Cref{subsec:Mod} we introduce the governing equations that describe the thermal and mechanical problems involved in the PBF process, whereas in \Cref{subsec:Numerical approach}, we present the numerical approach used to simulate the process.
	
\subsection{Governing equations}
\label{subsec:Mod}
	
    \subsubsection{Thermal problem}
	\label{subsubsec:Th}
	Assuming that the material follows Fourier's law, the thermal problem is governed by the temperature-based heat transfer equation as follows \cite{burmeister1993convective}:
	\begin{equation}
		\rho c_p(T) \dfrac{\partial T}{\partial t} -  \nabla \cdot \left(k(T) \nabla T\right)=0 \s \mathrm{in} \s \Omega,
		\centering
		\label{eq: eq1}
	\end{equation}
	where $T$ is the temperature field, $\rho$ denotes the constant material density, $c_p$ is the temperature-dependent specific heat capacity at constant pressure, and $k$ is the temperature-dependent thermal conductivity.
	
	The thermal problem initial condition at time $t=0$ is set as:
	\begin{equation}
		T=T_0 \s \mathrm{in} \s \Omega,
		\centering
		\label{eq:initialCondition}
	\end{equation}
	whereas the Dirichlet and the Neumann boundary conditions on domain's boundary , $\partial \Omega = \partial \Omega_{Q} \cup \partial \Omega_{T} \cup \partial \Omega_{H}$, are defined as follows:
	\begin{align}
		&T=\bar{T} \s\s\s \,\,\,\,\, \mathrm{on}  \s \partial \Omega_{T} \subset \partial \Omega,\\
		&k \nabla T \cdot \boldsymbol n  =  \bar{q}   \s  \mathrm{on}      \s  \partial \Omega_{Q} \subset \partial \Omega,
	\end{align}
	where $\bar{T}$ is the temperature of the environment on the build plate lateral surface boundary $\partial \Omega_{T}$, and $\bar{q}$ denotes the heat loss through the free surface of normal $\boldsymbol n$ on the boundary lateral and upper surface of the beam $\partial \Omega_{Q}$; on the remaining portion of the domain boundaries ($\partial \Omega_{H}$) adiabatic conditions are imposed.  
	
	In the PBF simulation process at part-scale, the heat loss through the boundary can be described by means of two heat transfer mechanisms: heat loss by conduction through the powder, denoted by $q_{p}$, and heat loss by convection through the environment gas, denoted by $q_{g}$. Therefore, the heat loss $\bar{q}$ is split into two terms as follows:
	\begin{equation}
		\bar{q}=q_{p}+q_{g}.
	\end{equation}
	\rev{Since powder is not included in our model, both heat loss mechanisms can be modeled by means of two convection-like boundary conditions, each with a different heat transfer coefficients: a powder convection coefficient, $h_p$, and a gas convection coefficient, $h_g$, respectively. This simplifying assumption is also motivated by the fact that at part-scale it is difficult to distinguish between the two heat transfer mechanisms \cite{chiumenti2017numerical}.} Therefore, following Newton's law, we can formulate the two different heat loss terms as:
	\begin{align}
		\label{eq:hg}
		q_{p} = h_p(T-\bar{T}) \s\s\s \,\, \mathrm{on}  \s \Gamma_{p} \subset \partial \Omega_{Q},\\
		q_{g} = h_g(T-\bar{T})  \s\s\s \,\, \mathrm{on}  \s \Gamma_{g} \subset \partial \Omega_{Q},
		\label{eq:hp}
	\end{align}
	\rev{where $\Gamma_{p}$ is the interface between the powder and the component and $\Gamma_{g}$ is the interface between the environment gas and the component, with $\partial\Omega_{Q} = \Gamma_{p} \cup \Gamma_{g}$ and $\Gamma_{p} \cap \Gamma_{g}=\emptyset$.}
	
	\subsubsection{Mechanical problem}
	\label{subsubsec:Mc}
	The solution of the mechanical problem is given by solving the equilibrium equation:
	\begin{equation}
		\nabla \cdot {{\boldsymbol \sigma}} = \boldsymbol 0 \s \mathrm{in} \s \Omega,
		\label{eq: eq10}
 	\end{equation}
	with ${\boldsymbol \sigma}$ the Cauchy stress tensor defined as:
	\begin{equation}
		{\boldsymbol \sigma} = {\boldsymbol D}^{el} {\boldsymbol \varepsilon}^{el},
		\label{eq: eq11}
	\end{equation}
	where ${\boldsymbol D}^{el}$ is the isotropic elasticity tensor, depending on the Young's modulus of elasticity and Poisson's ratio, and ${\boldsymbol \varepsilon}^{el}$ is the elastic strain. The total strain in the material, ${\boldsymbol \varepsilon}^{tot}$, can be decomposed into elastic strain, ${\boldsymbol \varepsilon}^{el}$, thermal strain, ${\boldsymbol \varepsilon}^{th}$, and plastic strain, ${\boldsymbol \varepsilon}^{pl}$, as:
	\begin{equation}
		{\boldsymbol \varepsilon}^{tot} =  {\boldsymbol \varepsilon}^{el} + {\boldsymbol \varepsilon}^{th} + {\boldsymbol \varepsilon}^{pl} = \frac{1}{2} [\nabla \boldsymbol{u} + (\nabla \boldsymbol{u})^T], 
		\label{eq: eq12}
	\end{equation}
	with $\boldsymbol u$ the displacement vector. 
	
	In our thermomechanical model, the thermal strain acts as an external (thermal) load and is defined as follows:
	\begin{equation}
		{\boldsymbol \varepsilon}^{th}  = \alpha\Delta T \boldsymbol I ,
		\label{eq:thermalStrain}
	\end{equation}
	with $\alpha = \alpha(T)$ the temperature-dependent thermal expansion coefficient, $\Delta T$ the variation in time of the temperature field, and $\boldsymbol I$ the second-order identity tensor.
	
	Finally, the plastic strain rate $\dot{\boldsymbol \varepsilon}^{pl}$ is computed following the Prandtl-Reuss flow rule \cite{chiumenti2017numerical,carraturo2020numerical,carraturo2021immersed} as follows:
	\begin{equation}
		\dot{\boldsymbol \varepsilon}^{pl} = \dot{\gamma} \dfrac{\partial \Sigma}{\partial \boldsymbol \sigma}, 
		\label{eq: eq15}
	\end{equation}
	where $\gamma$ is the equivalent plastic strain, $ \Sigma= \sigma_{\text{vm}}- \sigma_y \leq 0$ is the yield function describing the material through the equivalent Von Mises stress, $\sigma_{\text{vm}}=\sqrt{\frac{3}{2}\boldsymbol{s}:\boldsymbol{s}}$ with $\boldsymbol{s}=\boldsymbol{\sigma}-tr(\boldsymbol{\sigma})\boldsymbol{I}$, and the temperature-dependent yield stress, $\sigma_{y}=\sigma_y(T)$ \cite{chiumenti2017numerical,carraturo2020numerical,carraturo2021immersed}.
	
	The mechanical problem is solved with the following Dirichlet boundary condition:
	\begin{equation}
		\boldsymbol u=\boldsymbol 0 \s\s\s \,\,\, \mathrm{on}  \s \partial \Omega_{U} \subset \partial {\Omega},
	\end{equation}
	with $\partial \Omega_{U}$ the bottom surface of the build plate where the fixed support is imposed; on the remaining part of the boundary $\partial \Omega$, we impose homogeneous Neumann boundary conditions.
	
	\subsection{Numerical approach}
	\label{subsec:Numerical approach}
	
	In the present work, we use Ansys2021-R2 software to simulate the beam PBF process, which adopts a weakly coupled thermomechanical approach based on the FE method \cite{chergui2020finite,chiumenti2017numerical,denlinger2017thermomechanical,carraturo2020modeling,keller2017application,ghosh2018simulation}. This means that the thermal and mechanical analyses are not fully coupled, i.e., they are not solved monolithically at each time step, but rather in a staggered way, allowing the calculations to be greatly speeded up. In particular, the thermal transient analysis is solved first throughout the printing time by storing the thermal field at each layer. These temperature fields are then transferred to the quasi-static mechanical problem solver, where they are used to evaluate the thermal strain of the problem (see \Cref{eq:thermalStrain}).
	
	\subsubsection{Meshing strategy}
	\label{subsub:GM}
	We adopt two different meshing strategies for the component and the build plate. In particular, for the discretization of the component, we use a uniform mesh with quadratic hexahedral finite elements of size $0.5$ $\text{mm}$; while for the build plate we choose a coarser mesh, with linear hexahedral finite elements of size $3$ $\text{mm}$. This choice is motivated by the fact that the build plate is modeled only to take into account the heat loss through the build plate and as a mechanical constraint for the component, so it does not require accurate meshing. Since the two meshes are non-conforming at the interface, their coupling is achieved by a contact element approach. \rev{Finally, in \Cref{tab:FE} we report more information about the meshing strategy for the component and the build plate.} 
	
	    \begin{table}[!h]
		\begin{center}
			\begin{tabular}{l c c c}
				\hline
				\hline
				\textbf{$ $}  &   \textbf{nb. nodes} & \textbf{nb. elements} & \textbf{nb. $z$-layers} \\ \hline 
				component	 &  ${141095}$ & $30340$ & $25$\\
				build plate & $5105$ & $952$ & $4$\\
				\hline
			\end{tabular}
		\end{center}
		\caption{\rev{Summary of the number of nodes (nb. nodes), the total number of elements (nb. elements) in the domain and the number of layers in $z$-direction (nb. $z$-layers) related to the meshing strategy chosen for the component and build plate.}}
		\label{tab:FE}
	\end{table}

	\subsubsection{Material Properties}
	\label{subsubsec:matmar}
\begin{figure}[h!]
		\centering
		\label{fig:Matpar}
		\subfigure[PARAMETRI-1][\s\s\s\s\s\s\s\s\s\s\s\s\s]
		{\includegraphics[width=0.41 \textwidth]{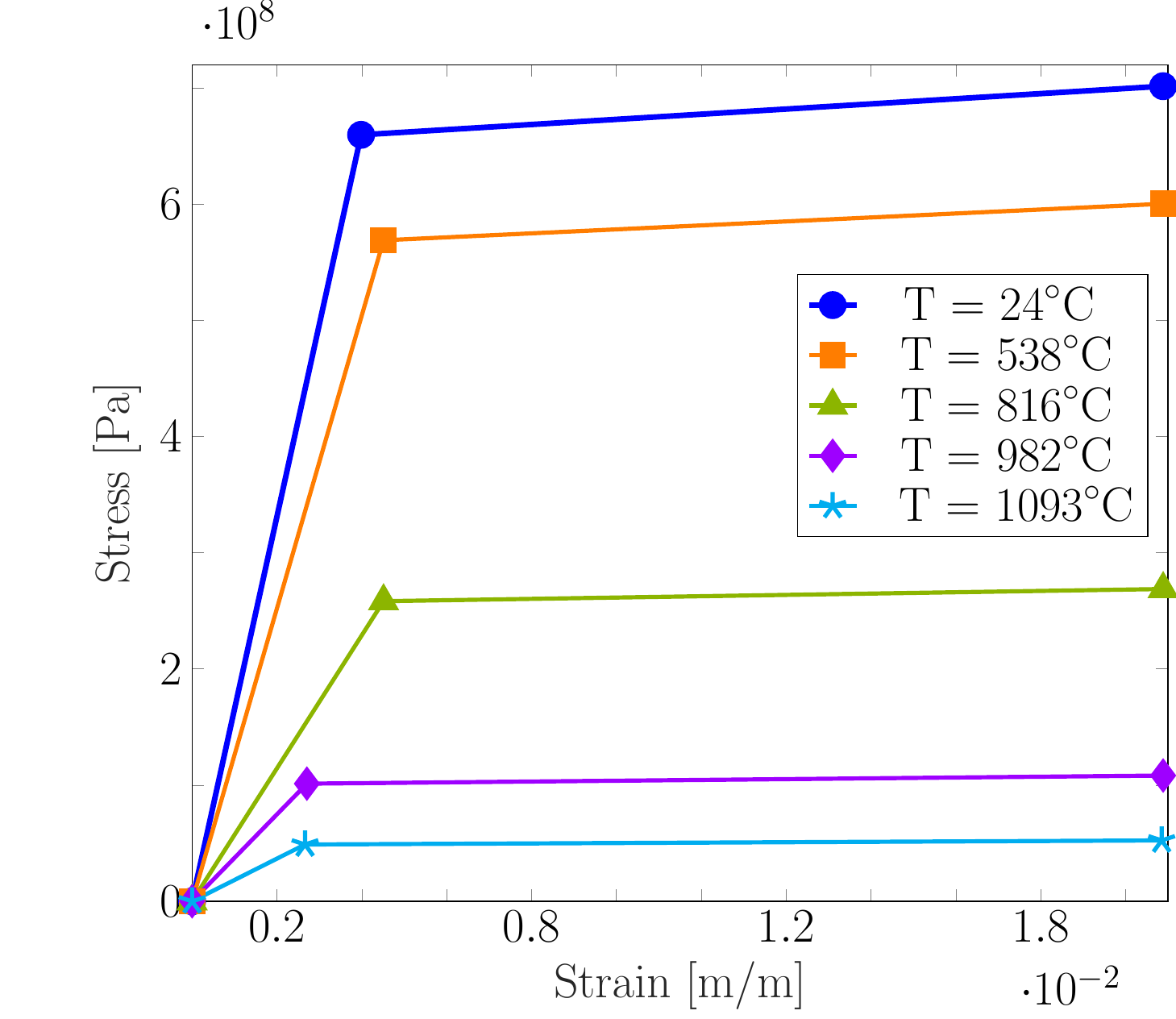} 
		\label{subfig:Hardening}} 
		\subfigure[PARAMETRI-2][\s\s\s\s\s\s\s\s\s\s\s\s\s]{\label{subfig:coefficientofthermalexpansion}\includegraphics[width=0.4 \textwidth]{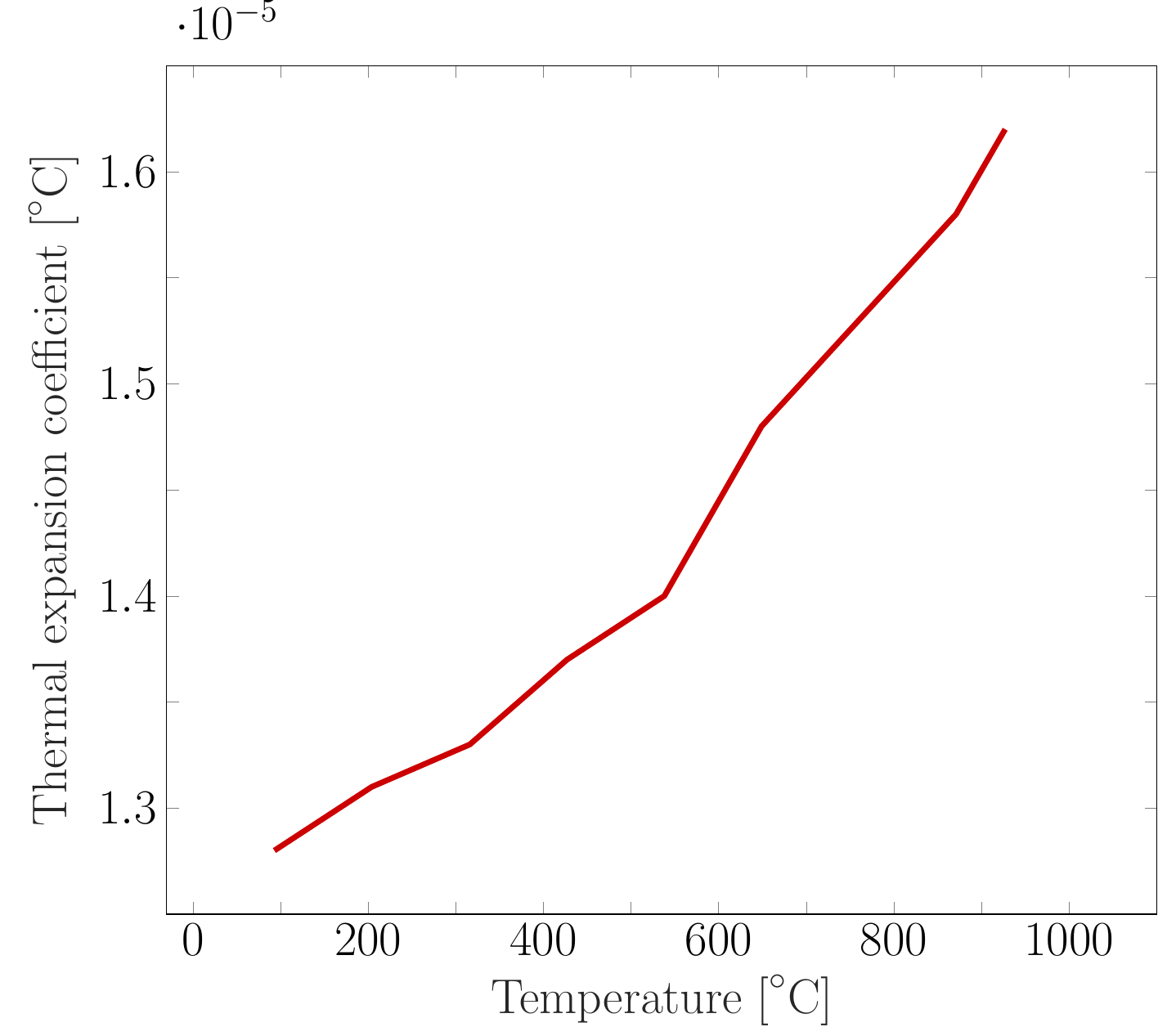}}
		\subfigure[PARAMETRI-3][\s\s\s\s\s\s\s\s\s\s\s\s]{\label{subfig:elastic modulus}\includegraphics[width=0.41 \textwidth]{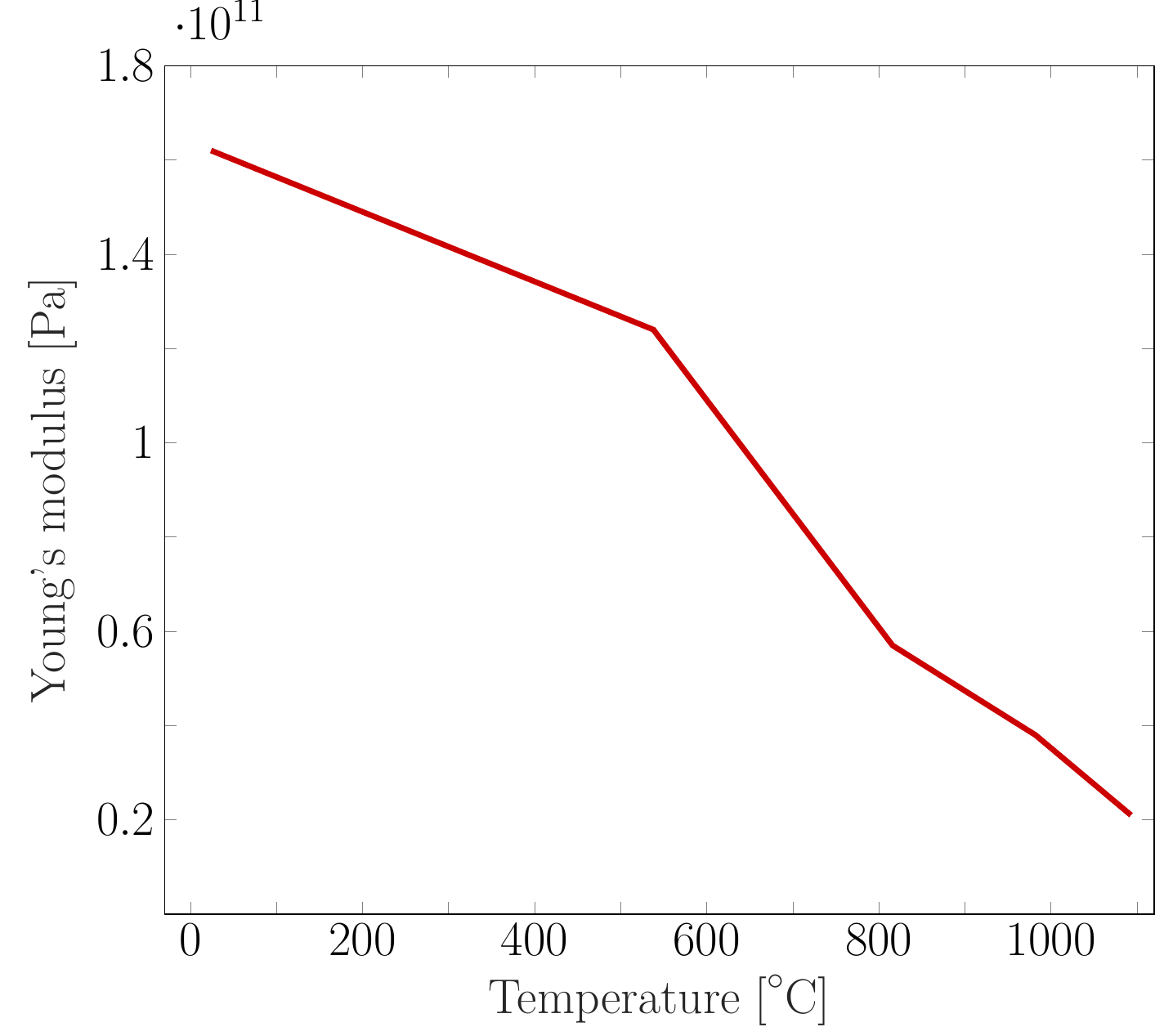}}
	   \subfigure[PARAMETRI-4][\s\s\s\s\s\s\s\s\s\s\s\s]{\label{subfig:Poissonsratio}\includegraphics[width=0.4 \textwidth,height=0.345 \textwidth]{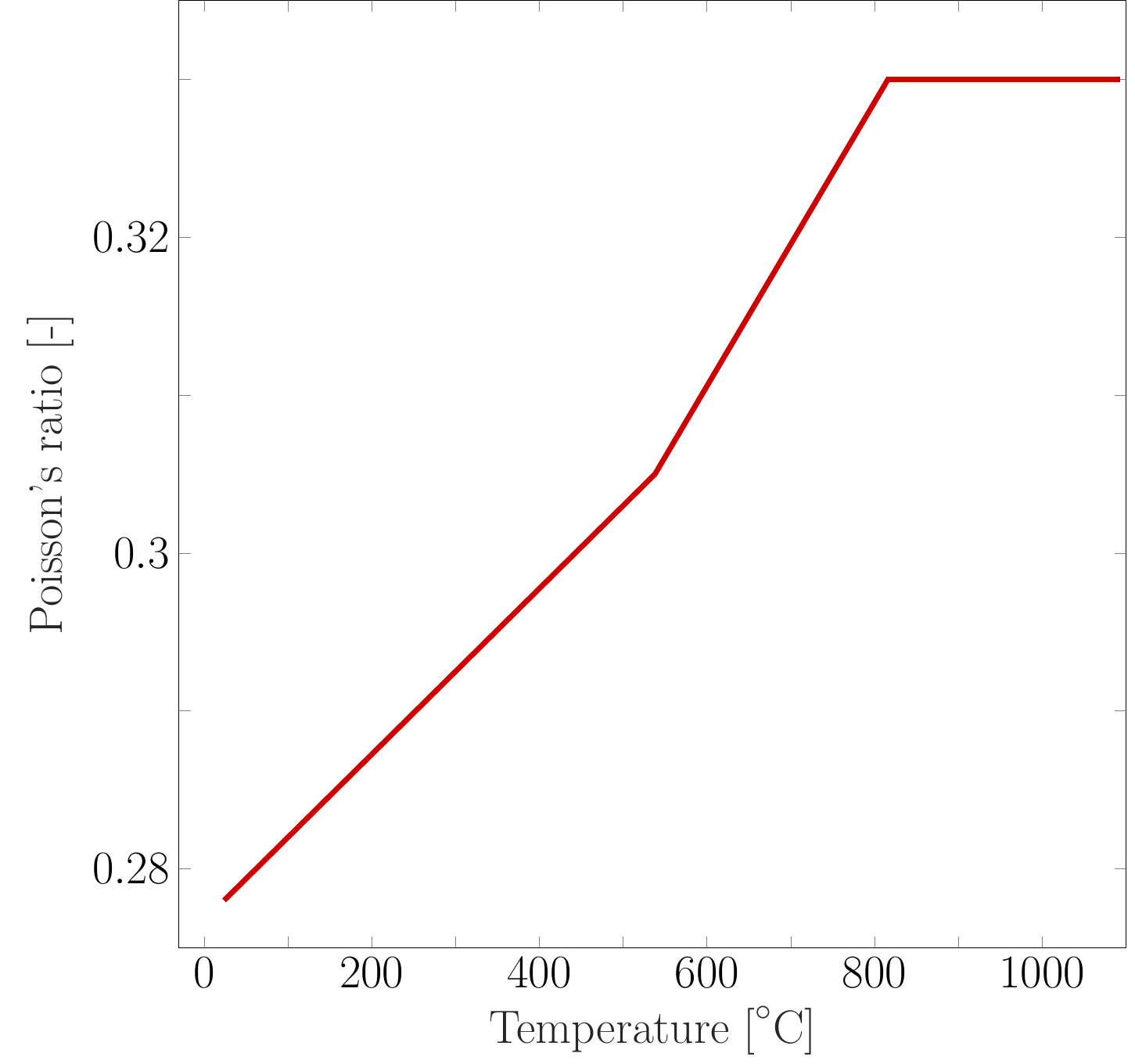}}
	    \subfigure[PARAMETRI-5][\s\s\s\s\s\s\s\s\s\s\s\s]{\label{subfig:specific heat}\includegraphics[width=0.405 \textwidth]{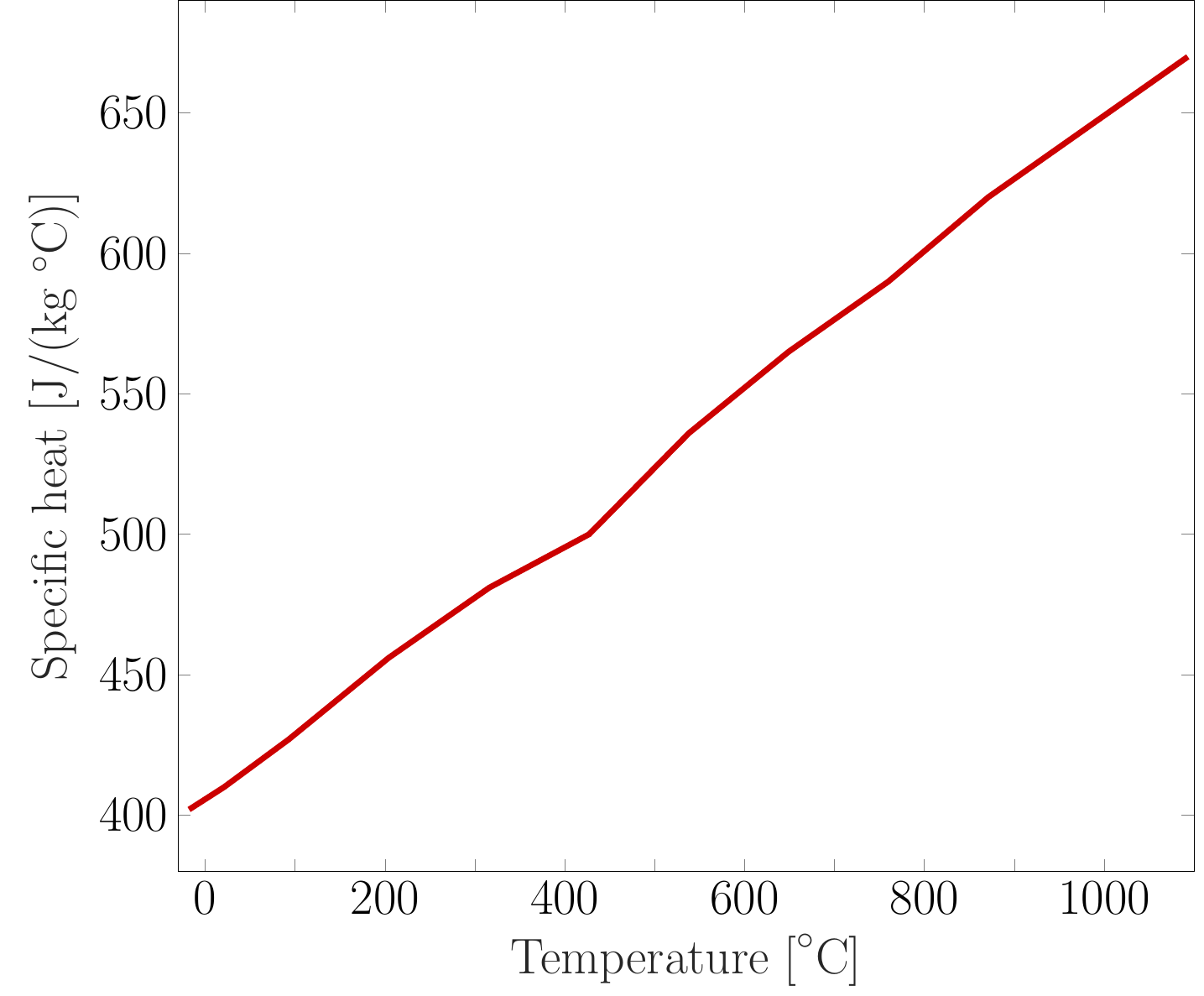}} 
		\subfigure[PARAMETRI-6][\s\s\s\s\s\s\s\s\s\s\s\s]{\label{subfig:thermalconductivity}\includegraphics[width=0.4 \textwidth]{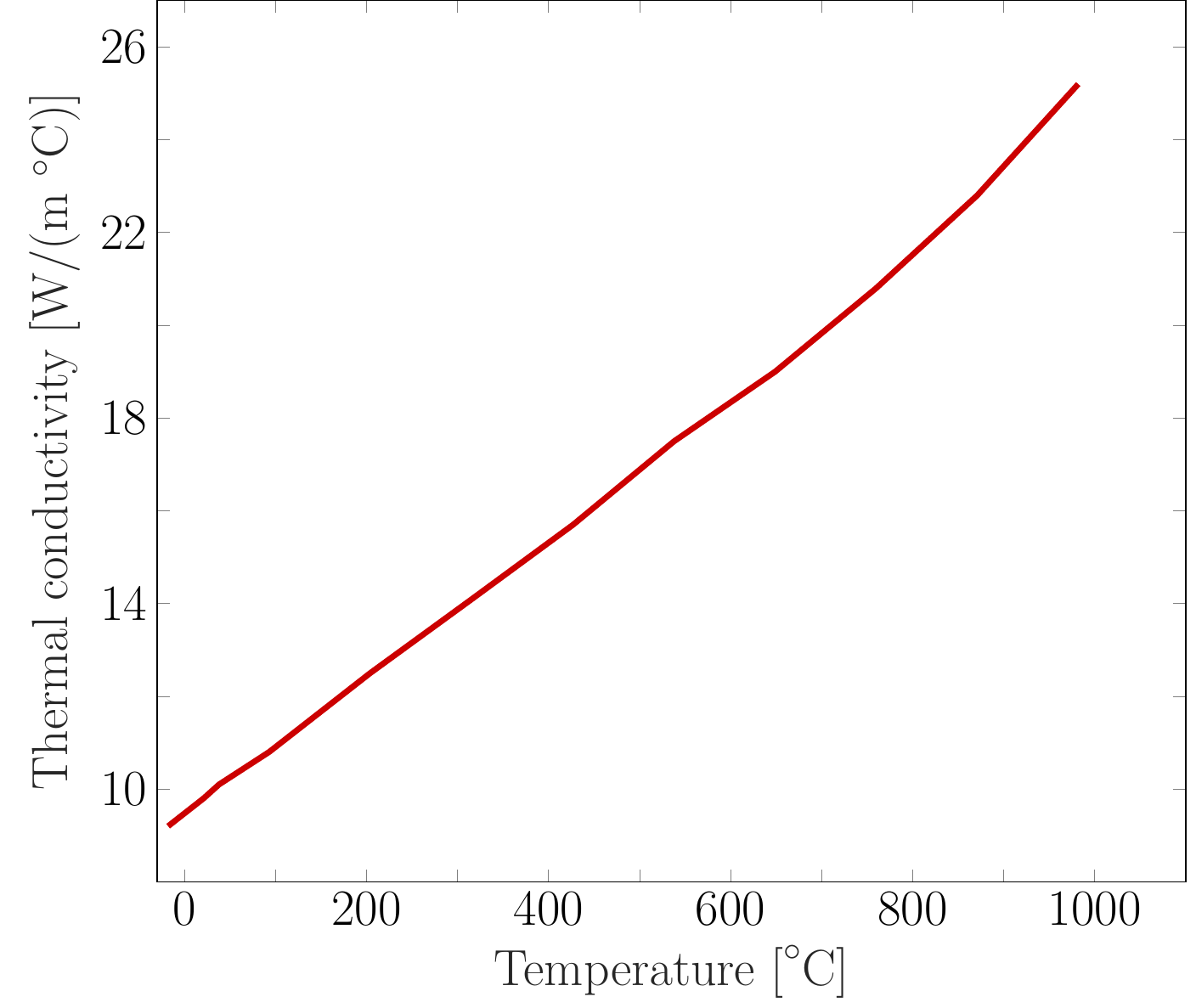}}
		\caption{\rev{Temperature-dependent Inconel 625 properties extrapolated from Ansys2021-R2: (a) Bilinear isotropic hardening; (b) Coefficient of thermal expansion; (c) Young's modulus; (d) Poisson’s ratio; (e) Specific heat; (f) Thermal conductivity. }}
	\end{figure}

	The material for the build plate and the component is set to be the nickel-based super-alloy  Inconel 625. During the PBF process -- due to the strong thermal gradients -- the material reaches temperatures ranging from chamber temperature to temperatures above the melting point. Therefore, we adopt a bi-linear isotropic plastic hardening model with a temperature-dependent yield behavior (\Cref{subfig:Hardening}). In \Cref{subfig:coefficientofthermalexpansion,subfig:elastic modulus,subfig:Poissonsratio,subfig:specific heat,subfig:thermalconductivity}, we report the temperature-dependent material properties used in the present work, whereas the temperature-independent material density and the melting temperature are set to $8440$ ${\text{kg}}/{\text{m}}^3$ and $1290$ $^\circ{\text{C}}$, respectively. \rev{We specify that both temperature-dependent and temperature-independent material properties used in the present work are extrapolated from Ansys2021-R2 software.}
     
	\subsubsection{Printing parameters}
	\label{subsub:amsimul}
	To simulate the melting and solidification process during the construction of a single layer, we assume a total layer process duration of 52 s in which, in the first 26 s, the entire layer is heated and, in the following 26 s, it is cooled down to a temperature of $20$ $^\circ{\text{C}}$. \rev{In \Cref{tab:procma} we provide the additional simulation parameters to simulate machine settings during the PBF process.}
	
    Due to the part-scale nature of our model, at each new layer activation, we do not model the localized laser heat source but, instead, we set the entire newly activated layer at a so-called activation temperature, $T_A$. 
	
	The material deposition process is modeled via the element birth and death technique \cite{michaleris2014modeling}. In such a technique, the FE models of both the component and the build plate are initially completely constructed, then all elements contained in the component model are deactivated. Deactivated elements remain in the FE model but contribute with only a very small conductivity value to the total system, i.e., their contribution to the problem is negligible. Only once a new powder layer is activated, all deactivated elements within that layer are switched to active elements, i.e., the actual material conductivity value is assigned to the element.
	More details can be found in \cite{chergui2020finite,michaleris2014modeling}. 
	
	To simulate the printing environment, we set the temperature of the top surface of the build plate to $80$ $^\circ{\text{C}}$ while the side surfaces are set to $40$ $^\circ{\text{C}}$ to contain thermal gradients. At the end of the printing process, we assume that the build plate is cooled down to the chamber temperature set at $20$ $^\circ{\text{C}}$.  
	Finally, we also assume that at the end of the printing process the component is partially removed from the build plate as depicted in \Cref{fig:stl}. 
	
	\begin{table}[!ht]
		\begin{center}
			\begin{tabular}{l c}
				\hline
				\hline
				\textbf{Parameters}  &   \textbf{Value} \\ \hline 
				deposition thickness	 &  ${20\mu \text{m}}$  \\
				hatch space & ${100\mu \text{m}}$  \\
				dwell time multiplier & 4\\
				number of heat sources & 1\\
				scan speed  &   ${900 \frac{\text{mm}}{\text{s}}}$  \\
				laser power &  ${100 \text{W}}$  \\
				\hline
			\end{tabular}
		\end{center}
			\caption{\rev{Summary of machine setting parameters for PBF thermomechanical numerical simulation of the Inconel 625 beam model.}}
			\label{tab:procma}
	\end{table}
	\newpage
\section{Uncertainty Quantification}
\label{sec:UQ analysis}
	
\rev{In this section, we present the details of the proposed UQ analysis workflow, starting from an in-depth description of the considered uncertain parameters of the model. We have chosen to use a sparse-grid surrogate modeling approach to perform the GSA, the inverse UQ analysis based on Bayesian inversion procedure, as well as the data-informed forward UQ analysis.}
	
	\subsection{Sources of uncertainty}
	\label{subsec:SU}
	
	The parameters that we consider as uncertain in our study are the logarithm (in base 10) of the powder convection coefficient, $\log h_p$ (see \Cref{subsubsec:Th}), the logarithm of the gas convection coefficient,  $\log h_g$ (see \Cref{subsubsec:Th}), and the activation temperature $T_A$ (see \Cref{subsub:amsimul}). As usual in UQ, these uncertain parameters are initially modeled as random variables. Moreover, since we want to enforce as little prior knowledge as possible on the parameters, we assume that each parameter is a uniform random variable over a suitable range, and that these three random variables are mutually independent; for the same reason, we take intervals larger than those typically used in literature
	\cite{an2017neutron, gan2019benchmark, arisoy2019modeling, li2019numerical, hong2021comparative},
	see \Cref{tab:PMat}. Note that, while we could have considered $h_p$ and $h_g$ as random variables themselves rather than their logarithm, this latter choice is more effective since it allows us to easily span a large range of values, giving at the same time equal importance in the numerical investigation to both the smaller end and the larger end of the interval where $h_p$ and $h_g$ live (specifically $[10^{-5},10^0]$); in other words, with this choice small values of $h_p$ and $h_g$ (say in the range from $10^{-5}$ to $10^{-1}$) are investigated as thoroughly as the larger values.
	
	The objective of the inverse UQ analysis that we will perform in this work is to compute a new PDF in which a subset of values of the parameters are \lq\lq more probable\rq\rq~than others, because they match better with the available data. In other words, we aim at incorporating the information from the data at disposal in the modeling of the uncertainty of the three parameters considered. 
	
	We collect the three uncertain parameters in a vector $\vv=(v_1,v_2,v_3) = (T_A, \log h_g, \log h_p)$ and introduce some notation that will be used in the following:
	\begin{itemize}
		\item $\Gamma_n = [a_n, b_n]$ is the range of each uncertain parameter $v_n$ with $n=1,2,3$;
		\item $\Gamma = \Gamma_1 \times \Gamma_2 \times \Gamma_3$ is the domain of $\mathbf{v}$, i.e., the hyper-rectangle
		$[a_1, b_1]  \times [a_2, b_2] \times [a_3, b_3]$;
		\item $\rho_n(v_n)$ is the PDF of $v_n$. Given the discussion above, we have $\rho_{n,prior}(v_n) = \frac{1}{b_n - a_n}$, where the subscript
		\lq\lq prior\rq\rq~denotes the fact that such PDF incorporates only the prior information in \Cref{tab:PMat}. After the inverse UQ procedure, these PDFs will
		be updated to a data-informed posterior PDF, indicated as $\rho_{n,post}$;
		\item $\rho(\mathbf{v})$ is the joint PDF of the vector $\mathbf{v}$. In particular, $\rho_{prior}(\mathbf{v})$ and $\rho_{post}(\vv)$
		are the joint prior and posterior PDFs of $\mathbf{v}$, respectively.
		Given the assumption that the three parameters are a-priori mutually independent,
		we have $\rho_{prior}(\vv) = \prod_{n=1}^3 \rho_{n,prior}(v_n)$; instead, we cannot assume at this stage
		that $\rho_{post}$ factorizes as $\rho_{post} = \prod_{n=1}^3 \rho_{n,post}(v_n)$,
		since we do not have (yet) information on the statistical independence of $v_n$ after the inversion;
		\item we will write $u(\xx,\vv)$ to denote the  displacement along the $z-$direction at $\xx \in \Omega$ corresponding to the value $\vv$ of the parameters;
		\item more generally, $\qoi(\vv): \Gamma \rightarrow \mathbb{R}^P$ denotes any QoI (output) of the simulation (displacements, residual strains)
		and emphasizes that such quantities are function of $\vv$. When $P=1$ (scalar-valued QoI), we use the notation $\qoiscal(\vv)$.
	\end{itemize} 
	
	\begin{table}[h!]
		\begin{center}
			\begin{tabular}{ccc}
				\hline
				\textbf{random variables}    & \textbf{units} & \textbf{range} $[a_n,b_n]$ \\ \hline 
				Activation temperature ($T_{A}$) &  $^\circ{C}$ & $[1130;1450]$  \\
				$\log$ of gas convection coefficient ($\log h_{g}$) &  $- $  & $[-5,0]$\\
				$\log$ of powder convection coefficient ($\log h_{p}$) &  $- $  & $ [-5,0]$\\
				\hline
			\end{tabular}
		\end{center}
		\caption{Parameter spaces of the PBF process simulation chosen for the inverse UQ analysis.}
		\label{tab:PMat}
	\end{table}
	
	\subsection{Uncertainty Quantification workplan}
	
	The fundamental premise of UQ is the observation that since $\vv$ is uncertain and described by a random vector with an associated PDF, any output $\qoi(\vv)$ is also an uncertain quantity. Computing efficiently the PDF of $\qoi(\vv)$ is the final goal of the UQ analysis; in particular, the QoI that we will ultimately consider in this work are the residual strains along the beam. To this end, we will proceed accordingly to the workplan detailed below, whose steps will be described in detail in the following subsections; note in particular that we will assume of having at disposal data (measurements) about another QoI of the model, namely, the displacements of the beam. Accordingly, the workplan is as follows.
	\begin{enumerate}
		\item GSA (see \Cref{subsec:GSA}):
		we investigate how much each uncertain parameter $v_n$ contributes
		to the variability of the displacements. 
		The finding of this analysis is that the second parameter, i.e., the gas convection coefficient, has little impact on the displacements and therefore can be
		fixed to a constant value in the subsequent analyses.
		This reduces the dimensionality of the problem and therefore the computational costs of the next two steps.
		In particular, the new PDF to be considered is now $\rho_{prior,red}(\vv) = \rho_1(v_1) \rho_3(v_3)$ and the reduced parameter space is
		$\Gamma_{red} = \Gamma_1 \times \Gamma_3$.
		\item Inverse UQ (see \Cref{subsec:IBayes}):
		by relying on Bayesian inversion techniques, we update the initial PDF $\rho_{prior,red}$ of the two remaining uncertain parameters,
		incorporating the information coming from available data on displacements. The result is a new PDF, $\rho_{post}$,
		tailored to the data at hand. Such new PDF has a \lq\lq reduced uncertainty\rq\rq, i.e., a smaller variance, compared to $\rho_{prior,red}$.
		\item Data-informed forward UQ (see \Cref{subsec:Forward UQ}): 
		we sample $\Gamma_{red}$ according to the PDF $\rho_{post}$ just derived and evaluate
		the corresponding residual strains. From these values, we finally compute the PDF of the residual strains. 
	\end{enumerate}

	\subsection{Speeding up uncertainty quantification by sparse-grid surrogate modeling}
	
	All the steps in the workplan above require repeatedly solving the PBF model for different values of $\vv$ to evaluate the corresponding QoIs $\qoi(\vv)$ (displacements, residual strains). 
	Specifically, in \emph{GSA} we need to assess how much changing the value of each $v_n$ impacts the value of the displacements. In the \emph{inverse UQ} we instead need to test the compatibility of each value of $\vv$ with the available displacement data: intuitively, if for certain values of $\vv$ the solution of the model is \lq\lq far\rq\rq~from the data, such values of the parameters are \lq\lq unlikely\rq\rq,
	and therefore the corresponding values of $\rho_{post}(\vv)$ must be small. Finally, in \emph{data-informed forward UQ} we need to obtain values of the residual strains to compute their PDF.
	
	To reduce the computational burden, in the following we replace the values of $\qoi(\vv)$ resulting from the evaluation of the PBF model with suitable approximations,
	obtained building so-called surrogate models of $\qoi(\vv)$. Surrogate models are typically obtained with a two-step procedure.
	In the first step (\lq\lq offline step\rq\rq / \lq\lq training step\rq\rq), we evaluate the PBF model for a handful of judiciously chosen values of $\vv$,
	say $\qoi(\vv_1),\ldots\qoi(\vv_M)$, and then create an approximation of $\qoi(\vv)$ out of these $M$ values.
        In the second step (\lq\lq online step\rq\rq / \lq\lq evaluation step\rq\rq),
        whenever a new value of $\qoi(\vv)$ is needed, we evaluate the cheap surrogate model
	instead of the expensive PBF model, with considerable computational savings.
	In particular, in this work we consider the so-called \emph{sparse-grid surrogate models}, that are among the most popular surrogate modeling technique in UQ, see
	\cite{babuska.nobile.eal:stochastic2,xiu.hesthaven:high,bungartz2004sparse}.
	In this section we explain the basics of sparse grids; more details are provided in \ref{Surrogate Model}.
	
	Before giving the definition of sparse-grid surrogate models, we point out that
	the sparse-grid approximation of a QoI $\qoi(\vv)$ depends (among other things) on the number of uncertain parameters, $N$, and on their PDF;
	therefore, we will actually need three different sparse-grids surrogate models: the first one for the GSA (3 parameters, $\rho_{prior}$);
	the second one for the inverse UQ analysis (2 parameters with the restricted $\rho_{prior,red}$),
	the third one for the data-informed forward UQ analysis (2 parameters with $\rho_{post}$). \Cref{tab:grids} summarizes the properties
	used for the three models; the meaning of the different entries will become clearer as we progress with the explanation.
	
	\begin{table}[tbh]
		\centering
		\begin{tabular}{cllllllcl}
			\hline
			\textbf{sparse grid}& \textbf{used in} & $N$& \textbf{PDF}  & $\mcI$ 	& $w$ & \textbf{univariate nodes} & \textbf{points} 	& \textbf{surrogate for}\\
			\hline 
			1			& GSA 		& $3$ 	& $\rho_{prior}$ & $\mcI_{max}$ 	& $1$ & symm. Leja 		& 27 				& 11 $z-$displacements\\
			2			& inverse UQ 	& $2$ 	& $\rho_{prior,red}$& $\mcI_{sum}$ & $3$ & symm. Leja 		& 25 				& 
			9 $z-$displacements\\ 
			3			& forward UQ 	& $2$ 	& $\rho_{post}$ 	& $\mcI_{sum}$ 	& $3$ & symm. Leja ($\log h_p$)	& 25 				& 120 $x-$residual strains \\ 
			& 	 	&  	& 	 	& 	 	&     & symm. Gaussian Leja ($T_A$)&  				&  \\ 
			\hline
		\end{tabular}
		\caption{Properties of the three sparse-grid surrogate models used in this work.}
		\label{tab:grids}
	\end{table}
	
	In general, the sparse-grid surrogate model of a scalar-valued $\qoiscal(\vv)$ 
	for $\vv \in \Gamma \subset \mathbb{R}^N$ with associated PDF $\rho(\vv)$ that we denote by $ \mcS_{\mcI} \qoiscal(\vv)$ reads:
	\[
	\qoiscal(\vv)  \approx \mcS_{\mcI} \qoiscal(\vv) 
	= \sum_{\ii \in \mcI} c_{\ii} \mcU_{\ii}(\vv), \quad c_{\ii}: = \sum_{\substack{\jj \in \{0,1\}^N \\ \ii+\jj \in \mcI}} (-1)^{\lVert \jj \rVert_1},
	\]
	where:
	\begin{itemize}
		\item $\ii = [i_1,i_2,\ldots,i_N] \in \mathbb{N}^N_+$ is a multi-index,
		i.e., a vector of $N$ integer positive numbers, $i_n \geq 1$; 
		\item $\mcI$ is a collection of multi-indices, called multi-index set, $\mcI \subset \mathbb{N}^N_+$.
		It must satisfy a so-called \emph{downward-closedness} condition, see \ref{Surrogate Model};
		\item $\mcU_{\ii}(\vv)$ is a tensor Lagrangian interpolant of $\qoiscal(\vv)$, built over a Cartesian grid on $\Gamma$
		with
		\begin{equation}\label{eq:how-many-points-in-a-tensor}
			(2 i_1 - 1) \times (2 i_2 - 1) \times \cdots \times (2 i_N - 1)  
		\end{equation}
		points. In other words, the $n$-th component of $\ii$
		specifies how many values should be used for $v_n$ when constructing the Cartesian grid on $\Gamma$;
		\item $c_{\ii}$ are the so-called \emph{combination technique coefficients}. 
		Note that some $c_{\ii}$ might be null, in which case $\mcU_{\ii}(\vv)$ is not part of the final approximation.
	\end{itemize}
	
	The sparse-grid surrogate model of $\qoiscal(\vv)$ thus consists of a linear combination of several
	tensor Lagrangian interpolants of $\qoiscal(\vv)$, each built over a different Cartesian grid covering the parameter space $\Gamma$.
	The specific tensor Lagrangian interpolants that form this approximation are dictated by a set $\mcI$; the choice of $\mcI$ is therefore
	pivotal for the construction of a good sparse-grid surrogate model. The easiest choice of $\mcI$ is certainly
	\[
	\mcI_{max} = \{ \ii \in \mathbb{N}^N_+ : \max_{n=1,\ldots, N} (i_n -1) \leq w \},  
	\]
	for some integer $w$. In this case, letting $\ii_w =[w{+}1,\,w{+}1,\,\ldots]$, it can be shown that $\mcS_{\mcI} \qoiscal(\vv) = \mcU_{\ii_w}(\vv)$, i.e.,
	the sparse grid reduces to a tensor Lagrangian interpolant based on $(2w-1)^N$ points (all $c_{\ii}$ are zero other than $c_{\ii_w}=1$).
	This choice is however unfeasible even for moderate
	values of $w$ or $N$, since it would require too many evaluations of $\qoiscal(\vv)$.
	A classic (and more reasonable) choice is choosing the set 
	\[
	\mcI_{sum} = \{ \ii \in \mathbb{N}^N_+ : \sum_{n=1}^N (i_n -1) \leq w \}, 
	\]
	again for some integer $w$. In this case, the Lagrange interpolants $\mcU_{\ii}(\vv)$ would require to sample extensively only some of the parameters (since $\ii$ would be such that if one component is large,
	the other ones are small); however, combining them by the coefficients $c_{\ii}$,
	we recover a good approximation of $\qoiscal(\vv)$, using significantly less samples than the previous $(2w-1)^N$.
	Using the set $\mathcal{I}_{sum}$ ultimately results in a sampling of the parameter space which is structured but not Cartesian;
	examples can be seen in \Cref{subsec:SFSM for I} and \Cref{subsec:forwardUQ}.
	In the following we will use the set $\mcI_{max}$ for the GSA, and $\mcI_{sum}$ for the inverse and data-informed forward UQ analysis.
	
\begin{figure}[h!]
		\centering
		\, \subfigure[PARAMETRI-1][{$\ii = [1,\,3]$}, $c_{\ii} = -1$]{\label{fig:Sparsea}\includegraphics[width=0.32\textwidth]{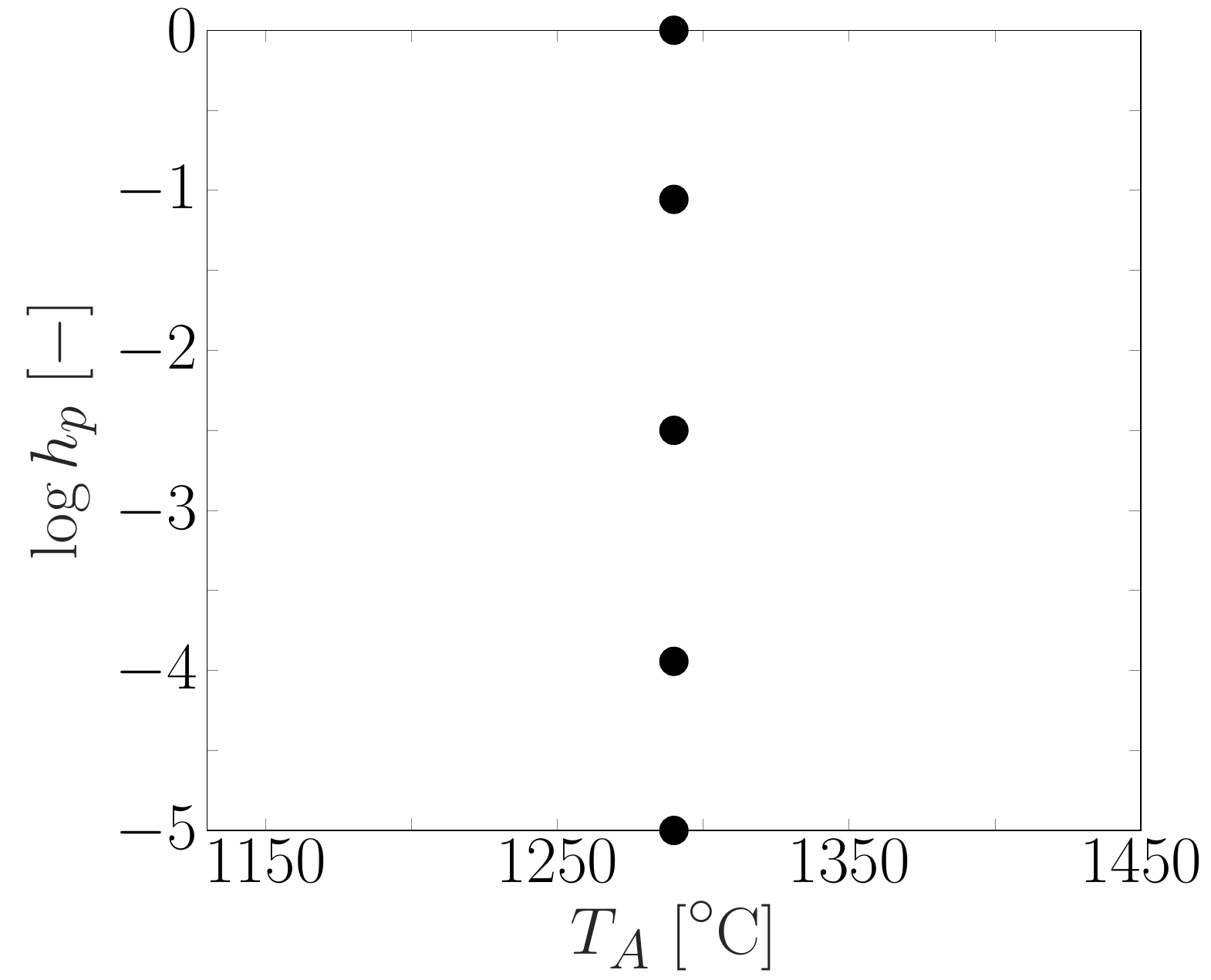}}
		\subfigure[PARAMETRI-2][{$\ii = [1,\,4]$}, $c_{\ii} =1$]{\label{fig:Sparseb}\includegraphics[width=0.32\textwidth]{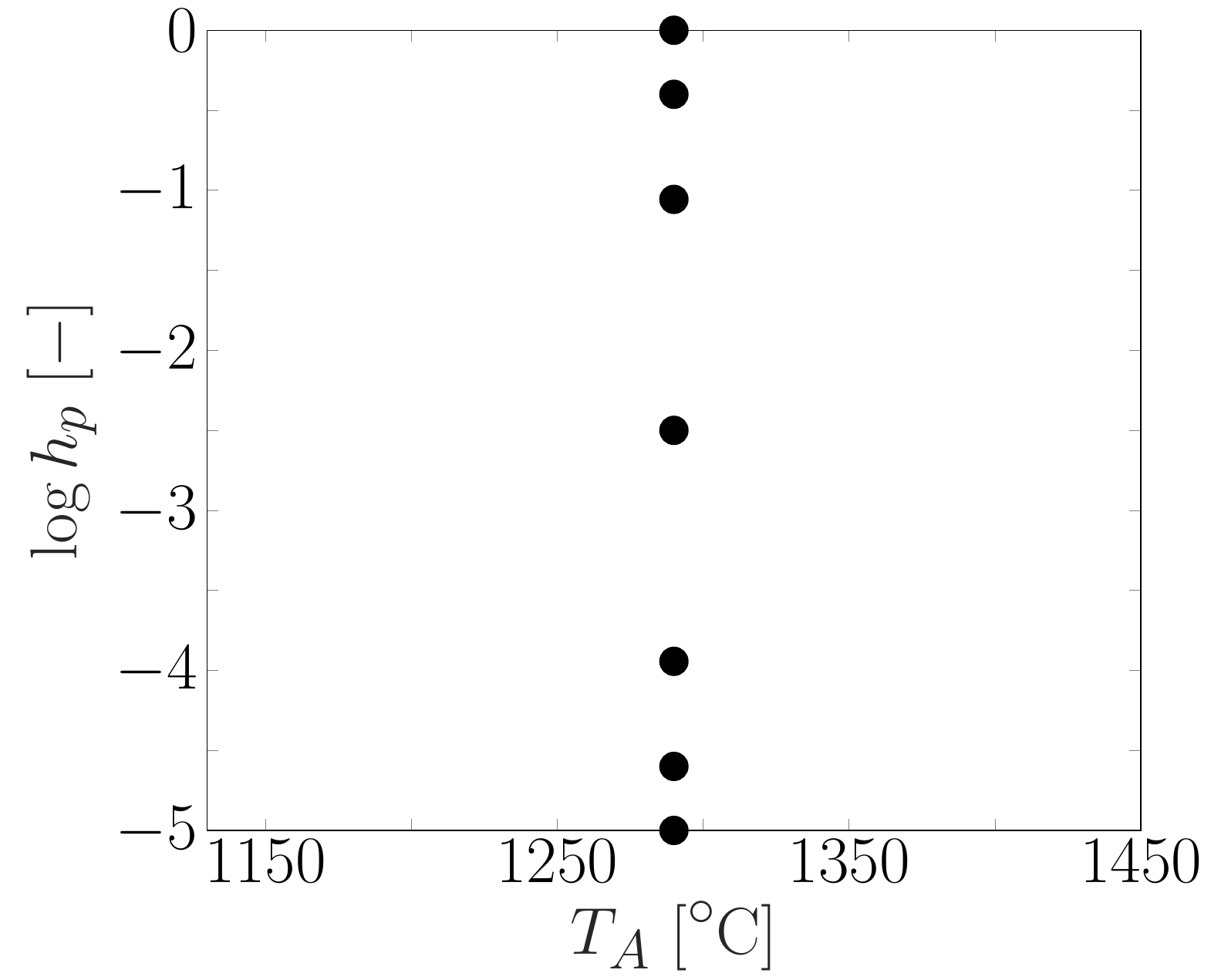}} 
		\subfigure[PARAMETRI-3][{$\ii = [2,\,2]$}, $c_{\ii} = -1$]{\label{fig:Sparsec}\includegraphics[width=0.32\textwidth]{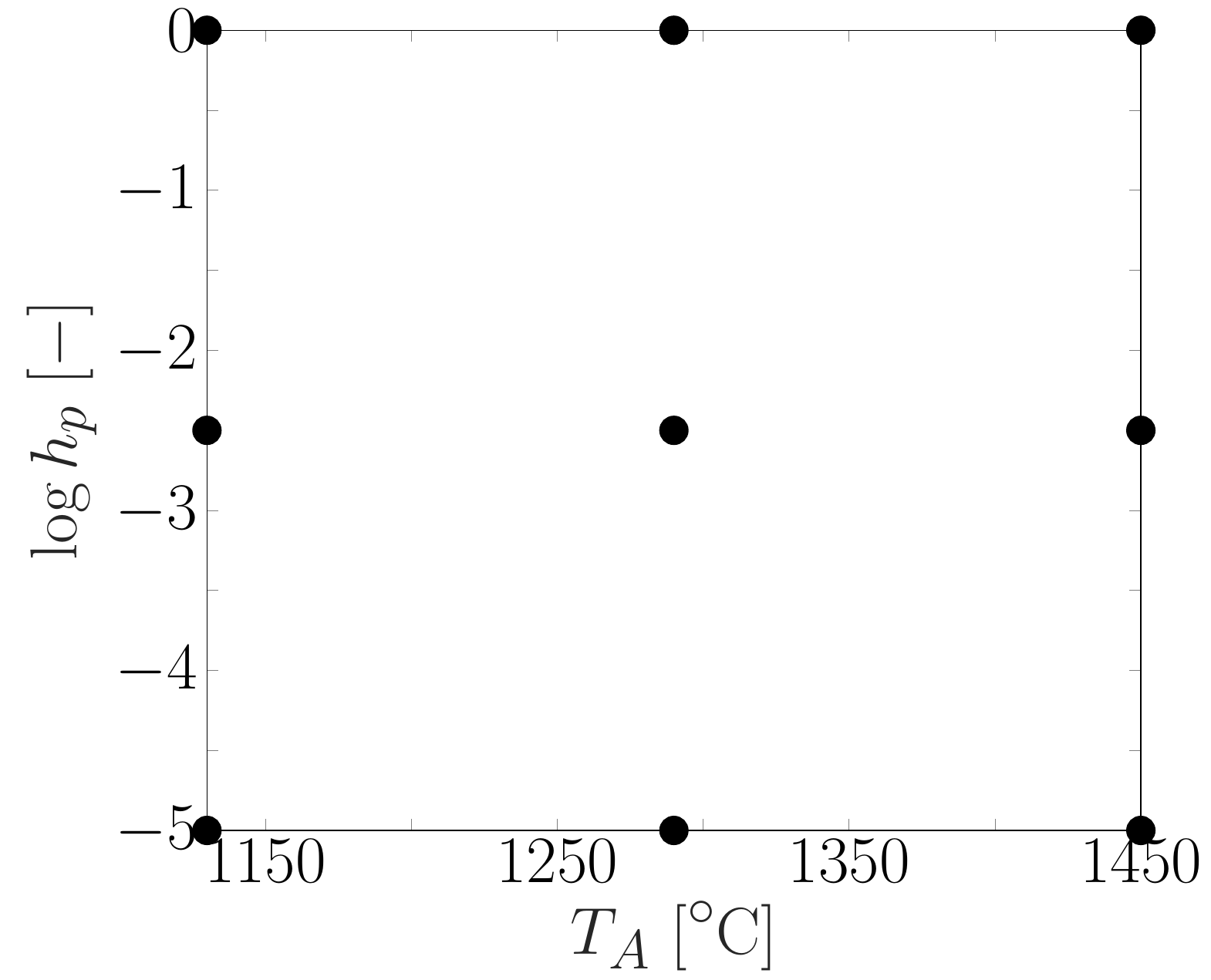}} \\
		\subfigure[PARAMETRI-4][{$\ii = [2,\,3]$}, $c_{\ii} = 1$]{\label{fig:Sparsed}\includegraphics[width=0.32\textwidth]{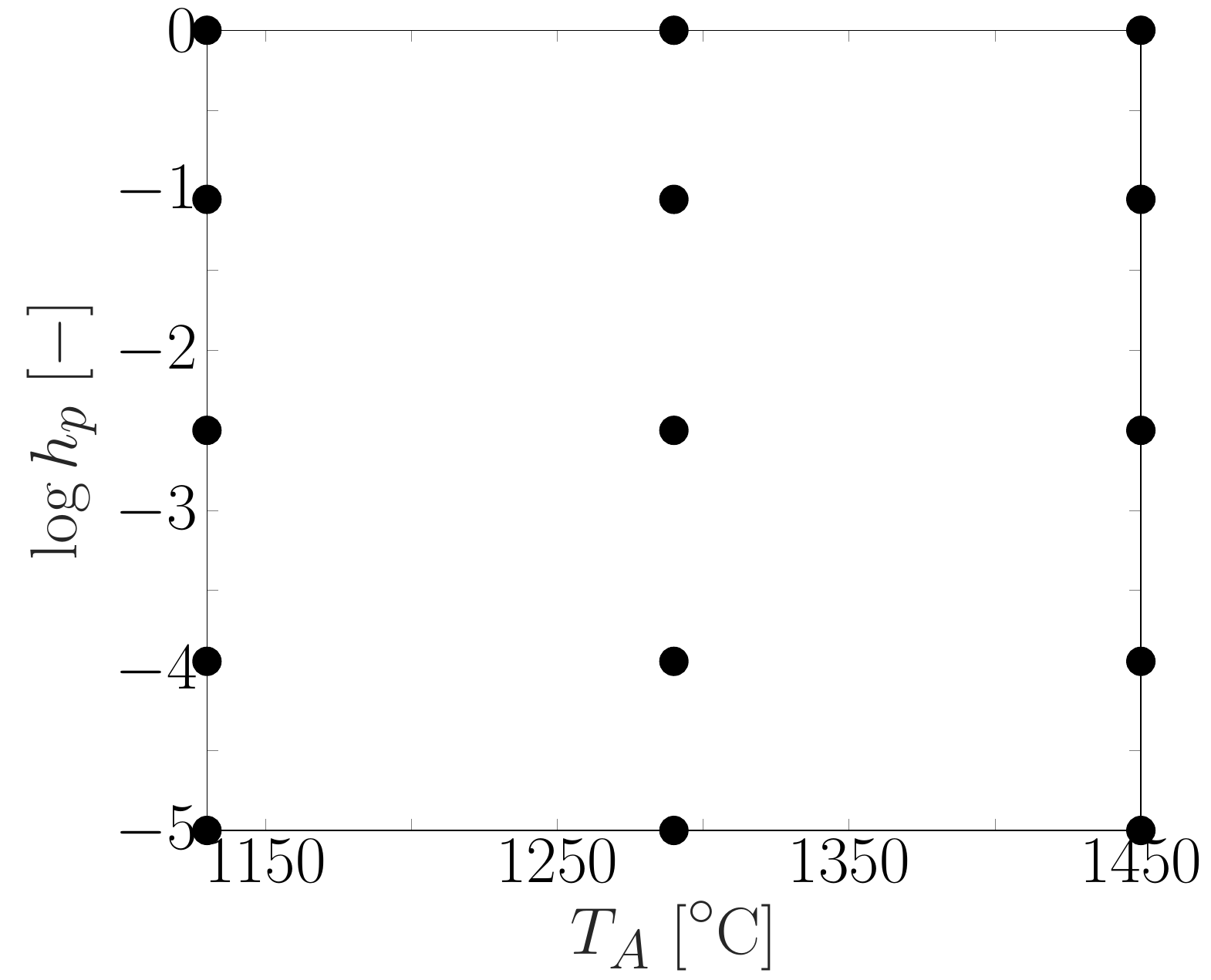}} 
		\subfigure[PARAMETRI-5][{$\ii = [3,\,1]$}, $c_{\ii} = -1$]{\label{fig:Sparsee}\includegraphics[width=0.32\textwidth]{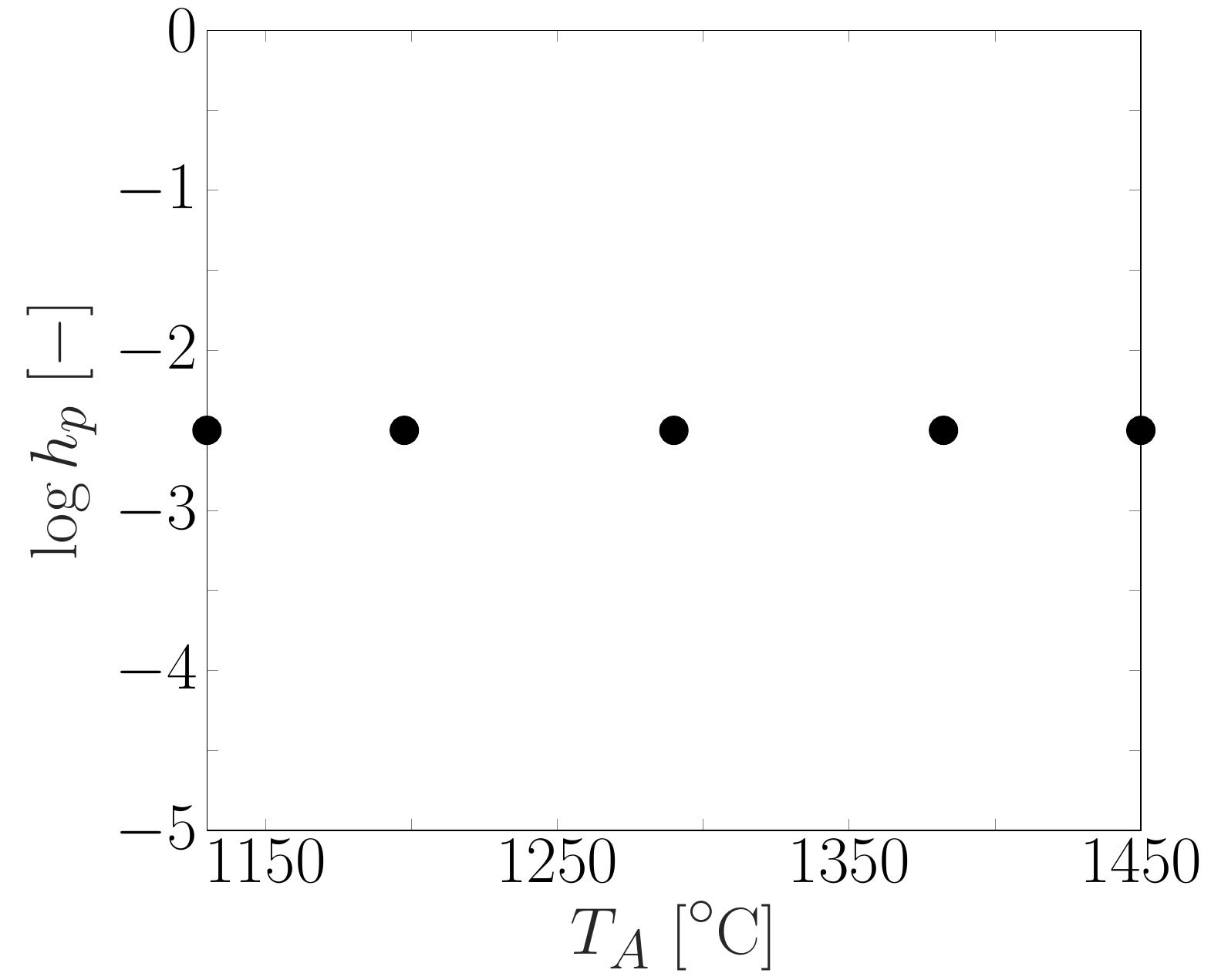}} 
		\subfigure[PARAMETRI-6][{$\ii = [3,\,2]$}, $c_{\ii} = 1$]{\label{fig:Sparsef}\includegraphics[width=0.32\textwidth]{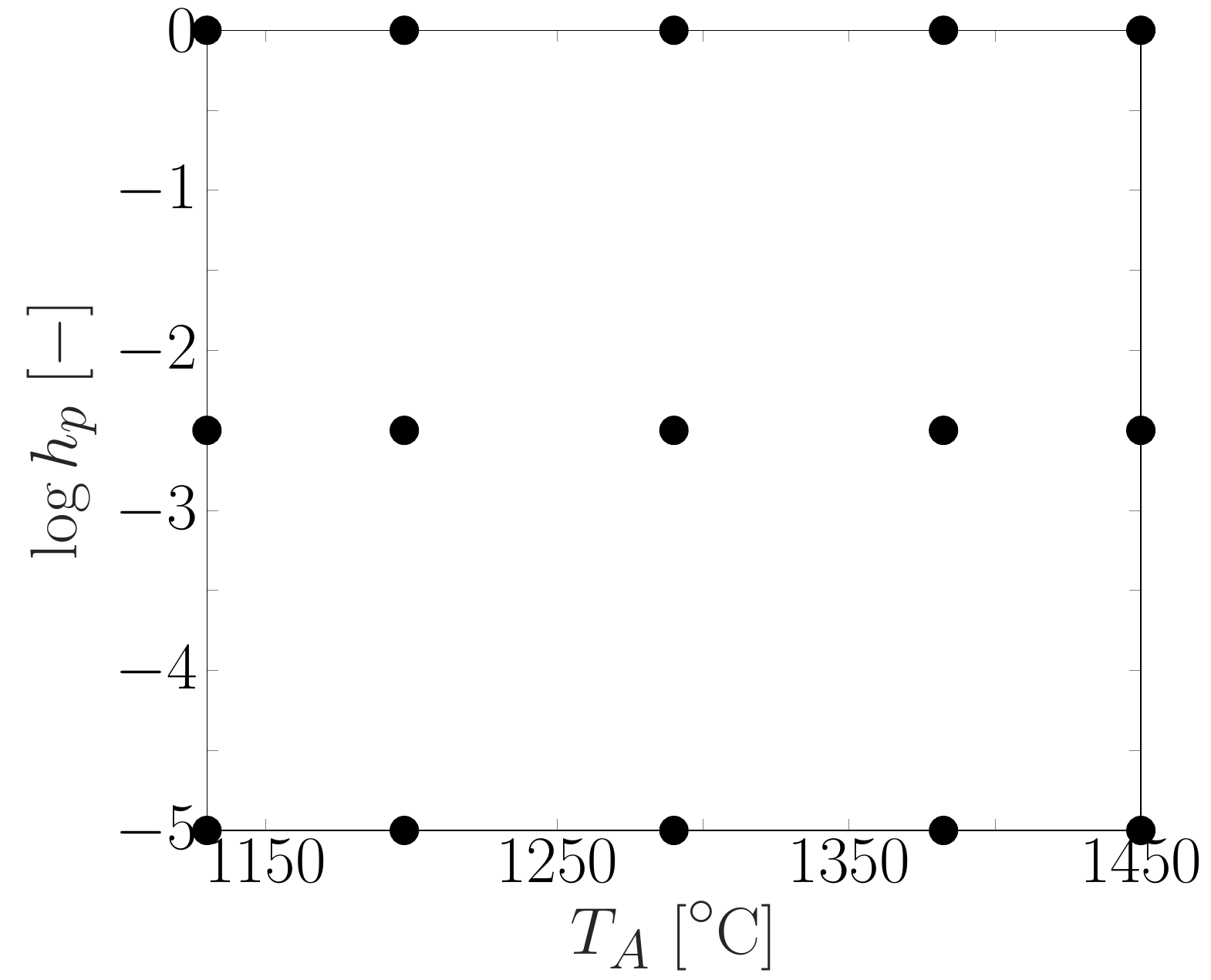}} \\
		\subfigure[PARAMETRI-7][{$\ii = [4,\,1]$}, $c_{\ii} = 1$]{\label{fig:Sparseg}\includegraphics[width=0.32\textwidth]{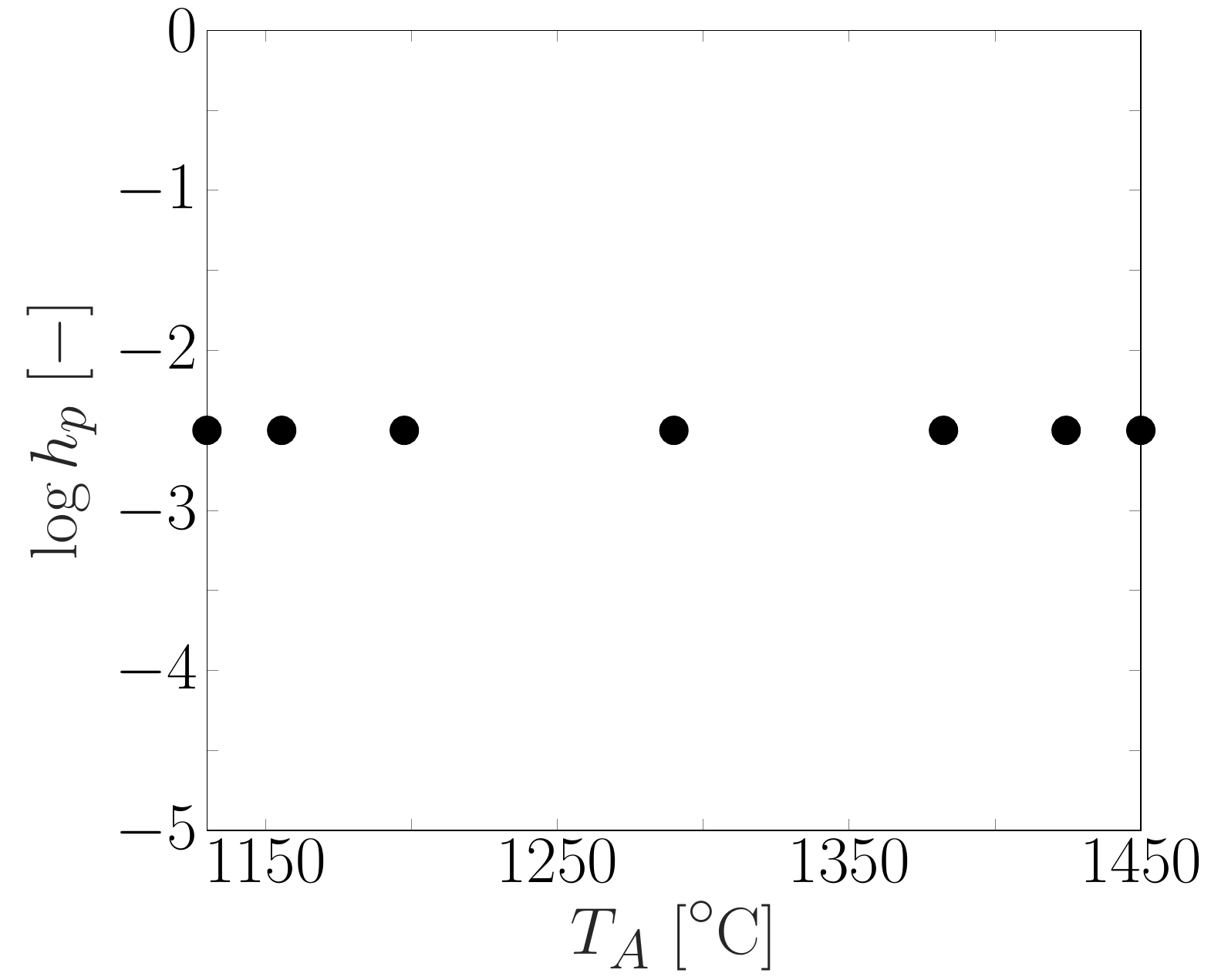}} 
		\subfigure[PARAMETRI-8][Final sparse grid]{\label{fig:Sparseh}\includegraphics[width=0.32\textwidth]{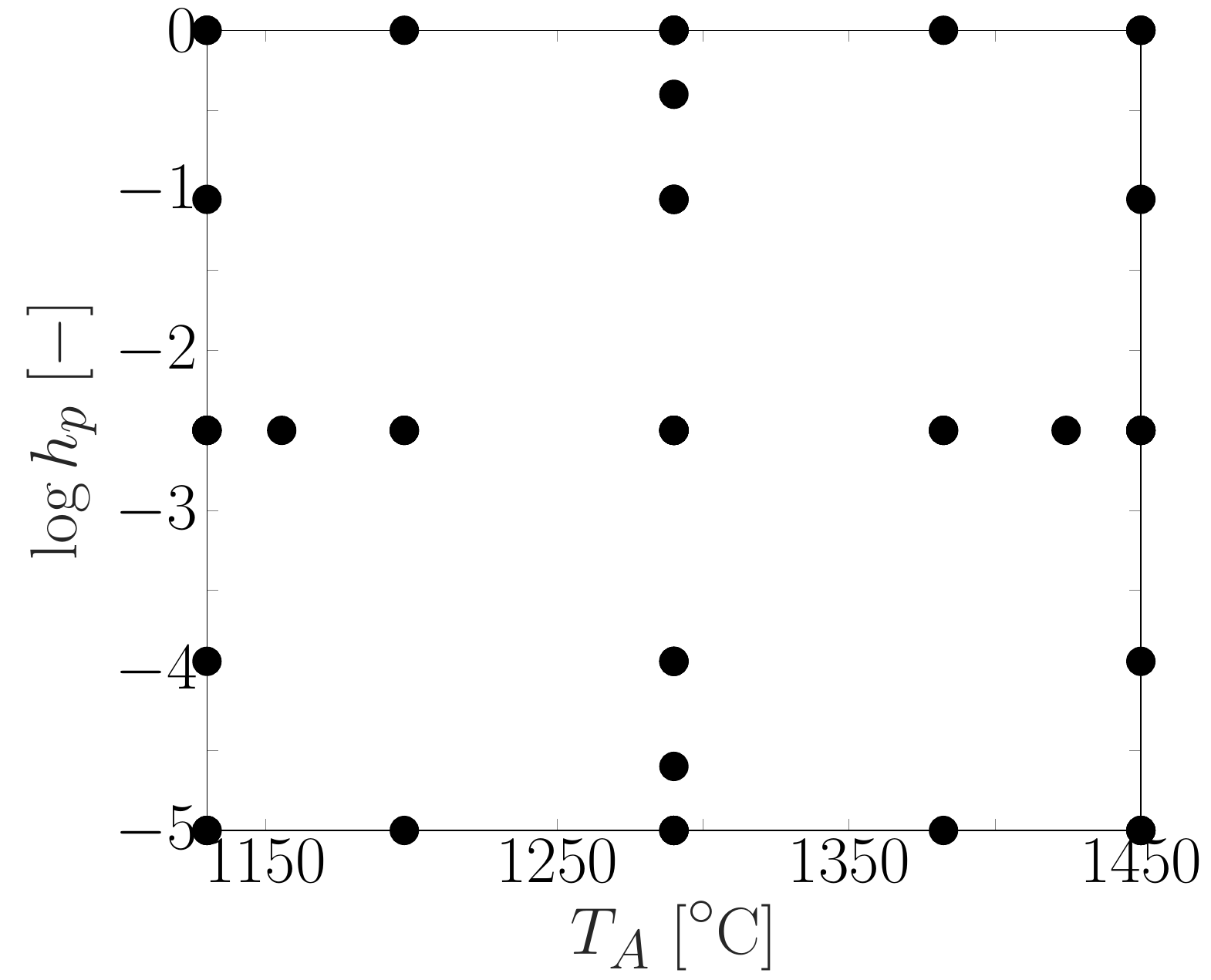}}
		\subfigure[PARAMETRI-9][\s\s\s\s]{\label{fig:LG}\includegraphics[width=0.31\textwidth]{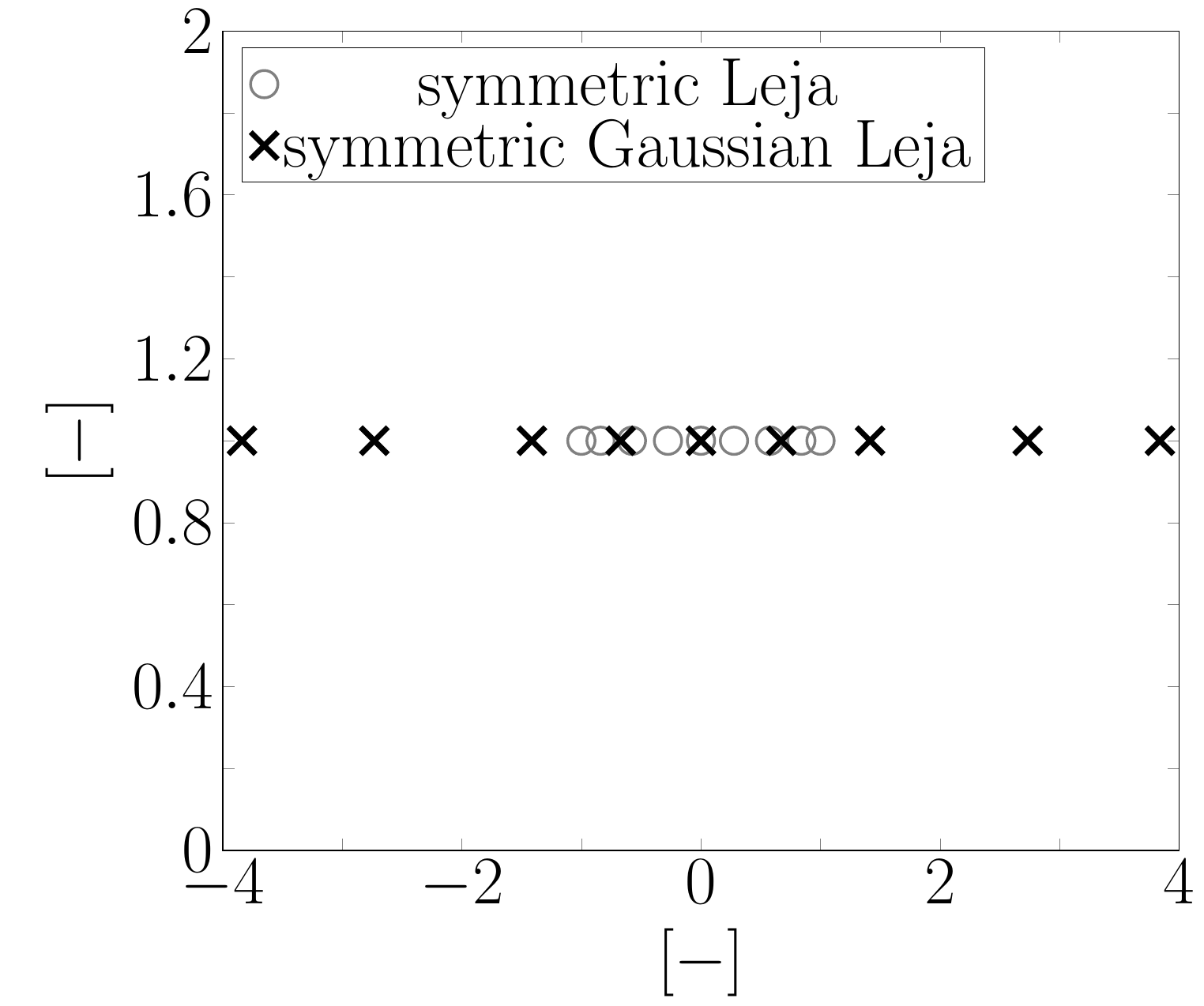}}
		\caption{\rev{Breakdown of the sparse grid (h) into the tensor grids composing it (a)--(g) and the different families of collocation points used in this work, i.e.,
				symmetric Leja points plotted here for a uniform PDF in $\Gamma_n=[-1;1]$
				and symmetric Gaussian Leja points plotted here for a Gaussian PDF with zero mean and unit standard deviation (i)}.}
		\label{fig:SGbreakdown}
	\end{figure}

	The set of all points where an evaluation of $\qoiscal(\vv)$ is required to build the sparse-grid surrogate model
	(i.e., the union of all the Cartesian grids of the $\mcU_{\ii}(\vv)$ with $c_{\ii} \neq 0$) is actually called \emph{sparse grid}.
	\Cref{fig:SGbreakdown} shows an example of sparse grid over the space $\Gamma_{red}$: in particular,
	the final sparse grid is reported in \Cref{fig:Sparseh}, while
	\Cref{fig:Sparsea,fig:Sparseb,fig:Sparsec,fig:Sparsed,fig:Sparsee,fig:Sparsef,fig:Sparseg}
	show the breakdown of the sparse grid in \Cref{fig:Sparseh} into the tensor grids composing it.
	
	The final step is to specify how to choose the univariate collocation points over the ranges $\Gamma_n$, i.e., the values of $v_n$ to be used to generate the Cartesian grids over which $\mcU_{\ii}(\vv)$ are built
	(note that the multi-index $\ii$ specifies \emph{how many} points are required for each $v_n$ but not their \emph{location}).
	Since $\mcU_{\ii}(\vv)$ are Lagrangian interpolants, using equispaced points is not a good idea, due to
	the well-known Runge phenomenon \cite{quarteroni2010numerical};
	moreover, the points for each $v_n$ should be chosen according to the corresponding PDF for efficiency reason
	(it is recommended to put more points in regions of high probability).
	In view of this, we use different families of collocation points for each of the three sparse grids
	that we use in this work, see \Cref{tab:grids}. Specifically, the sparse grids for the GSA and for the inverse UQ need to be built according
	to the fact that $\rho_{prior}$ and $\rho_{prior,red}$ consist of uniform PDF, for which we elect to 
	use the so-called symmetric Leja points (see \ref{sec:Leja_points} for details on their definitions). Instead, for the final forward UQ based on $\rho_{post}$, we use symmetric Gaussian Leja points for $v_1$ (i.e., $T_A$; see again \ref{sec:Leja_points})
	and symmetric Leja points for $v_3$ (i.e., $\log h_p$). As detailed in \ref{sec:Leja_points}, both these variants of Leja points are non-equispaced points, obtained 
	minimizing the corresponding Lebesgue constant, and thus are good for interpolation and hence surrogate modeling. \Cref{fig:LG} shows a set of 9 symmetric Leja points and 9 symmetric Gaussian Leja points. 
	
	Finally, we conclude this overview of sparse grids  pointing out that extending the sparse-grid surrogate model construction to vector-valued
	QoIs $\qoi(\vv)$ is straightforward. In fact, we need to apply the same procedure to each component of $\qoi(\vv)$.
	In particular, this means that the same set of collocation points (i.e., the same sparse grid) can be used to compute
	the sparse-grid surrogate model of each component of a vector-valued $\qoi$.
	
	Now that we have presented the framework of the UQ workflow and the basics of sparse grids, we are ready to explain how each of the three steps of the workplan is performed.
	
	\subsection{Global Sensitivity Analysis}
	\label{subsec:GSA}
	In the present work, we use the Sobol decomposition method \cite{sobol2001global,saltelli2004sensitivity,saltelli2008global,sudret:sobol,iooss2015review}
	to assess the influence of the three uncertain parameters on the displacements.
	The method is similar to the classical ANOVA decomposition of the variance and consists of computing two sets of
	indicators, namely the principal and total Sobol indices.
	The first ones measure the \emph{individual} contribution of each parameter to the variance of the QoI,
	while the second ones quantify the contributions of each parameter \emph{combined with the others}.
	
	As already mentioned, the first step to perform the GSA  for the displacements is to build a sparse-grids surrogate model for them.
	We use the specifics listed in the first row of \Cref{tab:grids}, which result in a Cartesian sampling of $\Gamma$ with
	$3 \times 3 \times 3 = 27$ symmetric Leja point, meaning that we need to run $27$ PBF full model simulations. Note that 3 symmetric Leja points over an interval $\Gamma_n =[a_n, b_n]$ means in practice considering the following three values
	for the parameter $v_n$: $v_n = a_n, b_n, (a_n+b_n)/2$ (see \ref{sec:Leja_points}).
	Once having the sparse grid at hand, deriving the Sobol indices is a relatively easy but quite technical operation, see \cite{feal:compgeo} for details.
	
	The results discussed in \Cref{subsec:GSAres} show that the principal and total Sobol indices for $\log h_g$ are small compared to the others
	and thus $\log h_g$ can be neglected in the following
	steps of the UQ workplan; in practice, this means that from now on, $\log h_g$ can be fixed to a convenient value.
	As a consequence for the inverse and forward UQ analysis we consider a reduced parameter space $\Gamma_{red} = \Gamma_1 \times \Gamma_3$ and
	the new PDF $\rho_{prior,red}(\vv) = \rho_1(v_1) \rho_3(v_3)$. Note that with a slight abuse of notation, in the following, $\vv$ denotes also the reduced
	vector $\vv=(T_A,\log h_p)$; the context will always make clear whether $\vv$ denotes the reduced $\vv=(T_A,\log h_p)$ or the original
	$\vv=(T_A,\log h_g,\log h_p)$.

	\subsection{Inverse UQ analysis}
	\label{subsec:IBayes}
	
	In the present work, we adopt a Bayesian inversion approach \cite{berger2013statistical,stuart:acta.bayesian,thanh-bui.gattas:MCMC} to
	perform the inverse UQ analysis.
	More specifically, Bayesian inversion consists in updating the PDF of the two remaining uncertain parameters $T_A,\log h_p$ from the uniform $\rho_{prior,red}$
	to a new posterior PDF $\rho_{post}$, that incorporates
	the fact that we have at disposal a set of measurements of displacements of the beam at $K$ positions $\xx_{1,meas},\ldots,\xx_{K,meas}$
	(see \Cref{subsec:SFSM for I} for details on the locations of $\xx_{k,meas}$).
 	
	Ideally, we would like to use actual experimental measures of displacements in the inverse UQ analysis.
	However, as a preliminary step towards future work, in the present manuscript
	we consider instead a set of imperfect (noisy) synthetic data $\tilde{u}$
	obtained by first running the part-scale thermomechanical model for a set of parameters of our choice denoted with $\bar{\vv}$ (\emph{target values}),
	and then adding to such displacement field a set of $K$ independent Gaussian noises to mimic measurement error, as follows:
	\begin{equation}
		\begin{cases}
			\label{eq:errsig}
			\tilde{u}_{k}=u(\xx_{k,meas},{\bar{\mathbf v}})+\mathbf \varepsilon_{k}, \quad k=1,\ldots,K. \\
			\varepsilon_{k} \sim \mathcal{N}(0,\sigma_{{{\mathbf \varepsilon_{k}}}}^{2}).
		\end{cases}
	\end{equation}
	This allows us to focus on the methodological aspect of the inversion, removing from the analysis any error due to the inadequacy
	of the computational model to represent reality.
	Furthermore, for simplicity, in the following 	$\sigma_{\varepsilon_{k}}$ is assumed to be constant, i.e.,\ $\sigma_{\varepsilon_{k}} = \bar{\sigma}\,\forall\,k$.
	
	To begin with the Bayesian inversion, we introduce the misfits ${M_{k}}(\mathbf v)$ between the synthetic data
	and the displacements predicted by the model when the parameters have value $\vv$, i.e.:
	\begin{equation}
		\label{eq:misfit}
		{M_{k}}(\vv):= \tilde{u}_{k}-u(\xx_{k,meas},\vv), \quad k=1,\dots,K.
	\end{equation}
	The posterior PDF $\rho_{post}$ is calculated using Bayes' theorem \cite{berger2013statistical,stuart:acta.bayesian} as follows:
	\begin{equation}
		\label{eq:rhopost}
		\rho_{post}(\mathbf v) = \mathcal{L}(\vv \,|\, \tilde{u}_{k}) \ \rho_{prior}(\mathbf v) \ \frac{1}{C},
	\end{equation}
	where $C$ is a normalization constant that makes $\rho_{post}(\mathbf v)$ actually a PDF (i.e., its integral equal to 1)
	and $\mathcal{L}(\vv \,|\, \tilde{u}_{k})$ is the so-called \emph{likelihood function},
	which quantifies the plausibility (\lq\lq likelihood\rq\rq) of $\vv$ given the displacement data, i.e.,
	the plausibility that the measured displacements were generated by $\vv$ rather than by the actual $\bar{\vv}$ (that in the general scenario would be unknown). As we are assuming independent Gaussian distributions for the measurement noises, see \Cref{eq:errsig},
	the likelihood $\mathcal{L}(\vv\,|\, \tilde{u}_{k})$ is 
	\begin{equation}
		\label{eq:LKhgh}
		\mathcal{L}(\vv \,|\, \tilde{u}_{k})=\prod_{k=1}^{K}\frac{1}{\sqrt{2\pi \sigma_{{{ {\varepsilon_{k}}}}}^{2}}}e^{-\frac{{{M_{k}}^{2}}(\mathbf v)}{2 \sigma_{{{ {\varepsilon_{k}}}}}^{2}}},
	\end{equation}
	i.e., the joint probability of observing the misfits $M_1,\ldots,M_K$ corresponding to the value $\vv$ of the parameters. 
	
	We now recall that the final goal of the UQ analysis (step 3 of the UQ workflow)
	is to sample extensively $\Gamma_{red}$ according to $\rho_{post}$, to obtain the PDF of the residual strains given the data (data-informed forward UQ).
	Sampling $\rho_{post}$ as in \Cref{eq:rhopost} can be done by  means of Markov-Chain Monte Carlo (MCMC) methods \cite{brooks2011handbook,marjoram2003markov},
	which can be however quite computational intensive.
	A significant reduction in computational costs can be obtained upon assuming that ${\rho_{post}}$ is well-approximated by a Gaussian PDF \cite{thanh-bui.gattas:MCMC},
	with appropriate mean vector and covariance matrix (we will discuss the validity of this assumption later), which is standard to sample from. 
	We therefore devote the next subsection to detailing how to compute the mean and covariance matrix of the Gaussian approximation of $\rho_{post}$: this is where the second sparse grid in
	\Cref{tab:grids} comes in handy.
	
	\subsubsection{Gaussian approximation of $\rho_{post}$}
	
	The mean of the Gaussian approximation can be taken as the maximum of the posterior PDF, $\mathbf v_{MAP}$ (Maximum A Posteriori), as follows: 
	\begin{equation}
		\label{eq:ymapgg}
		\mathbf v_{MAP} :=  \argmax_{\vv \in \Gamma_{red}} \rho_{post} = \argmax_{\vv \in \Gamma_{red}} \mathcal{L}(\vv \,|\, \tilde{u}_{k}),
	\end{equation}
	where the second equality is due to the fact that $\rho_{prior}$ and C are constants.
	It is easy to see that this maximization is in practice equivalent to the classical least-squares
	approach for the calibration of $\vv$, i.e., the minimization of the sum of squared errors $LS$:
	\begin{equation}
		\label{eq:ymap}
		\mathbf v_{MAP} =  \argmin_{\vv\in \Gamma_{red}} [-\log \mathcal{L}(\vv \,|\, \tilde{u}_{k})]=  \argmin_{\mathbf v\in \Gamma_{red}} LS(\mathbf v),
	\end{equation} 
	\begin{equation}
		\label{eq:LKs}
		\; LS(\mathbf v) =\sum_{k=1}^{K}{{{M_{k}}^{2}}(\mathbf v)}.
	\end{equation}
	The functional $[-\log \mathcal{L}(\mathbf v\,|\, \tilde{u}_{k})]$ appearing in the first equality of
	\Cref{eq:ymap} is known in the Bayesian literature as \emph{negative log-likelihood} functional.      
	Note that the minimization in \Cref{eq:ymap}
	requires evaluating multiple times the functional $LS(\mathbf{v})$,
	and thus running multiple times the PBF model to obtain the displacements $ u(\xx_{k,meas},\mathbf v)$ for different values of $\mathbf{v}$.
	To reduce this cost, we replace $u(\xx_{k,meas},\vv)$ by the
	sparse-grids surrogate model detailed in the second row of \Cref{tab:grids}, i.e., we modify the definition of the misfits as
	\[
	{M_{k}}(\vv):= \tilde{u}_{k}  -\mathcal{S}_{\mathcal{I}_{sum}} u(\xx_{k,meas},\mathbf{v}), \quad k=1,\dots,K. 
	\]
	As reported in \Cref{tab:grids}, we need 25 PBF evaluations to build this new surrogate model,
	which is much less than the number of evaluations of the model requested by the optimization procedure.
	Of course, replacing the exact $ u(\xx_{k,meas},\mathbf v)$ with its sparse-grid surrogate
	$\mathcal{S}_{\mathcal{I}_{sum}} {u}(\xx_{k,meas},\mathbf v)$ introduces an error, which needs to be small enough:
	in \Cref{subsec:SFSM for I} we present a numerical procedure to check that this is actually the case (surrogate model validation).
	
	Once $\vv_{MAP}$ has been computed, the final step is to derive the covariance matrix of the Gaussian approximation of $\rho_{post}$. Such a matrix can be computed as:
	\begin{equation}
		\label{eq:covariance}
		{\boldsymbol \Sigma_{post}} = {\bar \sigma_{MAP}^{2}} \left( J_u^T J_u  + \sum_{k=1}^K M_k H_{\tilde{u} _k} \right)^{-1},
	\end{equation}
	where:
	\begin{itemize}
		\item ${\bar \sigma_{MAP}^{2}}$ is an approximation of $\bar{\sigma}^2$ (typically unknown for actual experimental data),
		that we can obtain using the standard sample variance estimator:
		\begin{equation}
			\label{eq:sisi}
			\bar{\sigma}^2 \approx \bar{\sigma}_{MAP}^{2} =
			\frac{1}{K} \sum_{k=1}^K \Big( \tilde{u}_{k}  -u(\xx_{k,meas},\vv_{MAP})  \Big)^2
			= \frac{1}{K}LS\left(\vv_{MAP}\right);
		\end{equation}
		\item $J_u$ is the Jacobian matrix with respect to $\vv$ of the displacements at the measuring location evaluated at $\vv_{MAP}$, i.e., the following $K \times 2$ matrix: 
		\[
		J_u = \left[
		\begin{array}{l}
			\nabla_{\vv}^T u\left(\xx_{1,meas},\vv_{MAP}\right) \\
			\nabla_{\vv}^T u\left(\xx_{2,meas},\vv_{MAP}\right) \\
			\cdots \\
			\nabla_{\vv}^T u\left(\xx_{K,meas},\vv_{MAP}\right)
		\end{array}
		\right];
		\]
		\item $H_{\tilde{u}_k}$ is the Hessian of $u(\xx_{k,meas},\vv)$ with respect to $\vv$, evaluated at $\vv_{MAP}$ (i.e., a $2 \times 2$ matrix).
	\end{itemize}
	Also for these computations, it is helpful to use the sparse-grid surrogate model for $u(\xx_{k,meas},\vv)$, thanks to which
	we can cheaply compute finite difference approximations of the first and second partial derivatives of $u(\xx_{k,meas},\vv)$ with respect to $v_n$.

	\subsubsection{A mixed Gaussian-uniform approximation of $\rho_{post}$}\label{sec:mixed_approx_rho_post}
	
	As we have already mentioned, the validity of the Gaussian approximation of $\rho_{post}$ must be checked. Since in this work
	we consider only two parameters in the inverse UQ procedure ($T_A$ and $\log h_p$), this check can be easily done by plotting the isolines
	of the posterior PDF (or equivalently of the negative log-likelihood functional or of the least squares functional), and verifying that such isolines
	are shaped as ellipses in the proximity of $\vv_{MAP}$. In our case, this is unfortunately not true: in \Cref{fig:YTTTT},
        we can see that the isolines form a band/strip surrounding $\vv_{MAP}$.
	This suggests that a Gaussian approximation of $\rho_{post}$ is not appropriate;
	more numerical evidence and discussion about this fact is provided in \Cref{sec:inversion}.
	The shape of the isolines actually suggests a mixed approach, in which the posterior PDF of $T_A$ is taken as
	Gaussian and the posterior PDF for $\log h_p$ as uniform on a smaller interval than the prior one.
	Furthermore, the fact that the isolines are parallel to the axis of $\log h_p$ suggests
	that the two parameters can be still considered as statistically independent, implying that $\rho_{post}$
	can be finally taken as the product of the two new PDFs for $T_A$ and $\log h_p$. 
	More in detail, the mean of the  Gaussian PDF for $T_A$ can be taken as the first entry of $\vv_{MAP}$
	and the variance can still be taken as the element $(1,1)$ of the covariance matrix
	${\boldsymbol \Sigma_{post}}$ presented above (that is hopefully smaller than the variance in the prior distribution),
	while the new interval for $\log h_p$ can be chosen by heuristic considerations
	that we will present in \Cref{sec:inversion}.

	\subsection{Data-informed forward UQ analysis}
	\label{subsec:Forward UQ}
	
	The final goal of the present work is to perform a forward UQ analysis based on the (data-informed) posterior PDF of the parameters,
	to quantify the uncertainty in the prediction of the residual strains of the beam given the uncertainty on the
	parameters $T_A$ and $\log h_p$, now modeled by $\rho_{post}$.
	More precisely, our aim is to approximate the PDF of the residual strains at $L$ locations $\xx_{j,str}$
	along the x-direction (${\varepsilon}_{xx}(\xx_{j,str},\vv)$) of the beam 
	(see \Cref{subsec:forwardUQ} for details on the locations of $\xx_{j,str}$).
	
	To this end, we first generate a sparse-grid surrogate model for the residual strains
	$\mcS_{\mcI_{sum}}{\varepsilon}_{xx}(\xx_{j,str},\vv)$ at each of the $L$ locations,
	using the specifics listed in the third row of \Cref{tab:grids}. We then
	check that the accuracy with which such sparse-grid surrogate models approximate the full model residual strains ${\varepsilon}_{xx}(\xx_{j,str},\vv)$
	is enough for our purposes (see \Cref{subsec:SFSM_forward}). To do this,  we generate 10000 samples of $\vv$ according to $\rho_{post}$,
	and for each of these values we approximate the residual strains ${\varepsilon}_{xx}(\xx_{j,str},\vv)$ by evaluating the sparse-grid surrogates
	$\mcS_{\mcI_{sum}}{\varepsilon}_{xx}(\xx_{j,str},\vv)$. Finally, we approximate the residual strain PDF at each location
	by applying a kernel density estimate method \cite{parzen:kde,rosenblatt:kde} to the 10000 residual strain values obtained by the surrogate models at each location.
	
\section{Results and discussion}
\label{sec:Results and Discussion}
\rev{In the present section, we report the numerical results obtained for the previously described UQ workflow. In particular, we discuss results of the GSA, the inverse UQ analysis, and the data-informed forward UQ analysis presented in the \Cref{sec:UQ analysis}}. All the simulations of the PBF process discussed in this section are obtained on an HPC server equipped with a CPU with 128 AMD EPYC 7702@1.67 GHz cores and 376 GB RAM. The UQ analyses are implemented in Matlab, relying on the Sparse-Grids Matlab-Kit \cite{piazzola2022sparse}.
	
	\subsection{Global sensitivity analysis}
	
	\label{subsec:GSAres}
	
\begin{figure}[h!]
		\centering
		\subfigure[PARAMETRI-1][\s\s\s\s\s\s\s\s\s\s\s\s\s\s\s\s\s ]{\label{fig:SP}\includegraphics[width=0.4\textwidth]{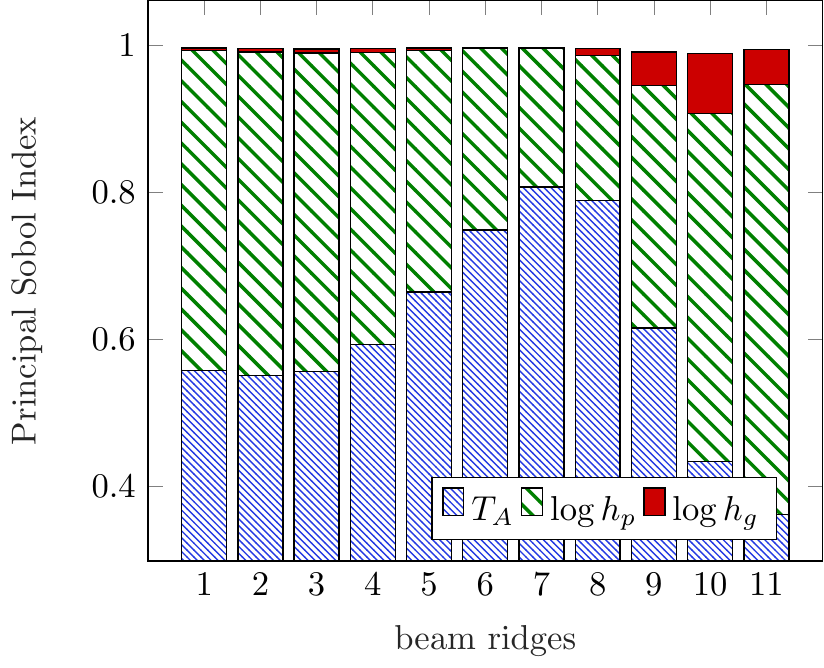}} \hspace{1.5em}
		\subfigure[PARAMETRI-2][\s\s\s\s\s\s\s\s\s\s\s\s\s\s\s\s\s]{\label{fig:ST}\includegraphics[width=0.4\textwidth]{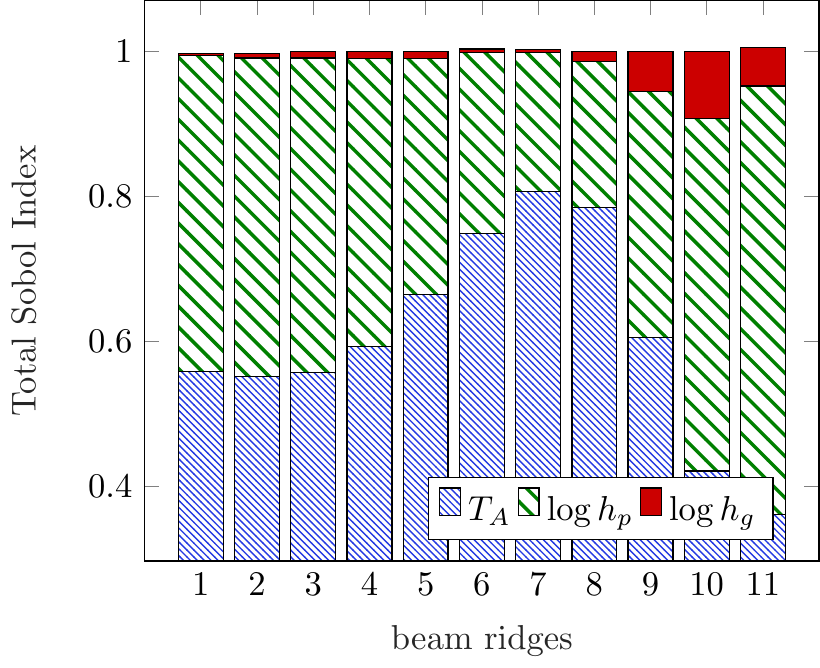}}
		\caption{\rev{Magnitude of Sobol indices
			based on the displacements evaluated at the 11 ridges of the beam (QoI) for the three-parameter model: (a) Principal Sobol Indices that measure the \emph{individual} contribution of each parameter to the variance of the QoI; (b) Total Sobol Indices that quantify the contributions of each parameter \emph{combined with the others}}.}
		\label{fig:Sobol_idices_3}
	\end{figure}

	The GSA analysis on the displacements described in \Cref{subsec:GSA} is performed considering as QoI the displacements
	at the centers of the eleven ridges of the beam (see \Cref{fig:stl}),
	i.e., a vector-valued QoI with components $\qoi_j(\vv) = u(\xx_{j,GSA},\vv)$ for $j=1,\ldots,11$,
	returning 11 sets of Sobol indices, as displayed in \Cref{fig:Sobol_idices_3}.
	The sets corresponding to the first 8 ridges further from the end of the removal area (\Cref{fig:stl})
	are similar, whereas the remaining 3 sets are unreliable due to their proximity to the end of the removal area
	and thus we neglect them in the subsequent analyses. \rev{This is motivated by the fact that on ridges 9, 10 and 11 the vertical displacements are essentially zero regardless of the values of the uncertain parameters due to the proximity to the removal area of the metal component from the build plate. Consequently, this would affect a correct evaluation of the Sobol indices.} The results indicate that the parameters with the greatest principal and total Sobol indices, i.e.\ with the greatest influence on the displacements,
	are $T_A$ and $\log h_p$ while $\log h_g$ has a negligible effect. This allows us to continue our study by treating as uncertain only the activation temperature and the powder convection coefficient,
	setting $\log h_g=-5$ (i.e., $h_g=10^{-5}$ ${\text{W}}/ ({\text{mm}}^{2}$ $ {^\circ{\text{C}}^{-1}})$) for subsequent analyses.
	
	\subsection{Inverse UQ }

	\label{subsec:Ires}
	
	\subsubsection{Surrogate model for inverse UQ}
	\label{subsec:SFSM for I}
\begin{figure}[t]
		\centering
		\subfigure[PARAMETRI-1][\s\s\s\s\s\s\s\s\s\s\s\s\s\s]{\label{fig:SparseTD}\includegraphics[width=0.33\textwidth]{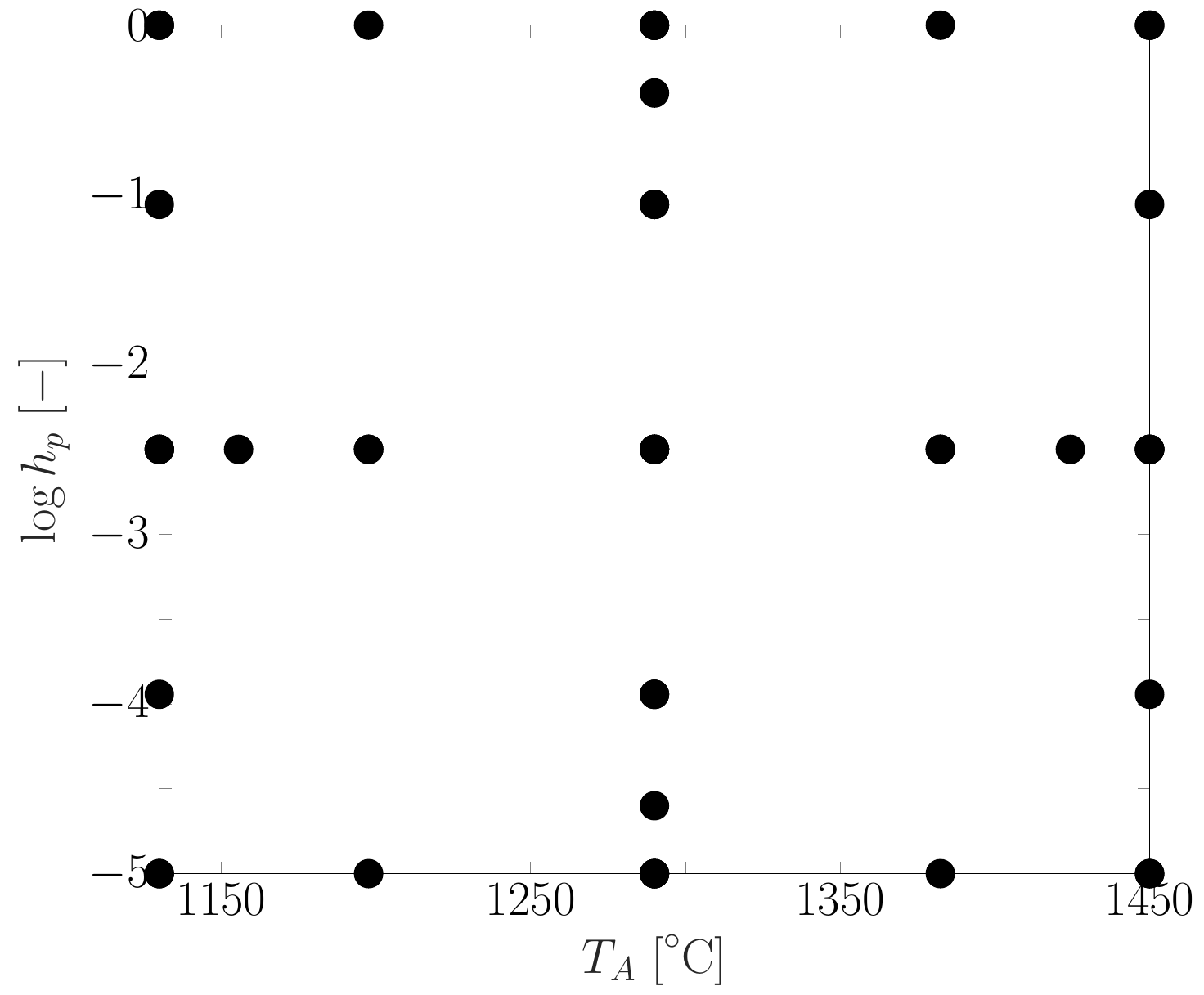}} \hspace{1.5em}
		\subfigure[PARAMETRI-2][\s\s\s\s\s\s\s\s\s\s\s\s\s\s\s\s]{\label{fig:x}\includegraphics[width=0.37\textwidth]{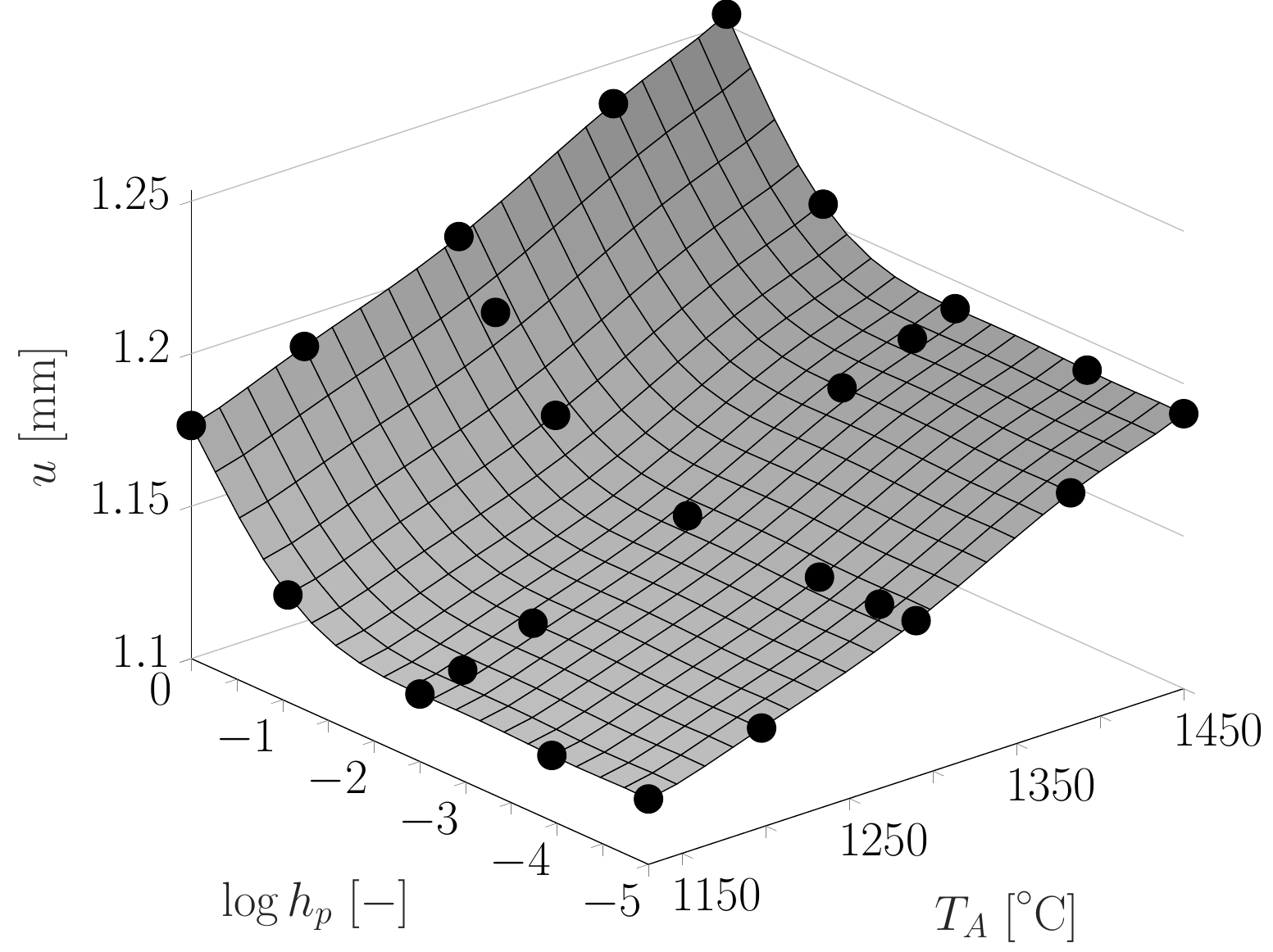}}
		\caption{\rev{Sparse-grid surrogate model construction for the inverse UQ analysis with $w=3$ (25 sparse-grid points) for the first ridge of the beam  (i.e., $\xx_{1, meas}=(0.5, 2.5, 12.5)$): (a) Sparse grid; (b) Surrogate model.}
			\label{fig:SparseGridtot_TD}}
	\end{figure}

	We then construct a new sparse-grid surrogate model depending on $T_A$ and $\log h_p$ only, to be used within the inverse UQ analysis.
	As already mentioned, the new sparse grid employs the set $\mathcal{I}_{sum}$ and is based on 25 evaluations of the PBF model  (cf. row 2 of \Cref{tab:grids}),
	corresponding to the points reported in \Cref{fig:SparseTD}. Note that also here we consider a vector-valued QoI, with components $\qoi_k(\vv) = u(\xx_{k,meas},\vv)$ for $k=1,\ldots,9$, i.e., we consider 9 measurement locations $\xx_{k,meas}$, that we set at the first 5 ridges and at the 4 midpoints between the respective ridges up to $x=28.5$ $\text{mm}$
	(see \Cref{fig:stl}; the 4 midpoints are those with labels from \lq\lq a\rq\rq~to \lq\lq d\rq\rq). 
	This choice guarantees that the measurement locations are at
	a sufficient distance from the end of the removal area, set at $x=56$ $\text{mm}$ (\Cref{fig:stl}), whose results could be affected by numerical instabilities.
	The corresponding 9 surrogate models behave similarly, therefore in the rest of this section we show results about the first node (ridge) of the beam. 
	In \Cref{fig:x} we report the surrogate model $\mathcal{S}_{I_{sum}}u(\xx_{1,meas},\vv)$,
	which shows a monotonically increasing behavior with respect to both parameters.
	
\begin{figure}[tbp]
		\centering
		\subfigure[PARAMETRI-1][\s\s\s\s\s\s\s\s\s\s\s\s\s\s\s\s\s\s]{\label{fig:Linf_IUQ}\includegraphics[width=0.31\textwidth]{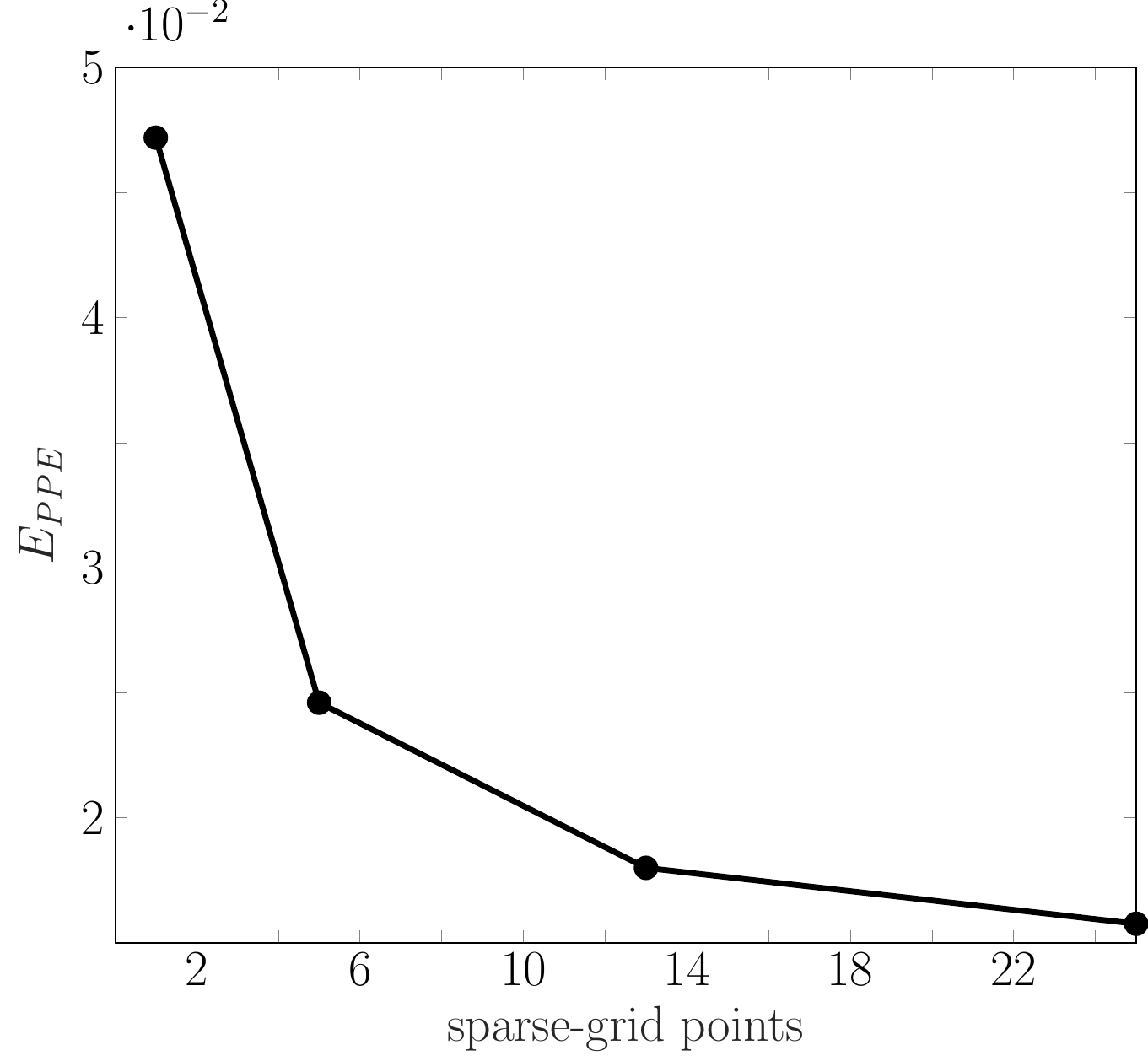}}\hspace{3em}
		\subfigure[PARAMETRI-2][\s\s\s\s\s\s\s\s\s\s\s\s\s\s\s\s\s]{\label{fig:L2_IUQ}\includegraphics[width=0.31\textwidth]{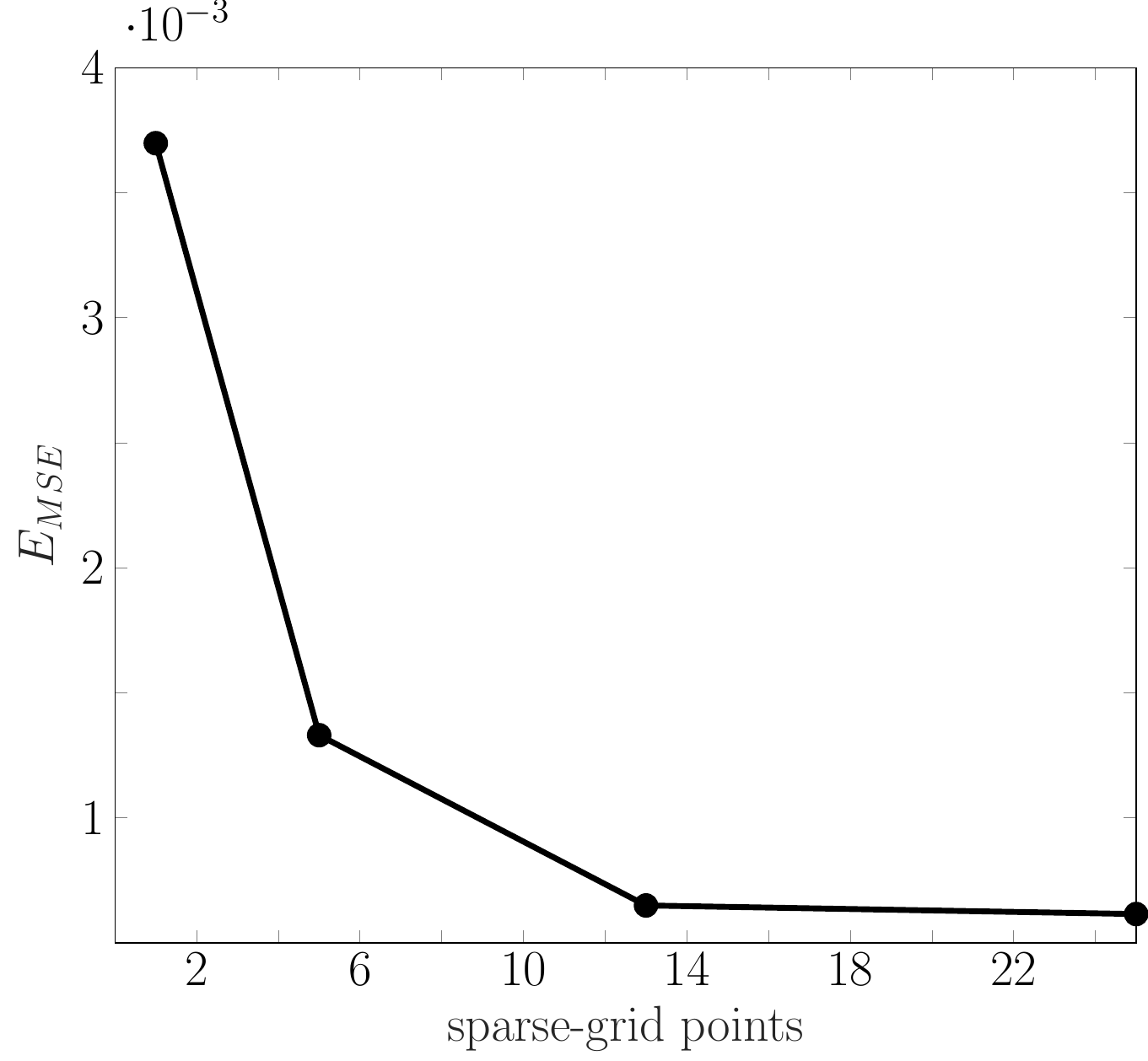}}
		\caption{\rev{Results of the sparse-grid surrogate model for inverse UQ analysis. Convergence test for the first ridge of the beam  (i.e., $\xx_{1, meas}=(0.5, 2.5, 12.5)$): (a) Pointwise prediction error $E_{PPE}$; (b) Root mean square error $E_{MSE}$.}}
		\label{fig:convergence_test}
	\end{figure}

\begin{figure}[tbp]
		\centering
	\subfigure[PARAMETRI-1][\s\s\s\s\s\s\s\s\s\s\s\s\s\s\s]{\label{fig:w1_uz}\includegraphics[width=0.39\textwidth]{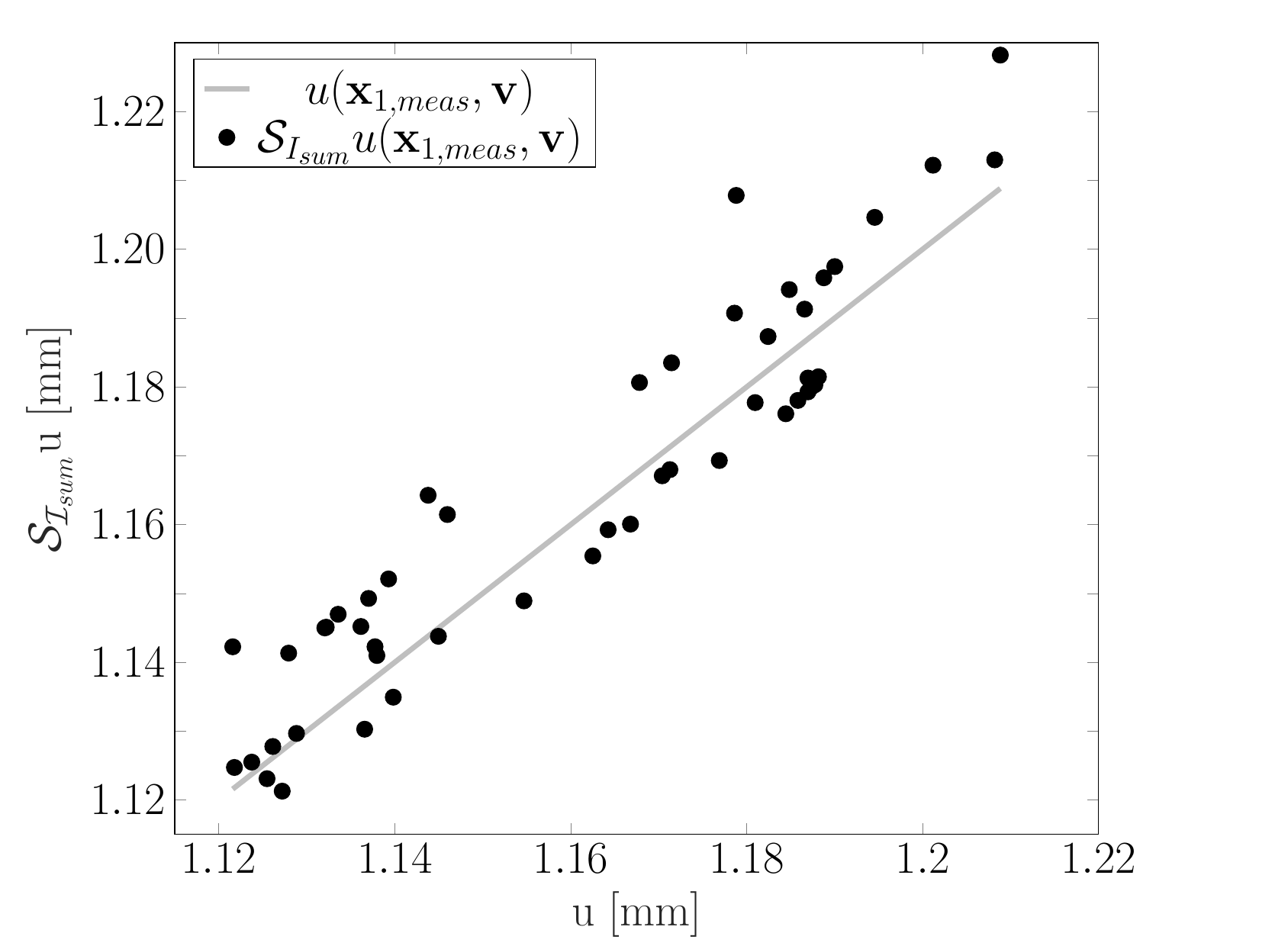}} 
		\subfigure[PARAMETRI-2][\s\s\s\s\s\s\s\s\s\s\s\s\s\s\s]{\label{fig:w3_uz}\includegraphics[width=0.39\textwidth]{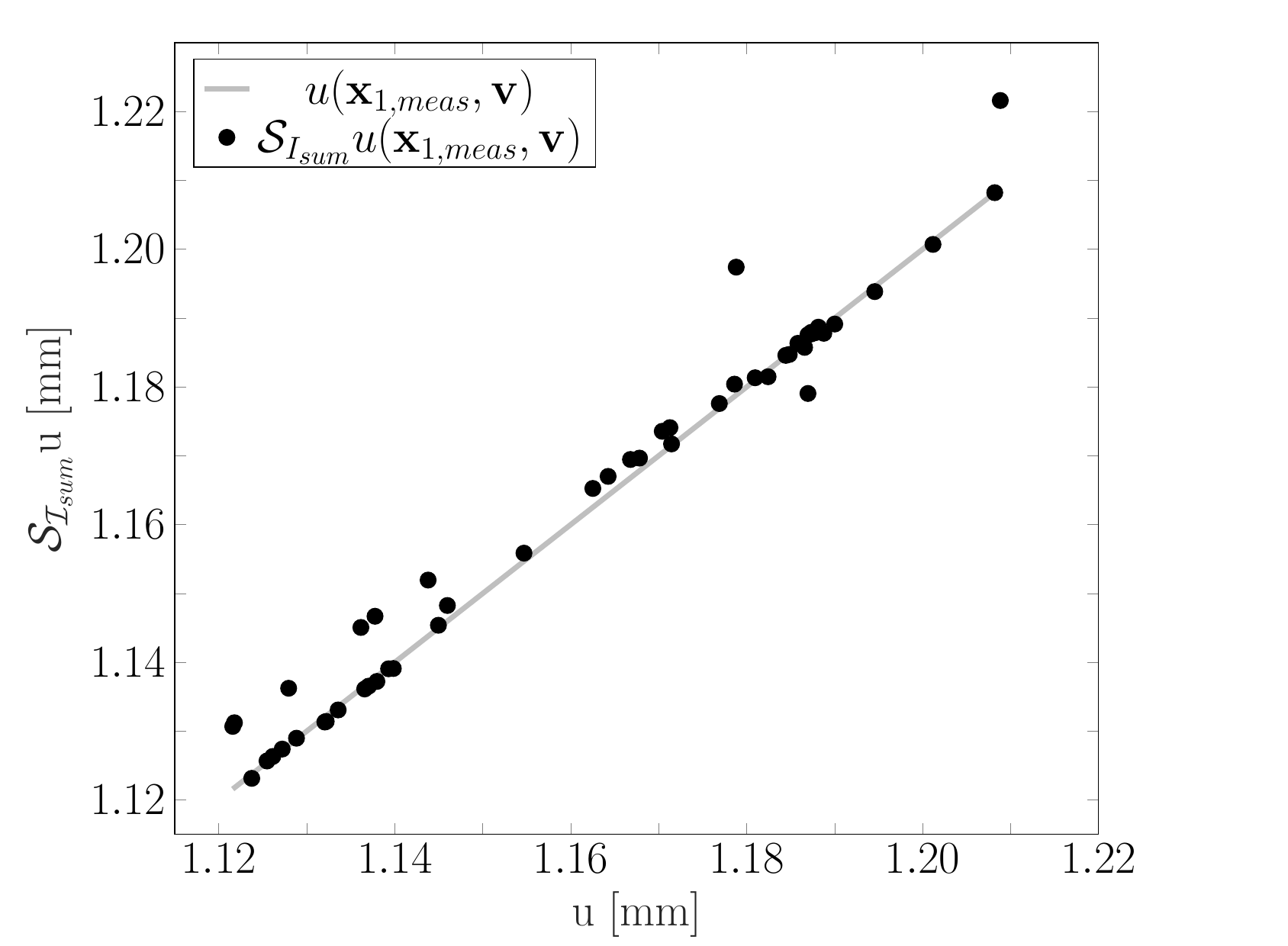}}
	   \caption{\rev{Results of the sparse-grid surrogate model for inverse UQ analysis. Comparison between sparse-grid surrogate model displacements and part-scale thermomechanical model displacements for the first ridge of the beam (i.e., $\xx_{1, meas}=(0.5, 2.5, 12.5)$): (a) Sparse grid with level $w=1$; (b) Sparse grid with level $w=3$.}}
		\label{fig:ansysSurr}
	\end{figure}

	Before proceeding with the inversion, we evaluate the quality of the sparse-grid surrogate model through a convergence test.
	In detail, we generate $M=50$ random couples $\vv=(T_A, \log h_p)$ and for each of them we compute the displacements at the 9 locations; then, we generate the sparse grids corresponding to the specifics in row 2 of \Cref{tab:grids} for increasing $w=0,1,2,3$. For each of
	these grids we compute the surrogate model predictions of the displacements at the same locations. Finally, we compute the
	following pointwise prediction errors, $E_{PPE}$, and the root mean square error, $E_{MSE}$.
	\begin{align}
		\label{eq:Linf}
		E_{PPE} &
		= \max_{i = 1,...,M} \frac{\left| {{ u}(\xx_{k,meas},\mathbf{v}_{i})- \mathcal{S}_{I_{sum}}{{u}}(\xx_{k,meas},\mathbf{v}_{i})}\right|}{\left| {{u}(\xx_{k,meas},\mathbf{v}_{i})}\right|};
		\\
		\label{eq:L2}
		E_{MSE}  &
		= \sqrt{{{ \frac{1}{M}} \sum_{i=1}^{M}} \frac{({u}(\xx_{k,meas},\mathbf{v}_{i})- \mathcal{S}_{I_{sum}}{u}(\xx_{k,meas},\mathbf{v}_{i}))^{2}}{{{ u}(\xx_{k,meas},\mathbf{v}_{i})}^{2}}}.
	\end{align}
	The results of the convergence test reported in \Cref{fig:Linf_IUQ,fig:L2_IUQ} show that the sparse-grid surrogate model with $w=3$ (25 sparse-grid points)
	can be considered suitable for the inverse UQ analysis, since the maximum relative error is approximately $1\%$, and the root mean square error is even smaller.
	The effectiveness of the surrogate model can be further appreciated by comparing the displacements obtained
	from the full model analyses with those obtained from the sparse-grid surrogate model, as shown in \Cref{fig:ansysSurr}. In fact, it can be seen that, as the number of sparse-grid points increases (i.e., increasing $w$), the displacements at the first ridge obtained from the surrogate model align with those obtained from the full model analyses to an extent that can be considered sufficiently accurate for our purpose.


 \begin{figure}[tbp]
	\begin{center}
		\includegraphics[width=0.35\textwidth]{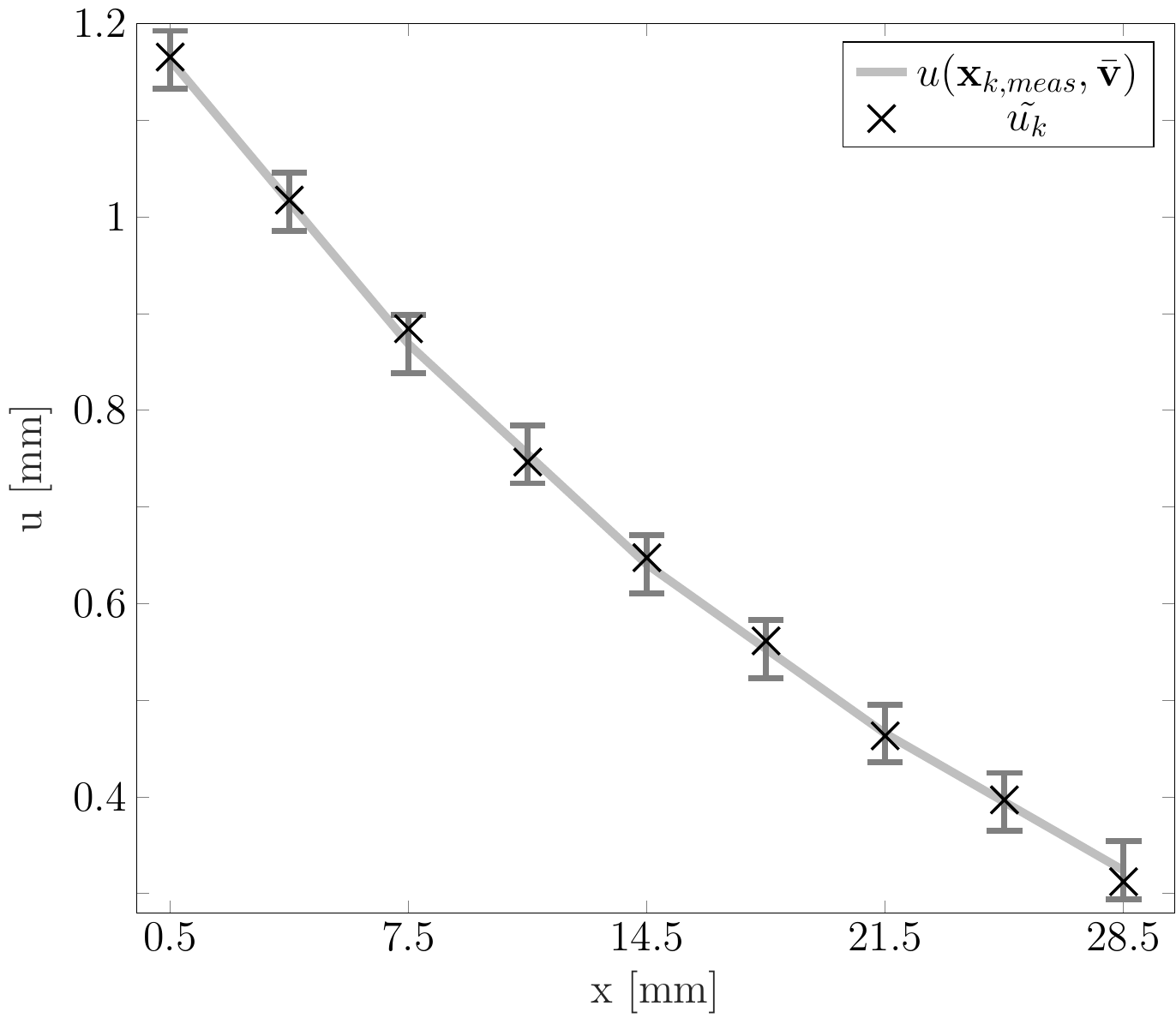}
		\caption{\rev{Displacements $u(\xx_{k,meas},\bar{{\vv}})$ obtained from part-scale thermomechanical analysis (gray curve)
				and synthetic displacements data $ \tilde{u}_k$ (black marker).
				We also report error bars associated with the measurements. Unlike standard error bar plots however,
				the error bars are centered at the exact (yet unknown) values on the displacements $u(\xx_{k,meas},\bar{\vv})$,
				and show the ranges within which the actual noisy measurements are most likely found, according to \Cref{eq:errsig}, i.e., $u(\xx_{k,meas},\bar{\vv}) \pm 3\bar{\sigma}$.}
		}
		\label{fig:u_withnoise}
	\end{center}
\end{figure}

	\subsubsection{Bayesian inversion}
	\label{sec:inversion}
	As discussed in \Cref{subsec:IBayes}, to perform the Bayesian inversion we consider as data the synthetic noisy displacements ${\tilde{u}_ {k}}$, $k=1,\ldots,9$ generated according to \Cref{eq:errsig} setting $\bar \sigma=10^{-2}$ and target value $\bar{\vv} = (T_A; \log h_p)=(1339. 8$ $^\circ{\text{C}}; -3.75$). \rev{The synthetic noisy displacement data are shown in \Cref{fig:u_withnoise} where we also show the displacement values obtained from part-scale thermomechanical simulation for the target value $\bar{\vv}$.} \rev{\Cref{fig:u_withnoise} also reports the error bars associated with the measurements. Unlike standard error bar plots however, here the error bars are centered at the exact (yet unknown) values of the displacements $u(\xx_{k,meas},\bar{\vv})$, and show the range within which the actual noisy measurement is most likely found, according to \Cref{eq:errsig}, i.e., $u(\xx_{k,meas},\bar{\vv}) \pm 3\bar{\sigma}$ (a Gaussian random variable takes values in such interval with probability 99\%).}
    The measuring locations $\xx_{k,meas}$ are the same as in the previous subsection. We consider these data sufficient for our purpose; in fact, additional tests with more data did not change the essence of the results shown below.

\begin{figure}[h!]
	\centering
	\subfigure[PARAMETRI-1][\s\s\s\s\s\s\s\s\s\s\s\s\s\s\s\s]{\label{fig:LS}\includegraphics[width=0.38\textwidth]{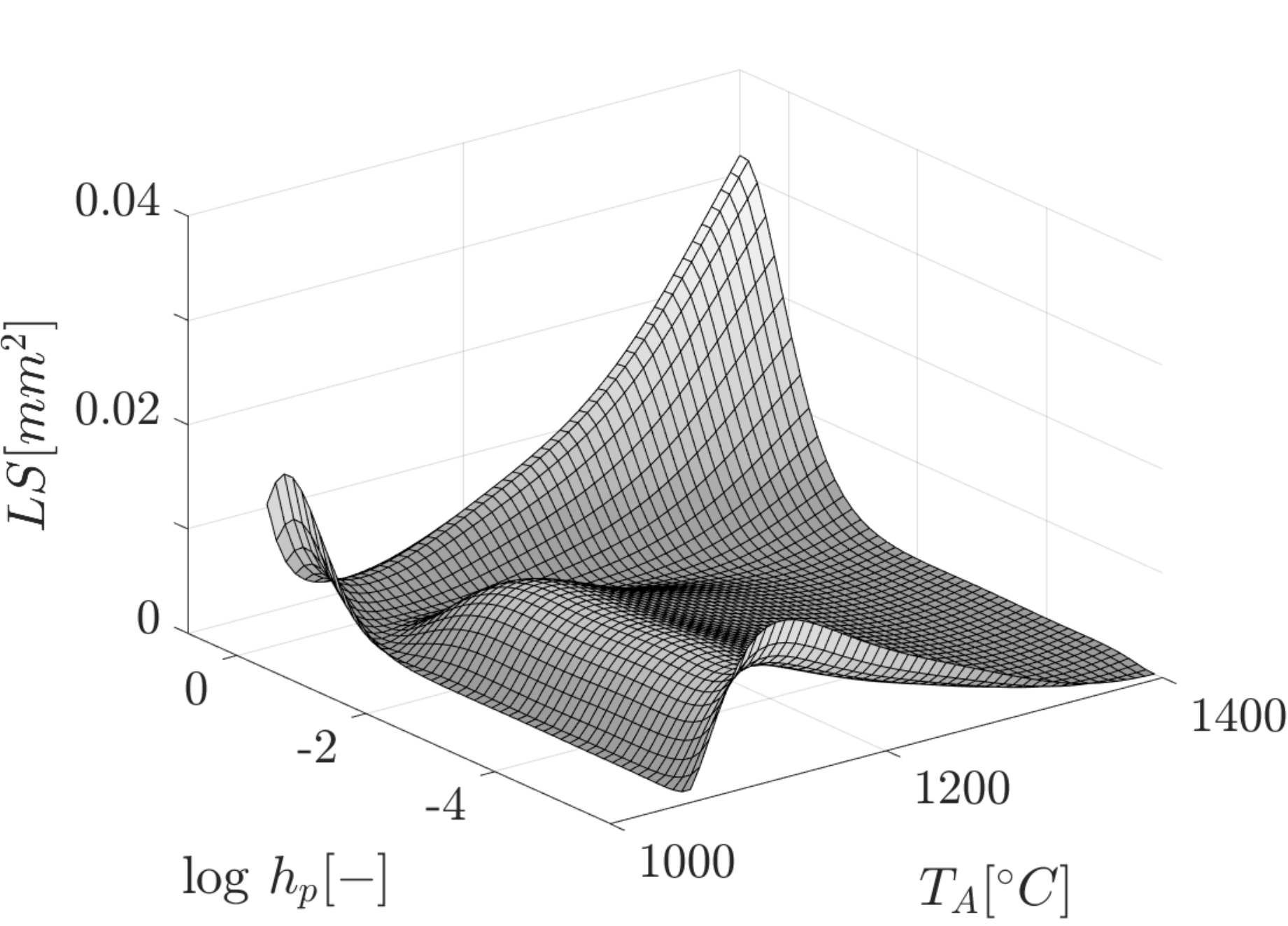}} \hspace{0.4cm}
	\subfigure[PARAMETRI-2][\s\s\s\s\s\s\s\s\s\s\s\s\s\s\s]{\label{fig:count}\includegraphics[width=0.345\textwidth]{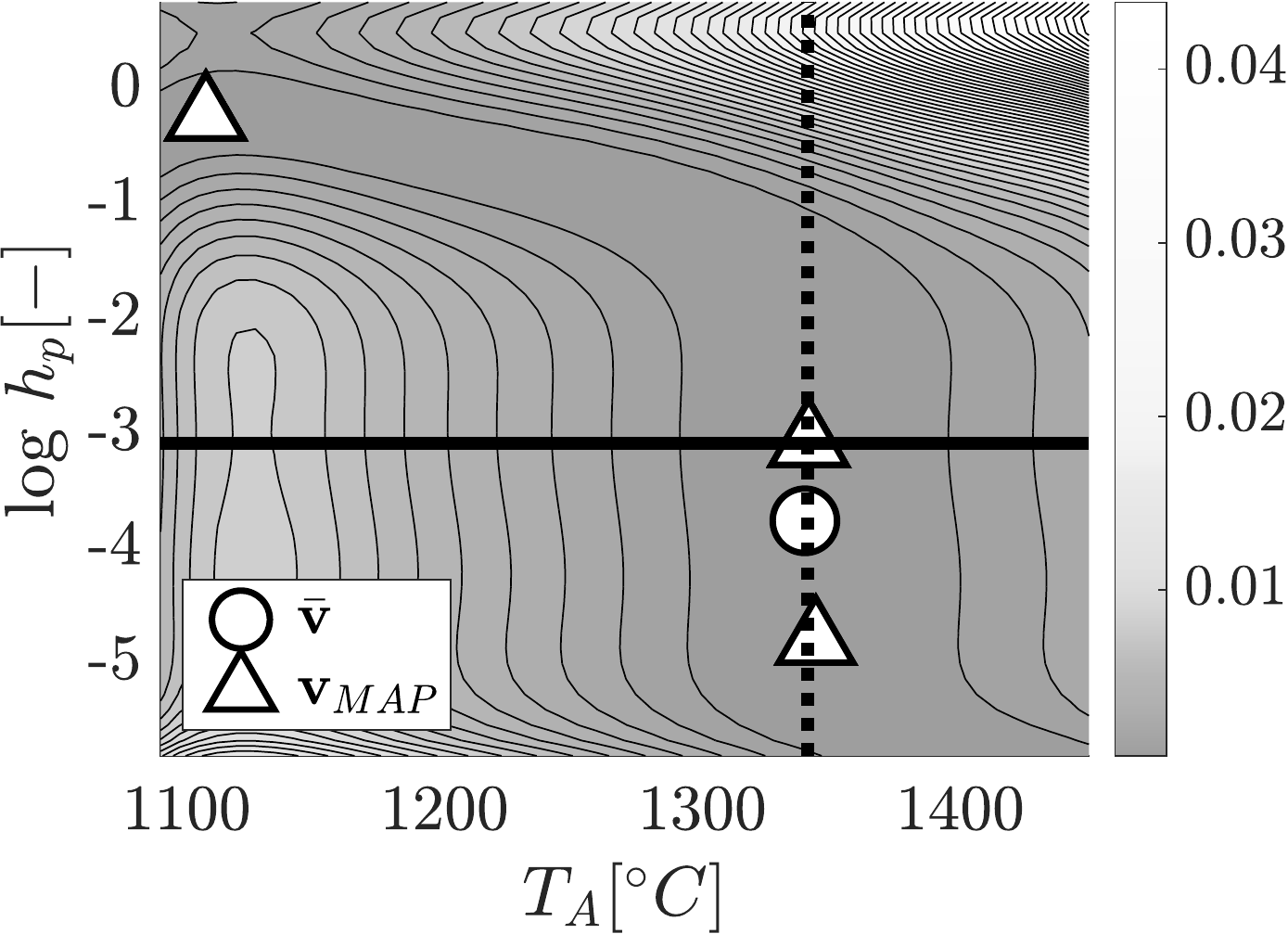}} 
	\subfigure[PARAMETRI-3][\s\s\s\s\s\s\s\s\s\s\s\s\s\s\s\s\s]{\label{fig:LSTA}\includegraphics[width=0.335\textwidth]{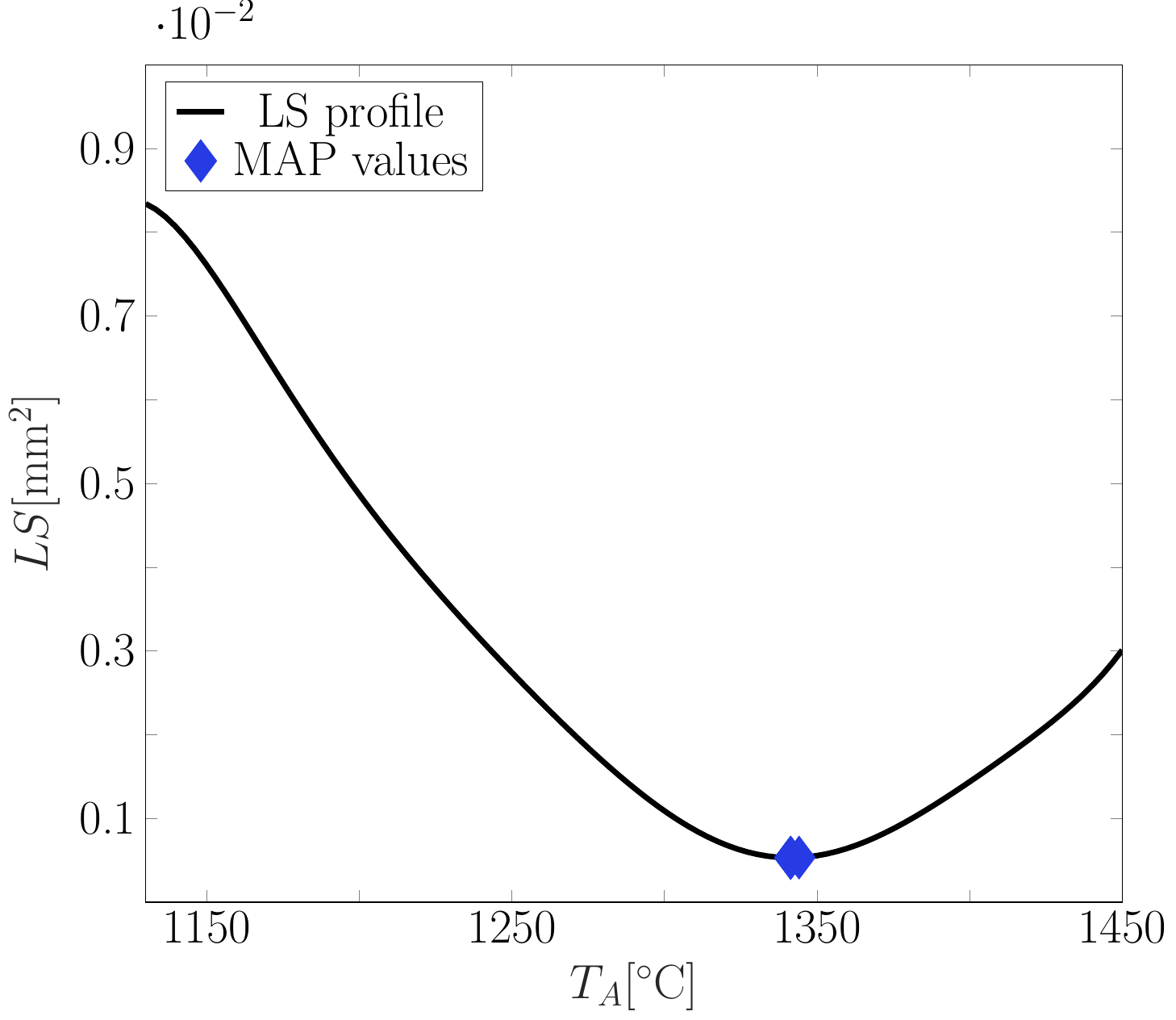}} 
	\subfigure[PARAMETRI-4][\s\s\s\s\s\s\s\s\s\s\s]{\label{fig:hpTA}\includegraphics[width=0.345\textwidth]{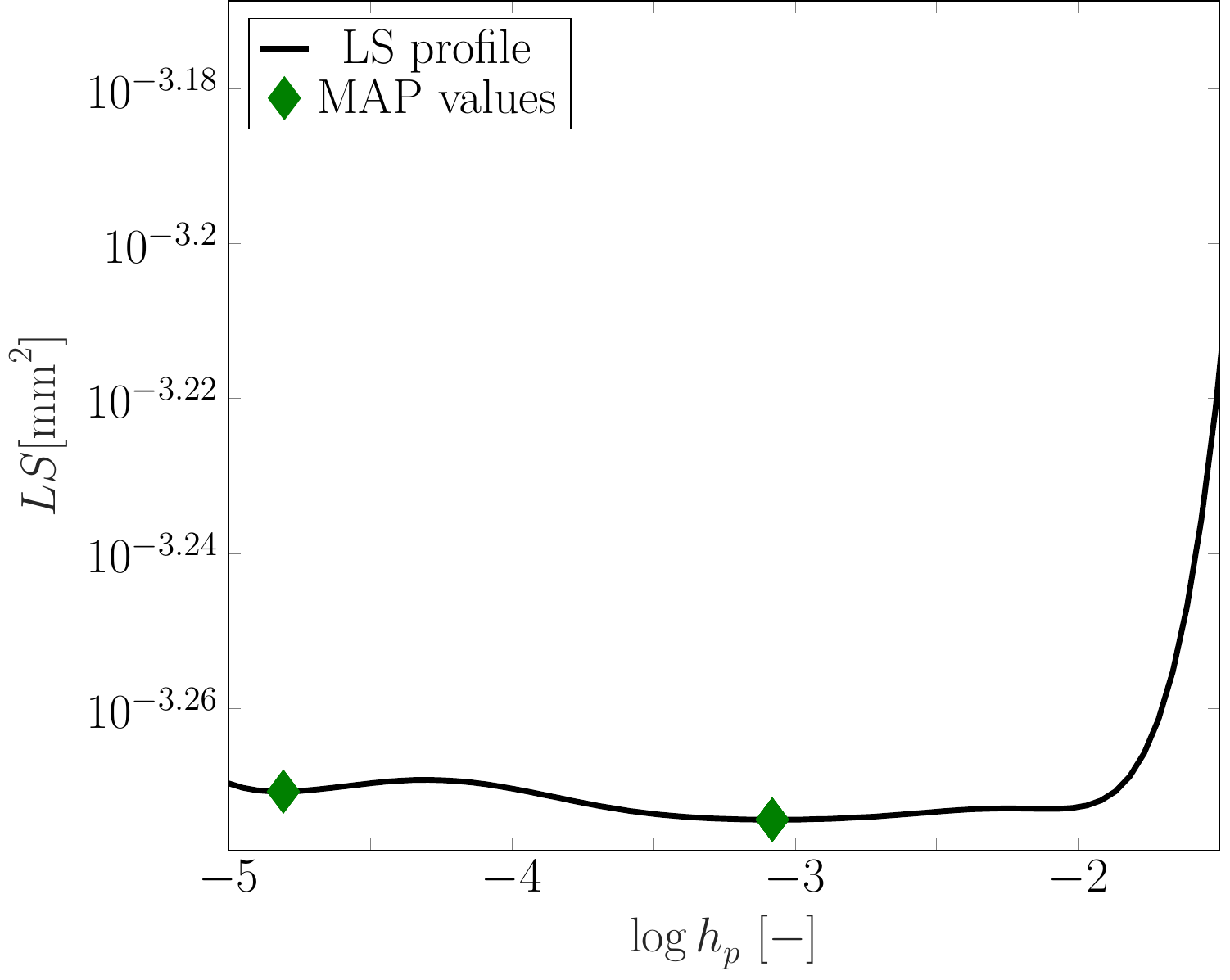}} 
	\label{fig:YTTTT}
	\caption{\rev{Results of the inverse UQ analysis: (a) Surface plot of the least-squares functional $LS(\vv)$; (b) Isolines of $LS(\vv)$, target value $\bar{\vv}$, the position of the cutting planes used to generate the one-dimensional plots in \Cref{fig:LSTA} marked with a solid black line and in \Cref{fig:hpTA} marked with a dashed black line, and the two MAP values computed by minimization of $LS(\vv)$; (c) $LS$ profile at $\log h_p=-3$ with the two MAP values; (d) $LS$ profile at $T_A=1341^{\circ}C$ with the two MAP values. }}
\end{figure}

	We begin by computing the mean of the Gaussian approximation of $\rho_{post}$, i.e., $\vv_{MAP}$ (see \Cref{eq:ymap}). This can be obtained by using, e.g., the
	gradient-free optimization algorithm Nelder-Mead (available in Matlab through the \texttt{fminsearch} command) to find the minimum of the $LS$ function (\Cref{fig:LS}).
	To make the result more robust, we repeat the optimization several times, for different initial points of the optimization algorithm.
        Proceeding in this way, we find three local minimum points, see \Cref{fig:count},
	one of which falls outside the chosen activation temperature range (see \Cref{tab:PMat}) and thus can be discarded.
	The two remaining minimum points have very similar values of $T_A \approx 1341$ ${^\circ{\text{C}}}$, but are very different in terms of $\log h_p$,
	one being approximately $-5$ and the other approximately $-3$. 
	This shows that a Gaussian approximation of the posterior  PDF of $\log h_p$ is inappropriate. 
	This consideration is further supported by inspection of the isolines of $LS$ in the area of the minima
        \rev{(see \Cref{fig:count}). In fact, by observing the $LS$ isolines, we can see that they form a band/strip in $\log h_p$ direction,
          as already mentioned in \Cref{sec:mixed_approx_rho_post}.
          Even further, we can plot the profile of $LS$ varying $T_A$ upon fixing $\log h_p = -3$ (\Cref{fig:LSTA})
          and conversely for varying values of $\log h_p$ upon fixing $T_A = 1341 \,^\circ\text{C}$ (\Cref{fig:hpTA}),
          i.e., one-dimensional cuts of $LS$ obtained by intersecting it with vertical planes (these planes can be seen in \Cref{fig:count}).
          We see that the former one has a parabolic profile, whereas the latter one is essentially flat in a large interval of values of $\log h_p$.
          This implies that a Gaussian approximation for $T_A$ is valid, whereas it is not for $\log h_p$.
          This kind of graphical analysis of the negative log-likelihood function is widely employed in the context of parameter estimation for dynamical systems,
          where is known as \lq\lq profile likelihood inspection\rq\rq, see e.g. \cite{Raue1}.}

\begin{figure}[h!]
	\centering
	\subfigure[PARAMETRI-1][\s\s\s\s\s\s\s\s\s\s\s\s\s\s]{\label{fig:TApost}\includegraphics[width=0.38\textwidth]{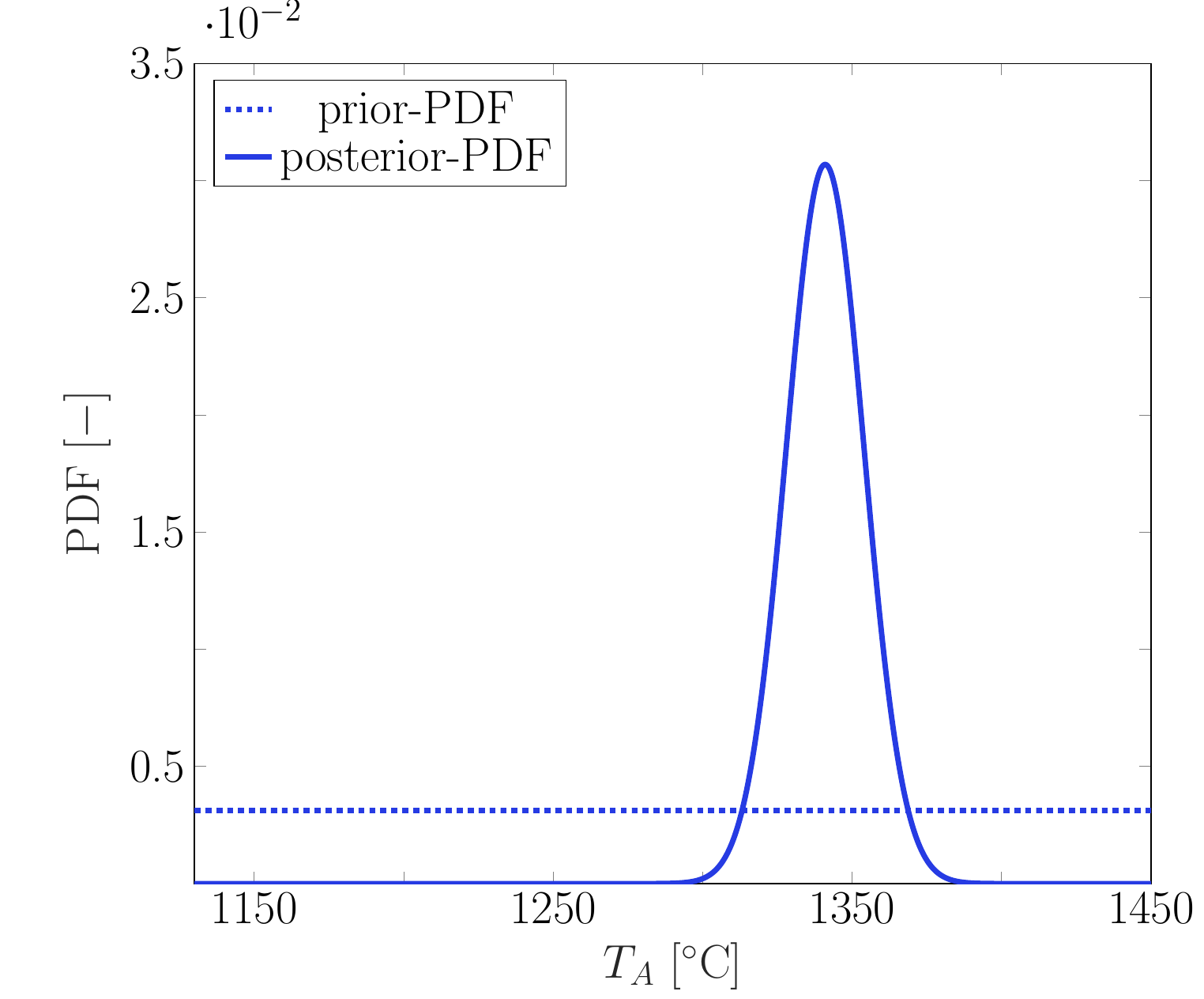}} 
	\hspace{1.5em}
	\subfigure[PARAMETRI-2][\s\s\s\s\s\s\s\s\s\s\s\s\s]{\label{fig:profLS}\includegraphics[width=0.348\textwidth]{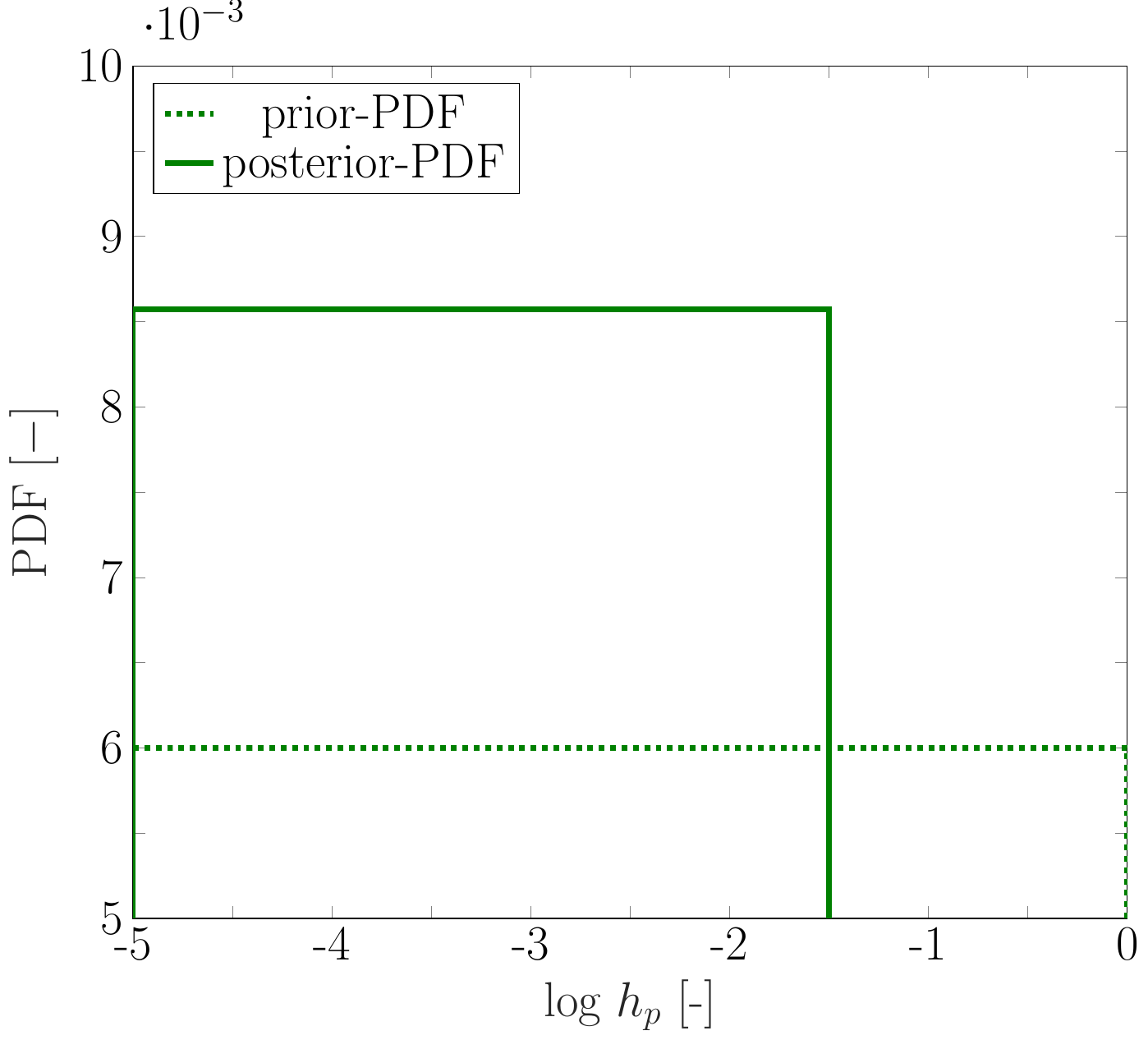}} 
	\label{fig:IUQ_res}
	\caption{\rev{Results of the inverse UQ analysis: (a) Uniform prior-PDF and Gaussian posterior-PDF for parameter $T_A$ before and after Bayesian inversion; (b) Uniform PDFs for the $ \log h_p$ parameter before and after Bayesian inversion.}}
\end{figure}

        \rev{Therefore, we employ the mixed approximation strategy already presented in  \Cref{sec:mixed_approx_rho_post}: we approximate the posterior PDF $\rho_{post}(\vv)$ as the product of a Gaussian PDF for $T_A$ and of a uniform PDF for $\log h_p$. In \Cref{fig:TApost,fig:profLS} we report the prior-PDFs and posterior-PDFs for $T_A$ and for $\log h_p$, respectively.} In details:
	\begin{itemize}
        \item the Gaussian PDF for $T_A$ (\Cref{fig:TApost}) is centered at the common value of the first component of the two minimum points of $LS(\vv)$, i.e.,
          at $T_A = 1341 \,^\circ\text{C}$, while the standard deviation $\sigma_{T_A,post}$ is taken as the square root of the
          entry $(1,1)$ of the covariance matrix ${\boldsymbol \Sigma_{post}}$ (see \Cref{eq:covariance}),
          resulting in  $\sigma_{T_A,post} \approx 13 \, ^\circ\text{C}$;
        \item as extrema of the uniform PDF for $\log h_p$ we employ $(-5; -1.5)$, since the profile of the likelihood functional (\Cref{fig:hpTA}) at $T_A=1341$ ${^\circ{\text{C}}}$ as a function of $\log h_p$ is substantially larger outside this interval.
	\end{itemize}
    The results are also summarized in \Cref{tab:PMatIUQ}. Finally, we verify that the approximation ${\bar \sigma_{MAP}}^2$ of ${\bar \sigma^2}$ is good  (see \Cref{eq:sisi}); in fact, ${\bar \sigma^2}$ is fixed at 0.01 and the value of ${\bar \sigma_{MAP}}^2$ is 0.00767.

	\begin{table}[h!]
          \begin{center}
            \begin{tabular}{ c  c }
              \hline
              \textbf{Random Variables}  & \textbf{posterior PDF} \\ \hline 
              \vspace{0.25cm}
              $T_{A}$ $[^\circ{\text{C}}]$ & Gaussian$(1341;13)$  \\
              $\log h_{p}$ $[-]$ & Uniform$(-5,-1.5)$  \\
              \hline
            \end{tabular}
          \end{center}
          \caption{PDF of the parameters resulting from the inverse UQ analysis and used as input for the data-informed forward UQ analysis.\label{tab:PMatIUQ}}
	\end{table}


	\subsection{Data-informed forward UQ for residual strains} \label{subsec:SFSM_forward}
	
	The final step of the UQ workflow consists in the data-informed forward UQ analysis. In particular, we focus on the residual strains ${\varepsilon}_{xx}(\xx_{j,str},\vv)$
	at $L=120$ positions $\xx_{j,str}$ in the central plane of the beam at $z=11$ $\text{mm}$ (see dotted green line in \Cref{fig:stl})
	obtained according to the PDF $\rho_{post}$ just derived (see \Cref{tab:PMatIUQ}).

	\subsubsection{Surrogate model for data-informed forward UQ}\label{subsec:forwardUQ}
\begin{figure}[tbp]
		\centering
	\hspace{0.35cm}	\subfigure[PARAMETRI-1][\s\s\s\s\s\s\s\s\s\s\s\s]{\label{fig:SparseTDFUQ}\includegraphics[width=0.33\textwidth]{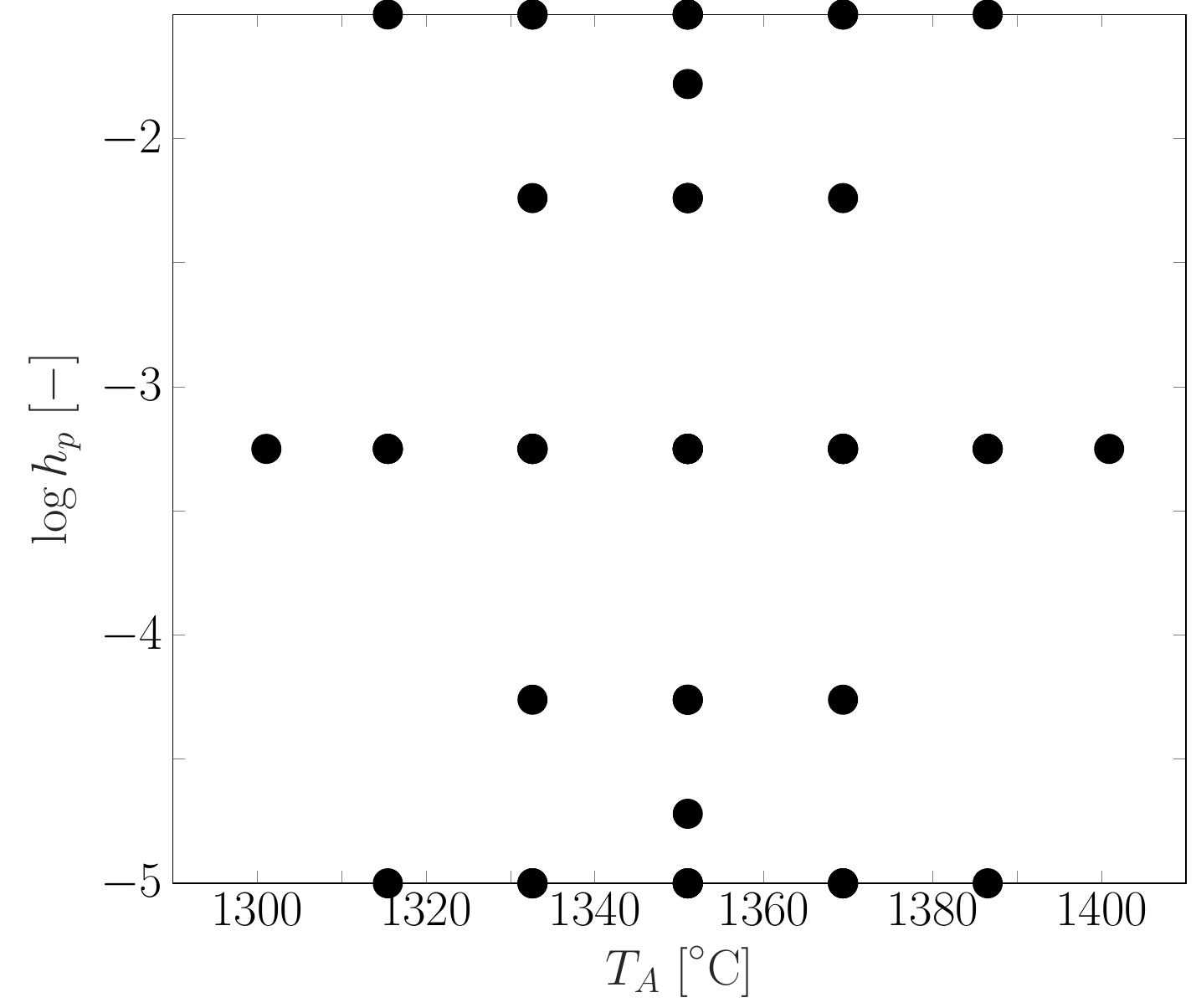}}  \hspace{1.5em}
		\subfigure[PARAMETRI-2][\s\s\s\s\s\s\s\s\s\s\s\s\s\s\s]{\label{fig:xFUQ}\includegraphics[width=0.37\textwidth]{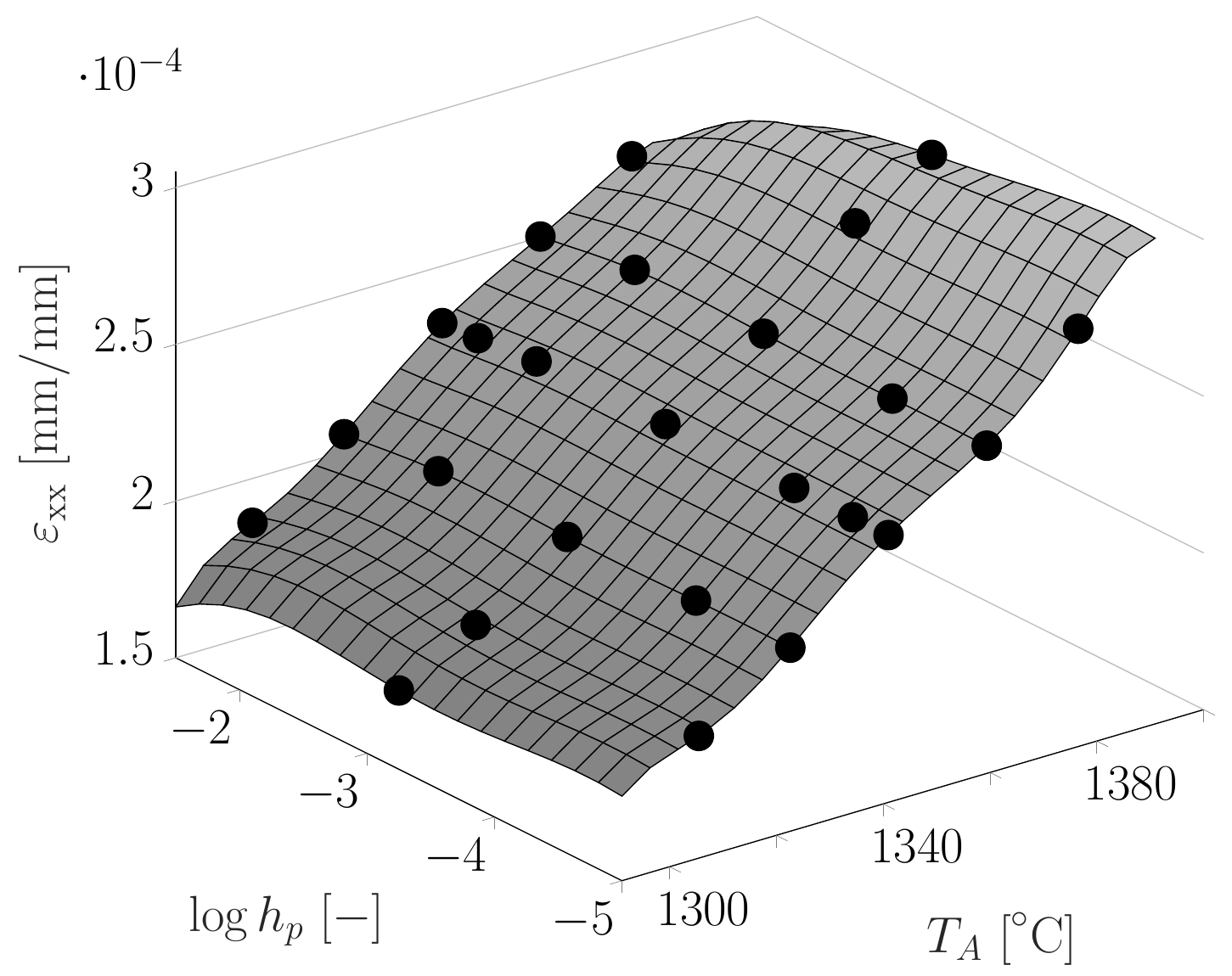}}
		\caption{\rev{Sparse-grid surrogate model construction for the data-informed forward UQ analysis with $w=3$ (25 sparse grid points) for the first position of the beam (i.e., $\xx_{1, str}=(1.5, 2.5, 11)$): (a) Sparse grid; (b) Surrogate model.}}
		\label{fig:surr_fuq_mod}
	\end{figure}

	Since the parameter PDF has changed (\Cref{tab:PMatIUQ}), we start by computing a new set of $120$ sparse-grid surrogate models, following row 3 of \Cref{tab:grids}.
	The sparse grid over which these surrogate models are based consists of 25 new collocation points in $\Gamma_{red}$
	and is shown in \Cref{fig:SparseTDFUQ}. \Cref{fig:xFUQ} shows instead the sparse-grid surrogate model for the strains at
	$\xx=(x,y,z) = (1.5, 2.5, 11)$ $\text{mm}$; 
	a similar interpolating surface is observed for all other locations.
	
\begin{figure}[tbp]
	\centering
	\subfigure[PARAMETRI-1][\s\s\s\s\s\s\s\s\s\s\s\s\s\s\s]{\label{fig:Linf}\includegraphics[width=0.31\textwidth]{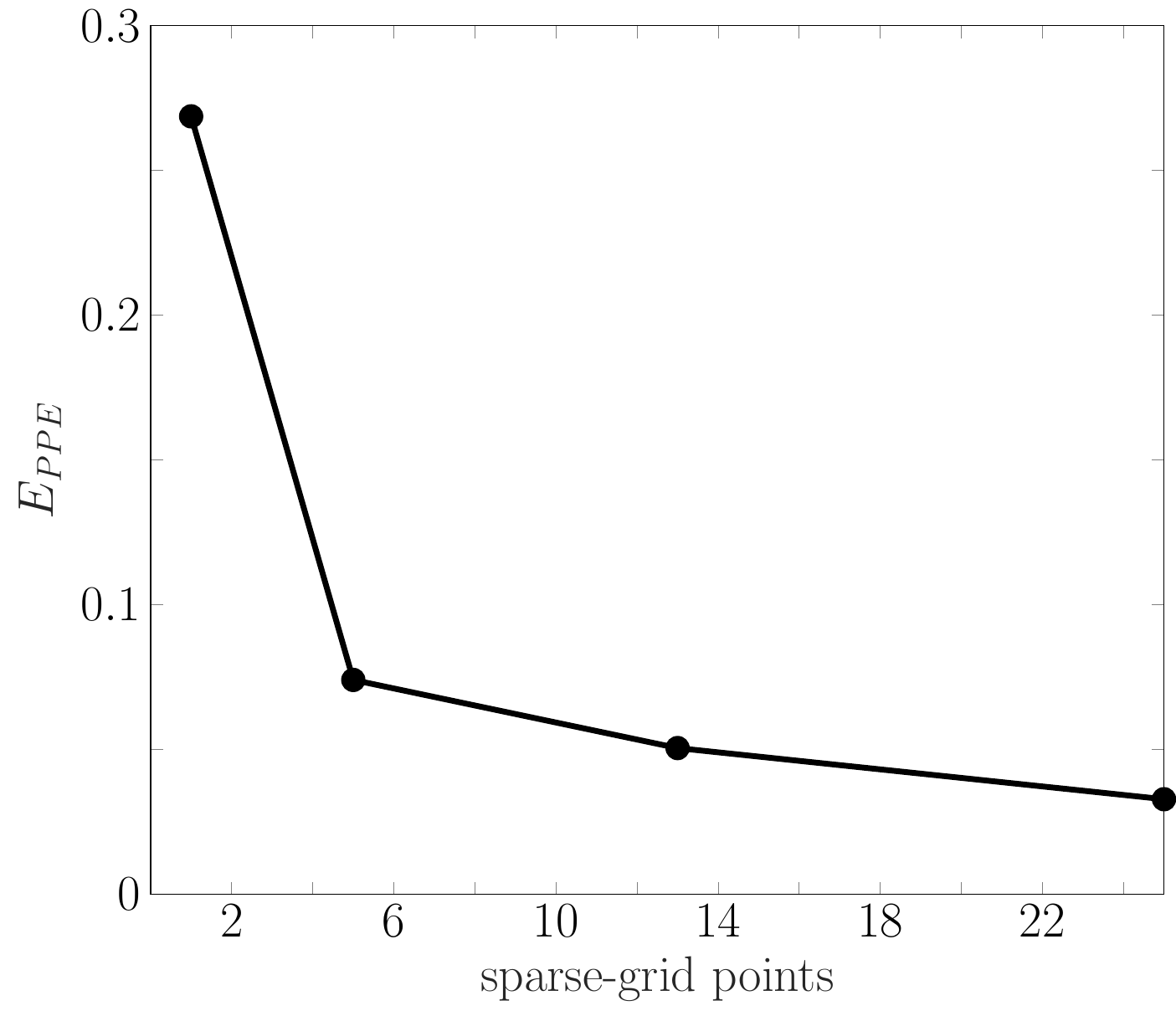}}   \hspace{2.5em}
	\subfigure[PARAMETRI-2][\s\s\s\s\s\s\s\s\s\s\s\s\s\s\s]{\label{fig:L2}\includegraphics[width=0.31\textwidth]{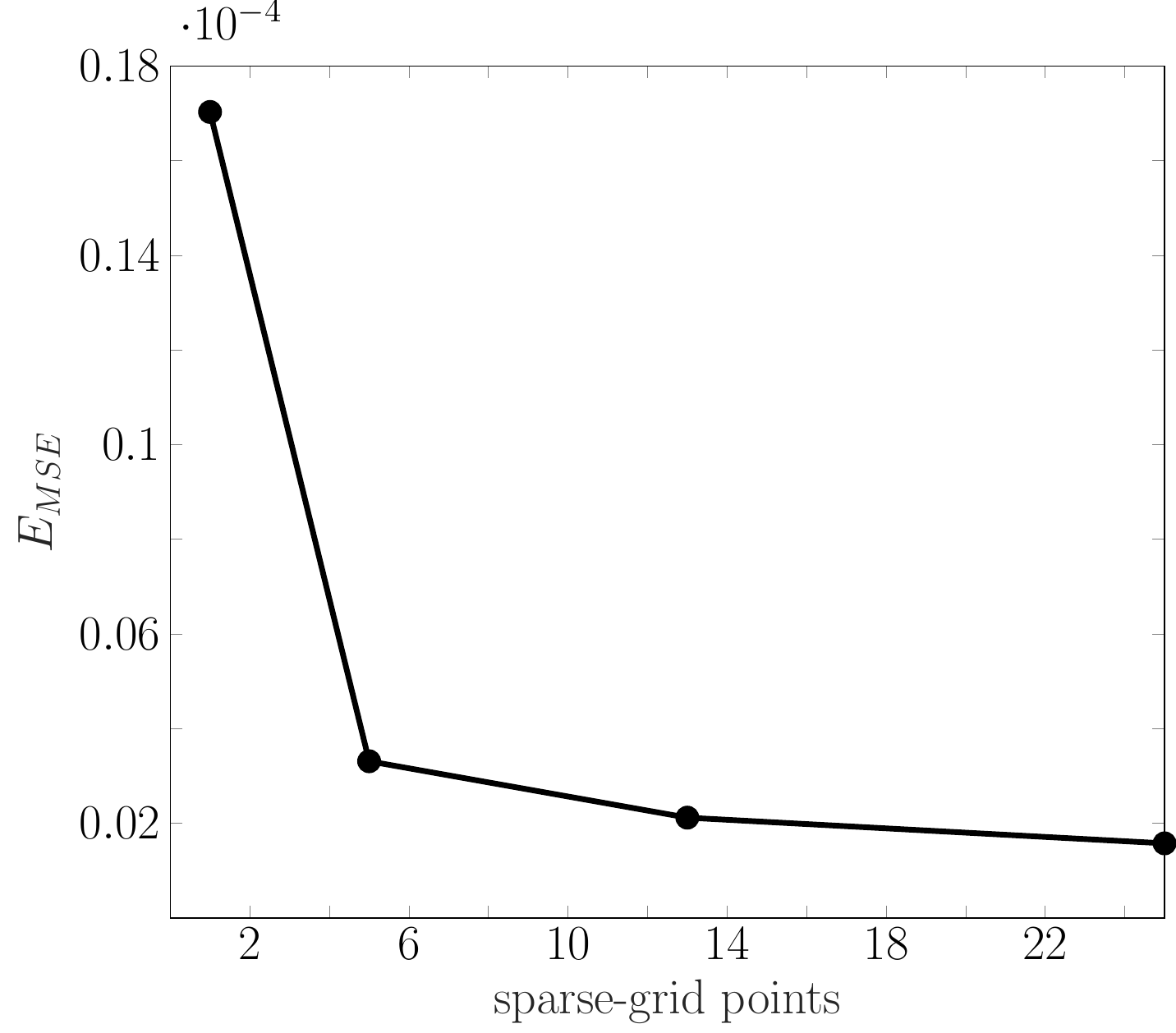}}
	\caption{\rev{Results of the sparse-grid surrogate model for data-informed forward UQ analysis. Convergence test for the first position of the beam  (i.e., $\xx_{1, str}=(1.5, 2.5, 11)$): (a) Pointwise prediction error $E_{PPE}$; (b) Root mean square error $E_{MSE}$.}}
	\label{fig:convergence_test_FUQ}
\end{figure}

\begin{figure}[tbp]
	\centering
	\subfigure[PARAMETRI-1] [\s\s\s\s\s\s\s\s\s\s\s\s\s\s\s]{\label{fig:w1_uz_FUQ}\includegraphics[width=0.31\textwidth]{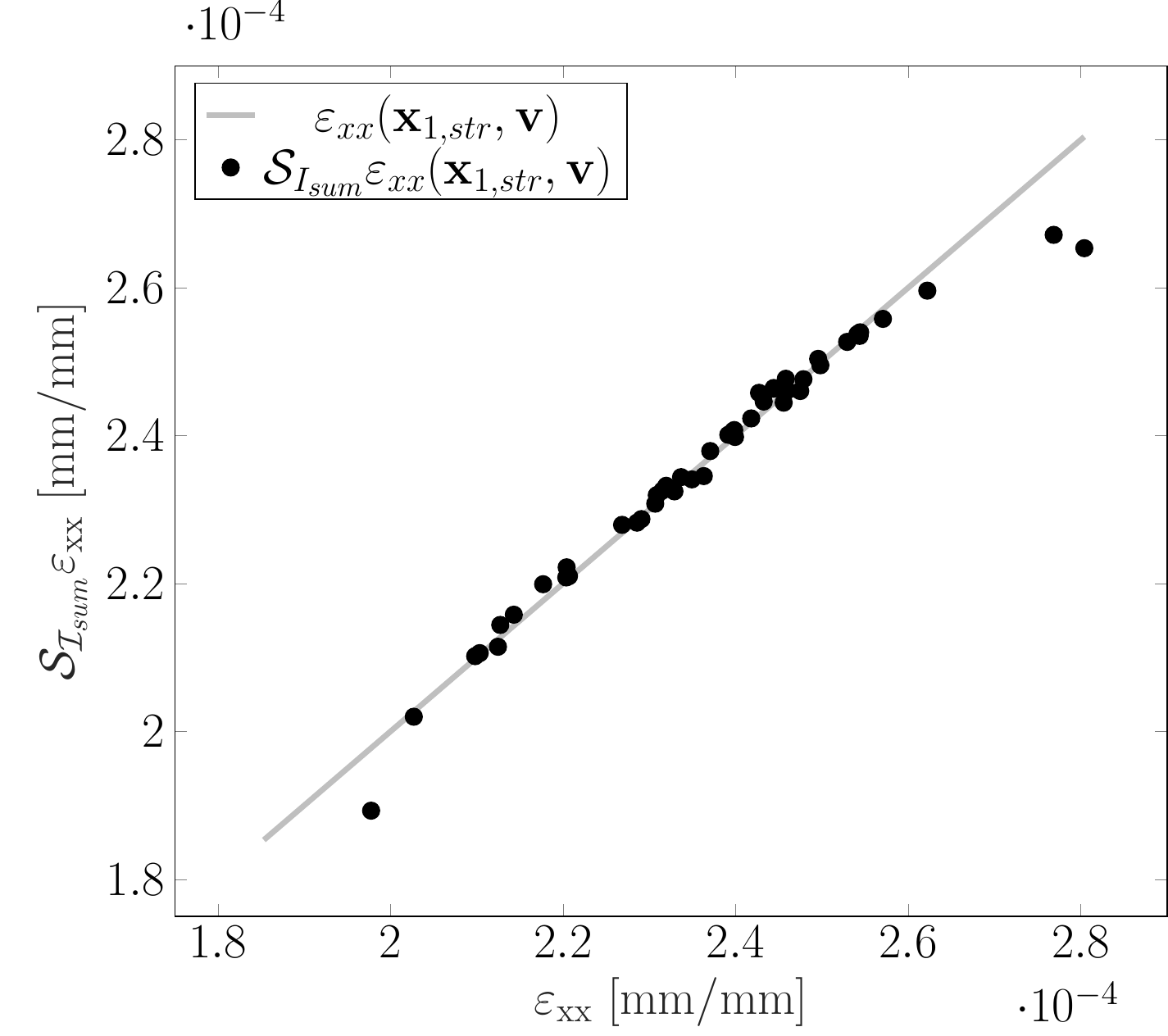}} \hspace{2.5em}
	\subfigure[PARAMETRI-][\s\s\s\s\s\s\s\s\s\s\s\s\s\s\s]{\label{fig:w3_uz_FUQ}\includegraphics[width=0.31\textwidth]{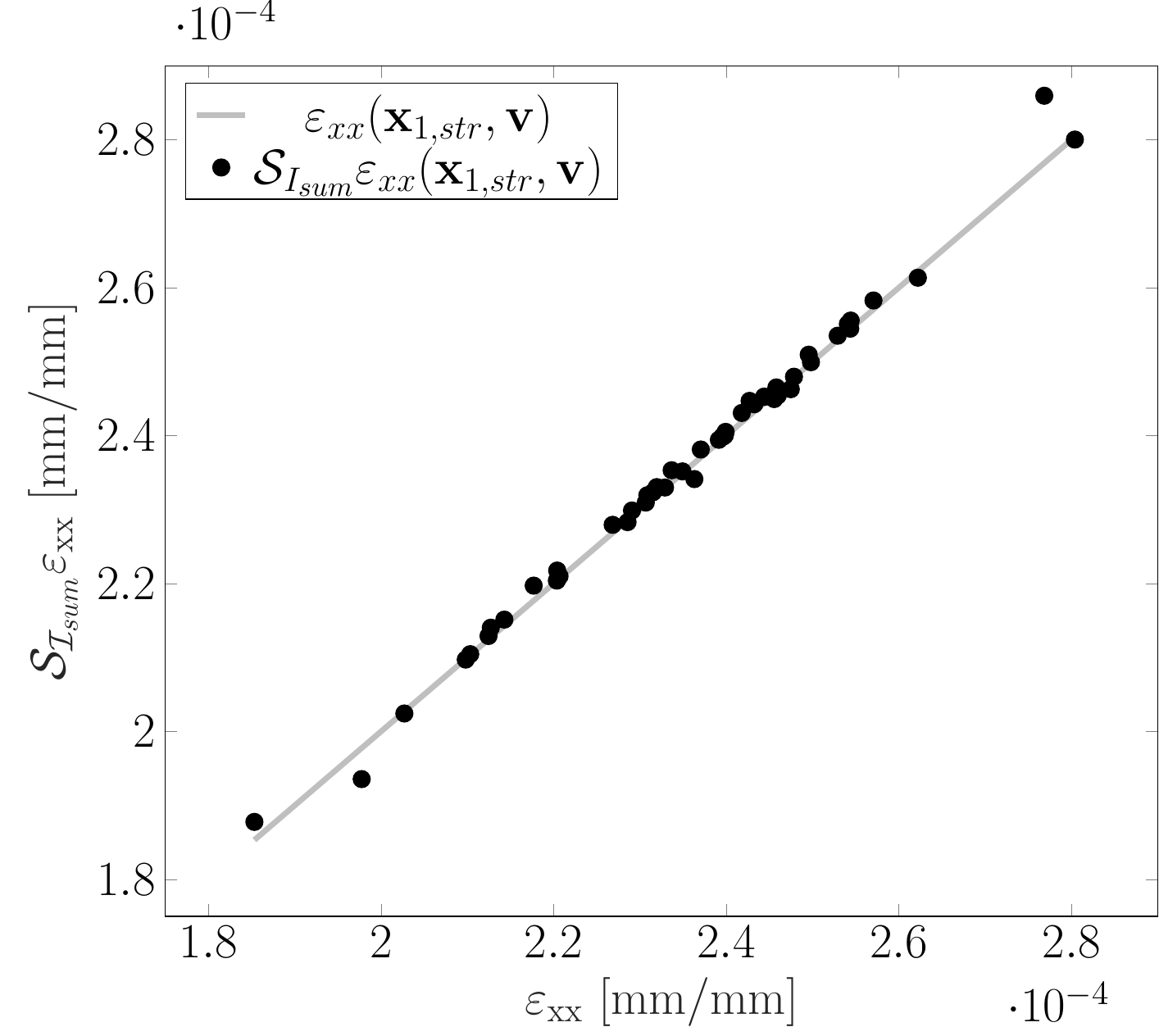}} 
	\caption{\rev{Results of the sparse-grid surrogate model for data-informed forward UQ analysis. Comparison between sparse-grid surrogate model residual strains and part-scale thermomechanical model residual strains for the first position of the beam  (i.e., $\xx_{1, str}=(1.5, 2.5, 11)$): (a) Sparse grid with level $w=1$; (b) Sparse grid with level $w=3$.}}
	\label{fig:ansysSurr_FUQ}
\end{figure}

	We also perform a convergence test similar to what done for the sparse-grid surrogate model used for the inverse UQ.
	We therefore evaluate the strains by the full model for $M=50$ new random values of  $\mathbf v=(T_A,\log h_p)$ according to $\rho_{post}$,
	and compare these residual strains with their approximations obtained by the sparse-grid surrogate models with $w=0,1,2,3$,
	obtaining the corresponding values for the pointwise prediction error, $E_{PPE}$ (\Cref{eq:Linf}) and the root mean square error, $E_{MSE}$ (\Cref{eq:L2}). As expected, the trend of errors $E_{PPE}$ and $E_{MSE}$ as $w$ increases is similar for all $120$ positions, so we report the result for $\xx_{1, str}=(1.5, 2.5, 11)$ mm, see \Cref{fig:convergence_test_FUQ}. As can be seen, the convergence test suggests that the
	surrogate model with $w=3$ (25 sparse-grid points) can be considered accurate enough for our purposes. 
	The same conclusion can be obtained by looking at \Cref{fig:ansysSurr_FUQ}, which shows that, as the number of sparse-grid points increases, the residual strains obtained from the surrogate model align with the residual strains obtained from the part-scale thermomechanical analyses. 
	
	\subsubsection{\rev{Data-informed PDF of residual strains}}
\begin{figure}[tbp]
       	\begin{center}
       		\includegraphics[width = 0.345\textwidth]{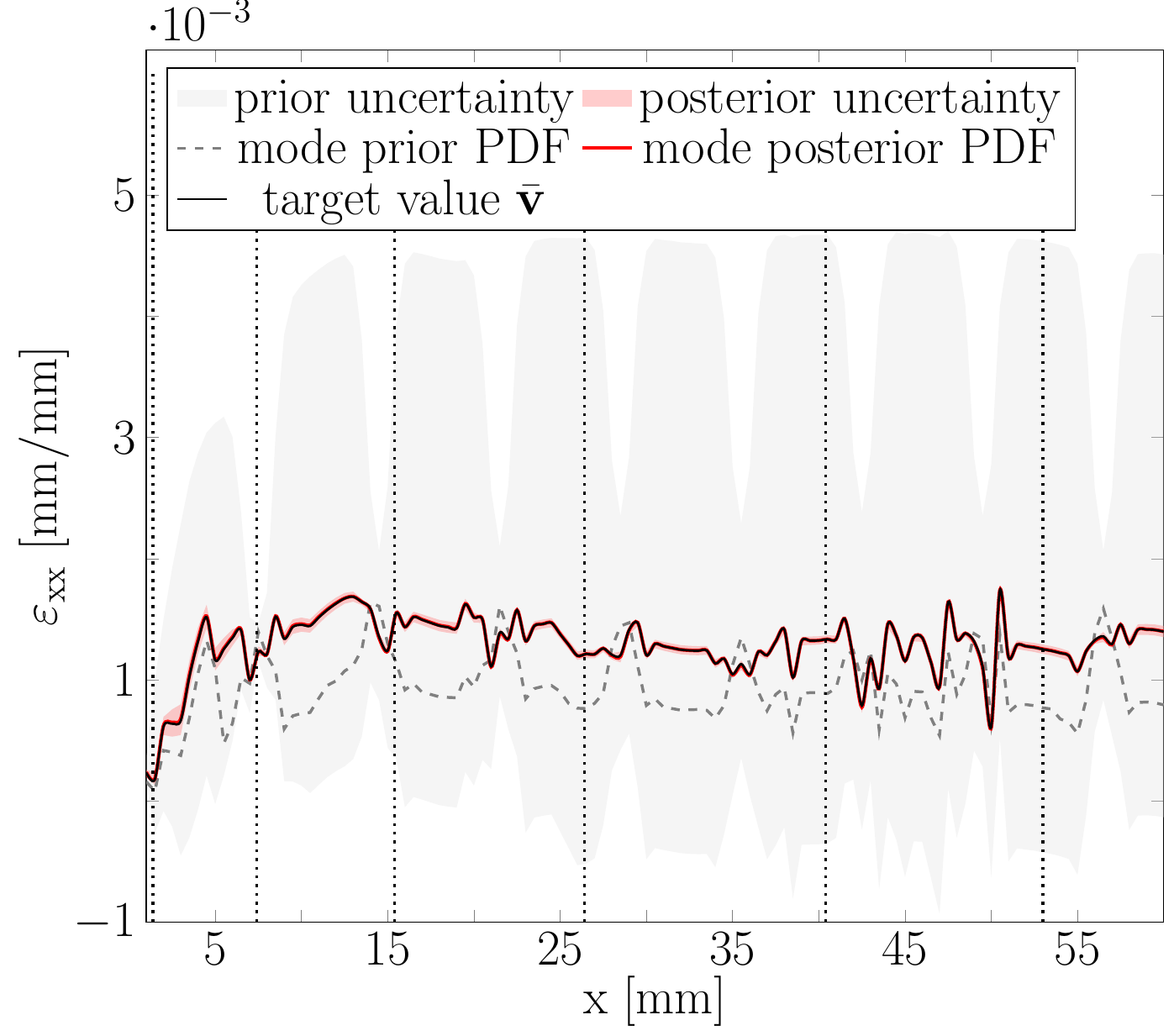}
       		\caption{\rev{Results of data-informed forward UQ analysis: i) the mode of the prior-based (dashed black line) and data-informed posterior (continuous red line)
       				PDFs of the residual strain $\varepsilon_{xx}$;
       				ii) the prior-based (gray area) and data-informed posterior (pink area) uncertainty bands for the residual strain $\varepsilon_{xx}$;
       				iii) the residual strain profile (continuous black line) obtained from part-scale thermomechanical analysis for the target value $\bar{\boldsymbol{\vv}}$; 
       				The vertical dotted lines represent the $6$ positions at which we report the prior and data-informed posterior PDFs in \Cref{fig:pdf40}.}}
       		\label{fig:ToTUQworkflow}
       	\end{center}
       \end{figure}

\begin{figure}[tbp]
		\centering
		\subfigure[PARAMETRI-1][$x = 1.5$ $\text{mm}$]{\label{fig:pdf5}\includegraphics[width=0.3\textwidth]{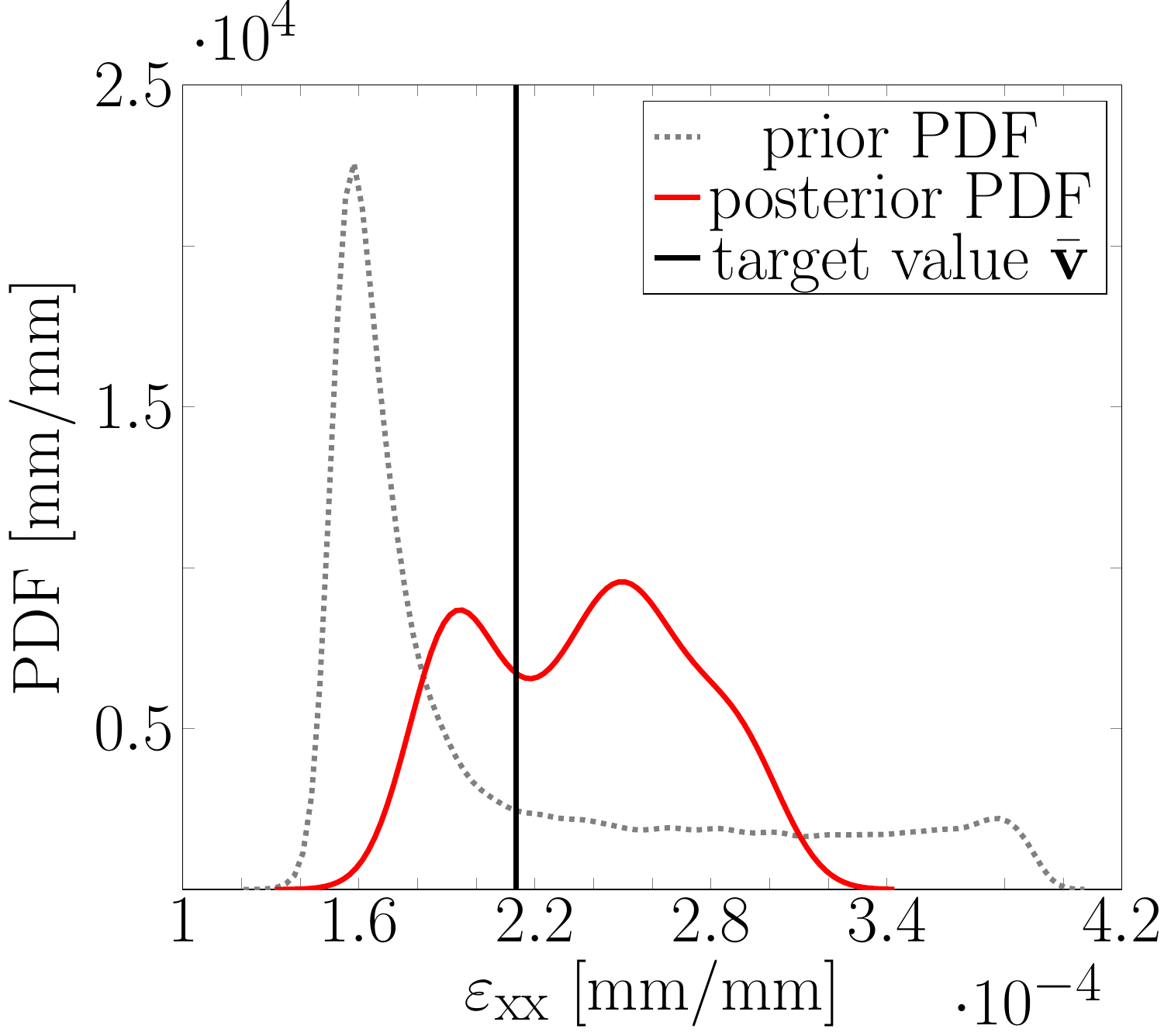}} 
		\subfigure[PARAMETRI-2][$x = 7.5$ $\text{mm}$]{\label{fig:pdf10}\includegraphics[width=0.3\textwidth]{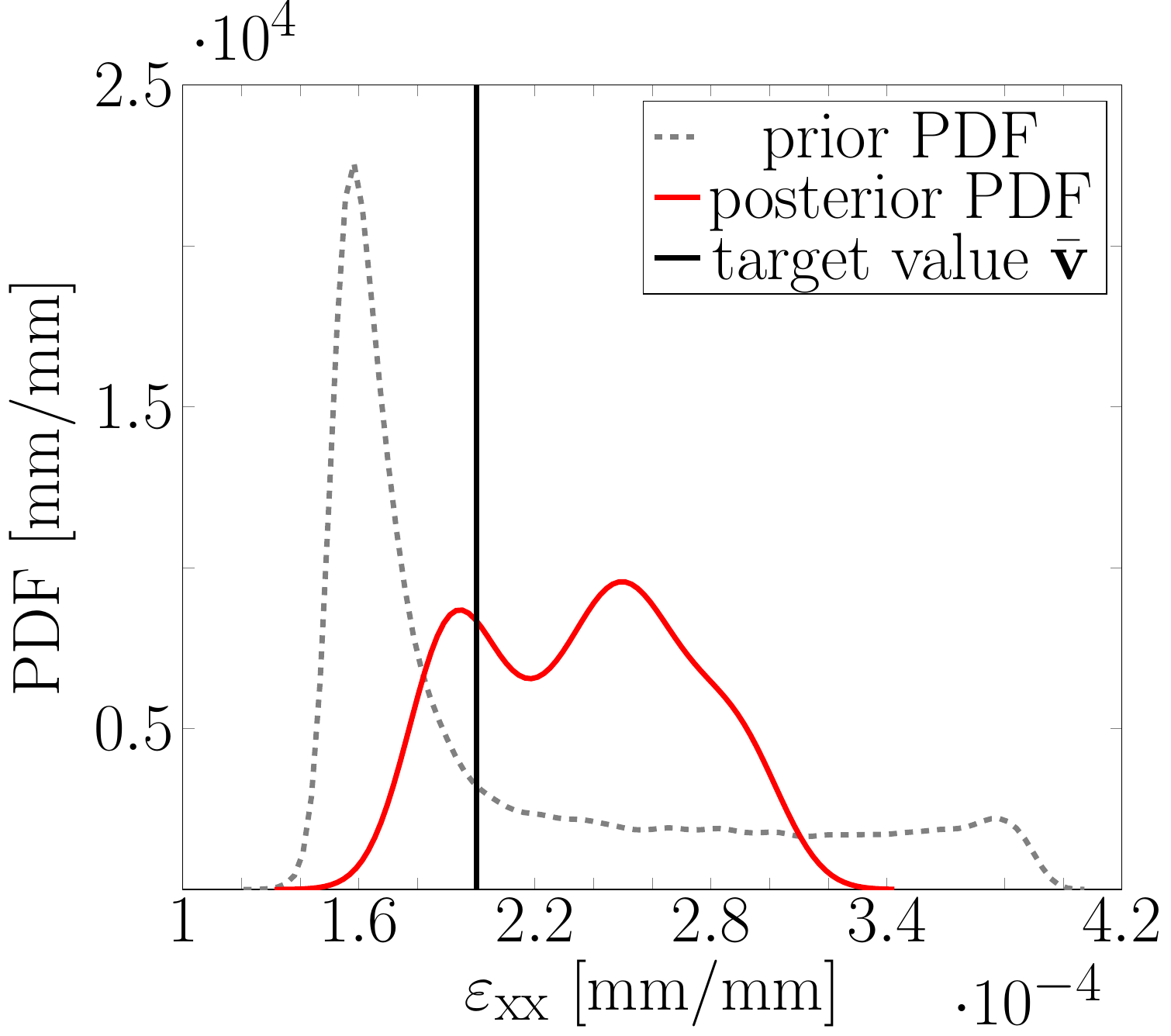}} 
		\subfigure[PARAMETRI-3][$x = 15.5$ $\text{mm}$]{\label{fig:pdf15}\includegraphics[width=0.31\textwidth]{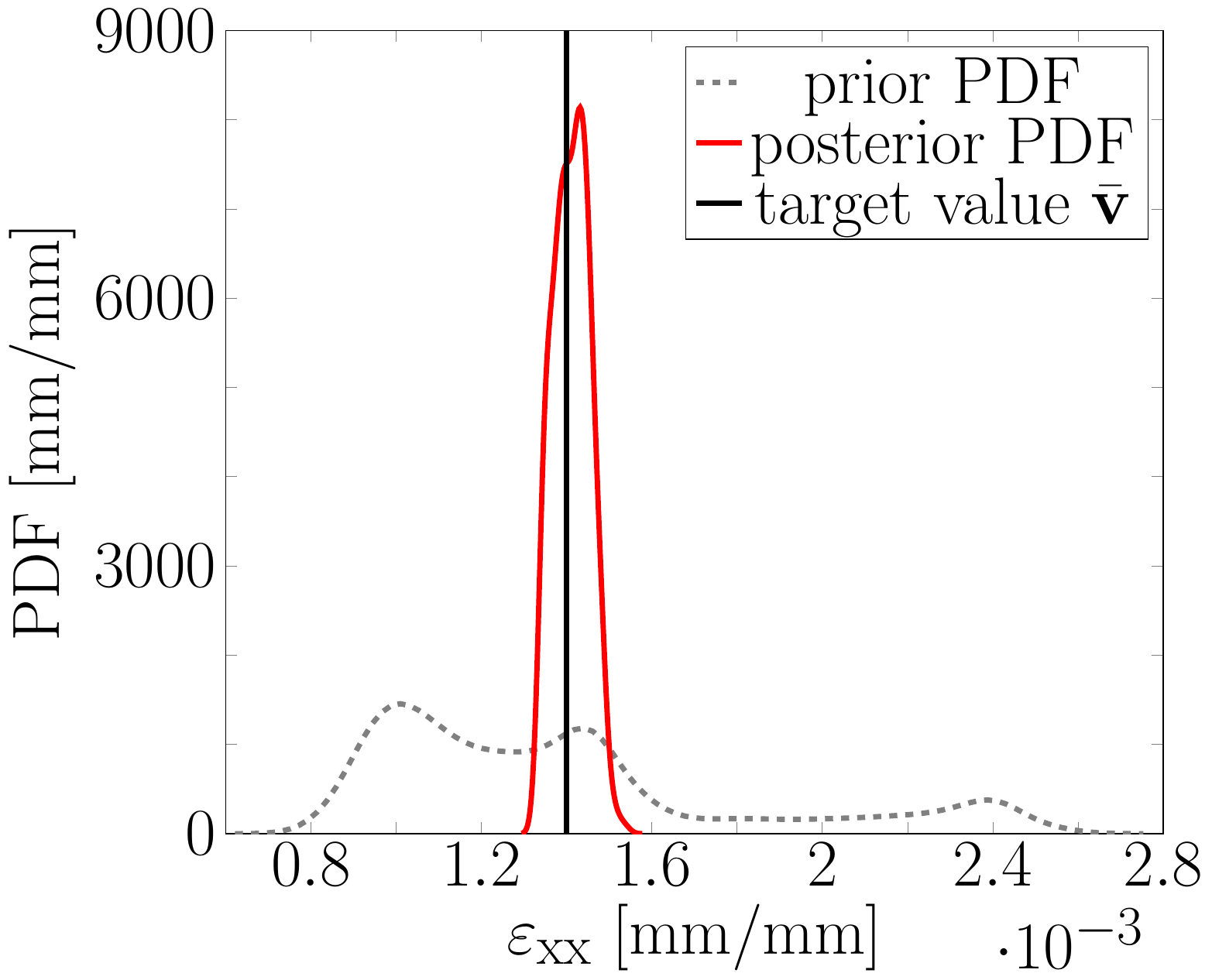}} 
		\subfigure[PARAMETRI-4][$x = 26.5$ $\text{mm}$]{\label{fig:pdf20}\includegraphics[width=0.29\textwidth]{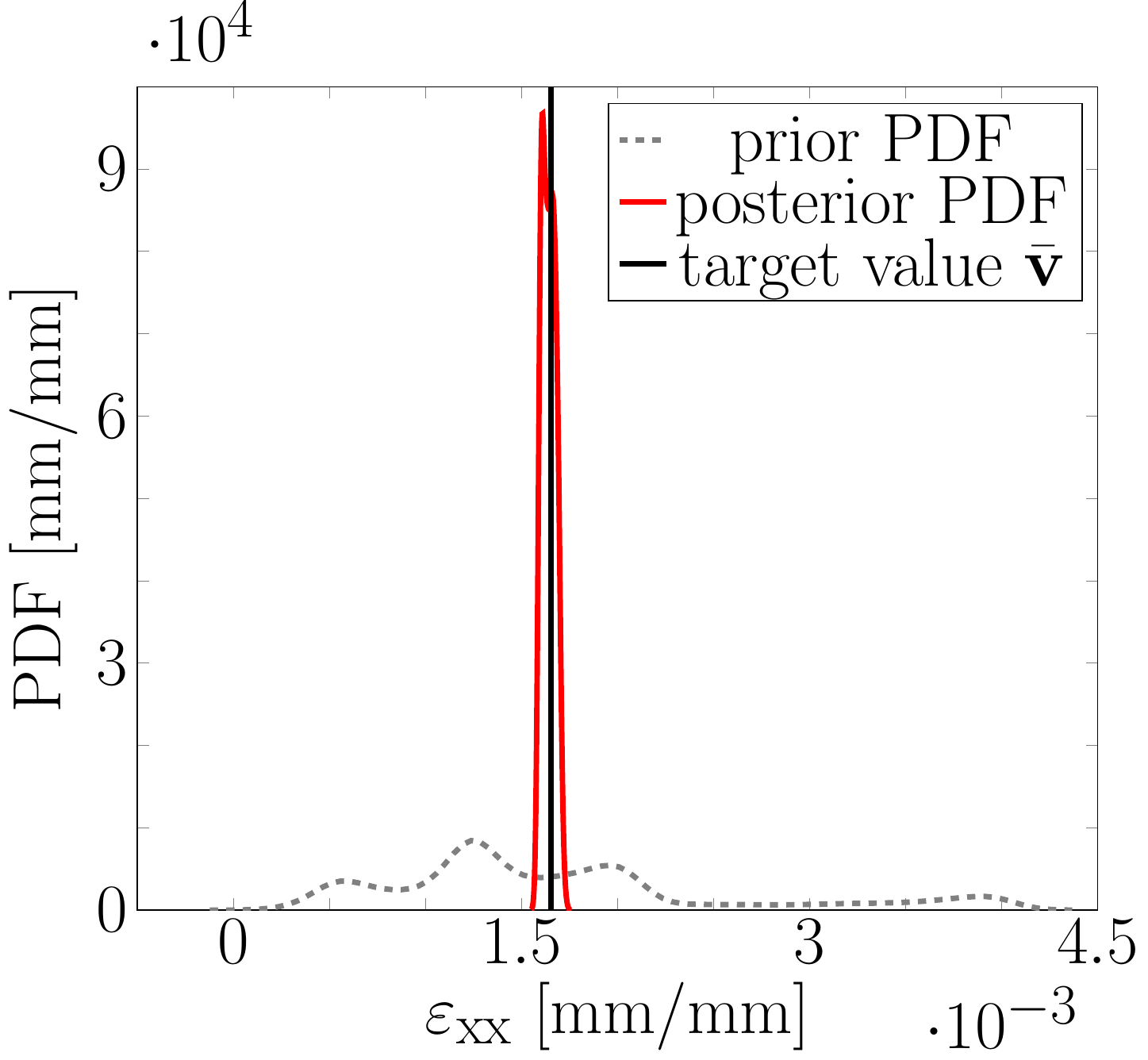}} 
		\subfigure[PARAMETRI-5][$x = 40.5$ $\text{mm}$]{\label{fig:pdf21}\includegraphics[width=0.3\textwidth]{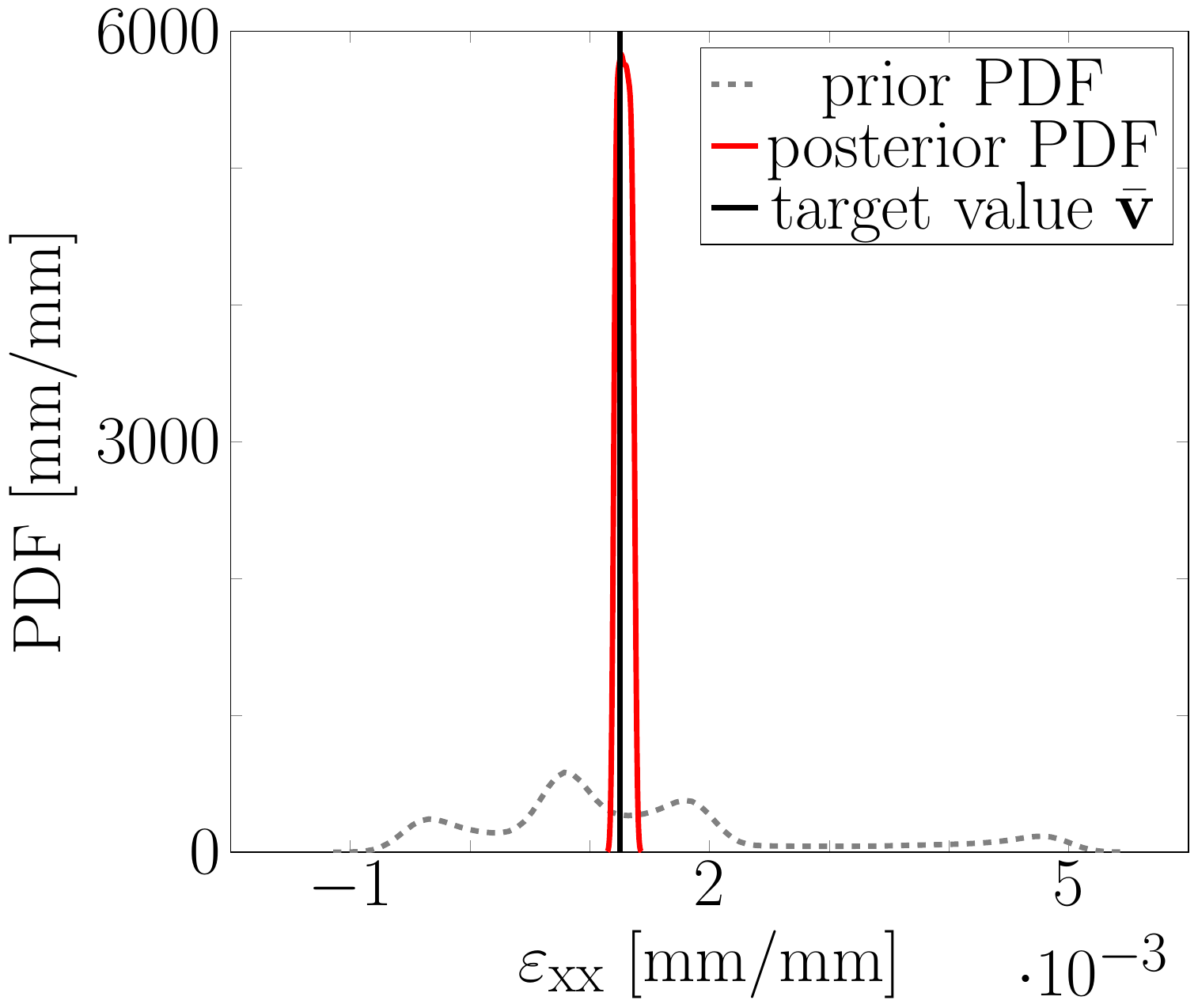}} 
		\subfigure[PARAMETRI-6][$x = 53$ $\text{mm}$]{\label{fig:pdf22}\includegraphics[width=0.3\textwidth]{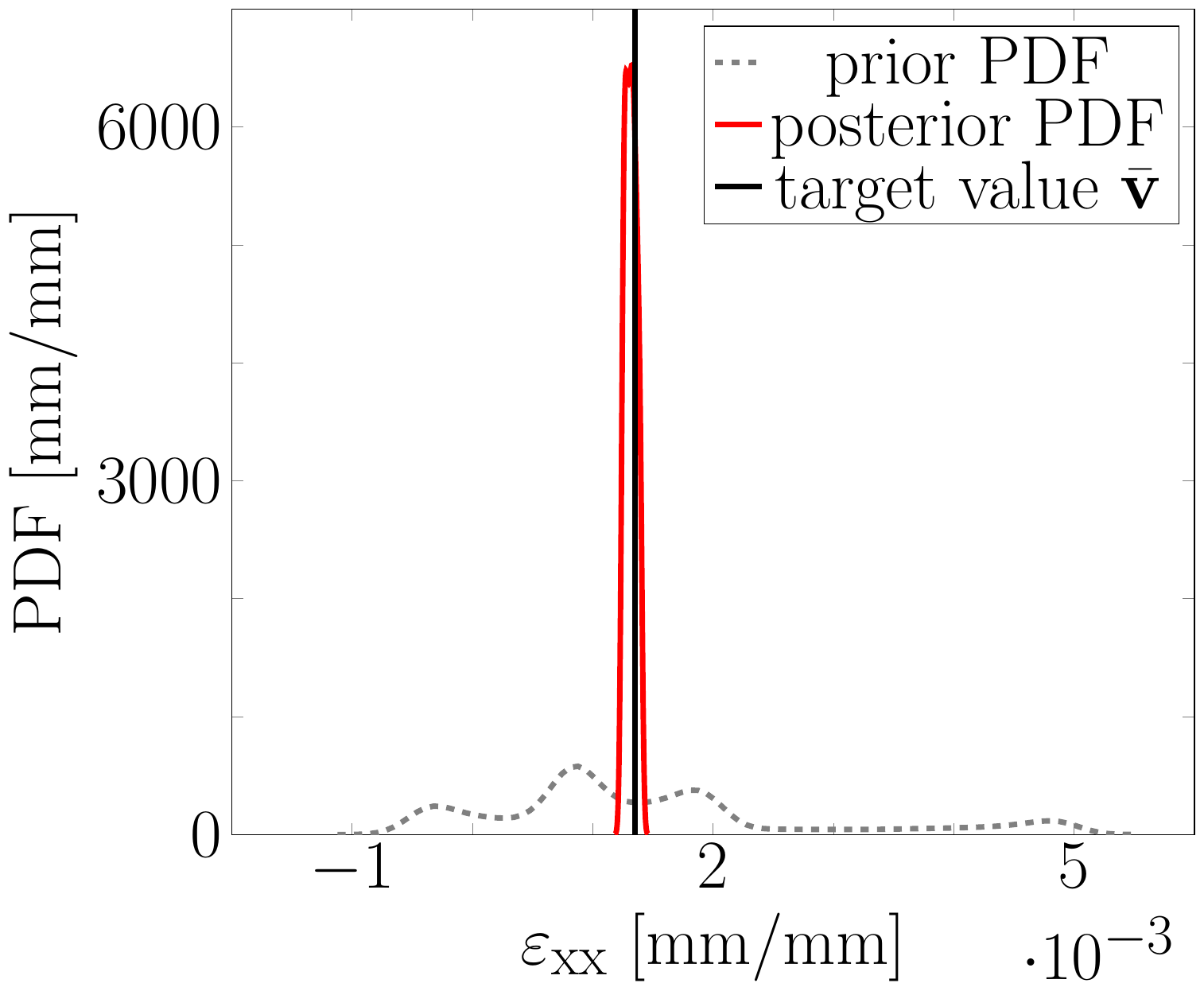}} 
		\caption{\rev{Results of data-informed forward UQ analysis. Prior-based and data-informed posterior PDFs of the residual strains of the beam
                    for the $x$ locations marked in \Cref{fig:ToTUQworkflow}
                    and residual strain value obtained from the part-scale thermomechanical numerical simulation for the target value $\bar{\vv}$.}}
		\label{fig:pdf40}
	\end{figure}

\begin{figure}[tbp]
    	\begin{center}
    		\includegraphics[scale=0.95]{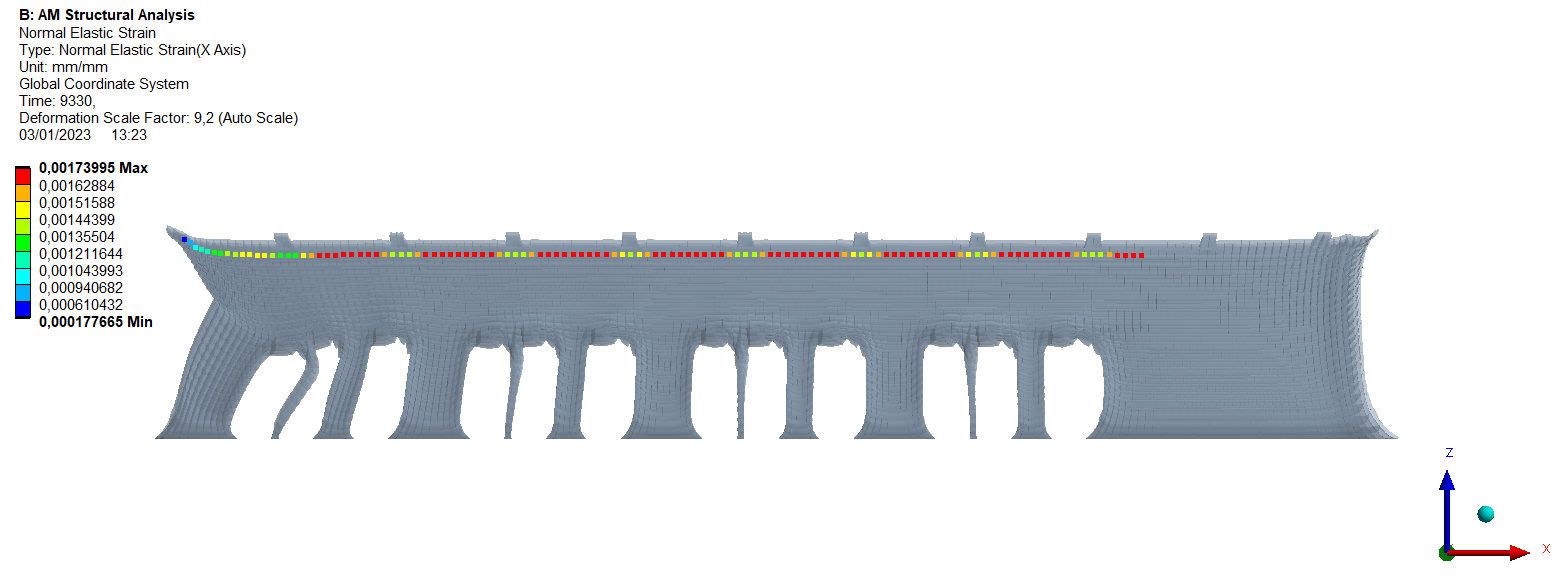}
    		\caption{\rev{Residual strains obtained from the part-scale thermomechanical analysis for the target value $\bar{\vv}$.}}
    		\label{fig:TargetAnsys}
    	\end{center}
    \end{figure}

	After validating the sparse-grid surrogate model, we proceed with the final step of the forward UQ, i.e., to compute the data-informed PDF of $\varepsilon_{xx}$,
	as explained in \Cref{subsec:Forward UQ}.
	To see to what extent the inverse UQ process allows us to reduce the uncertainty in the prediction of $\varepsilon_{xx}$,
        we also perform the forward UQ procedure based on the prior-information only, i.e.,
        we build a surrogate model for $\mathcal{S}_{I_{sum}}\varepsilon_{xx}$ according to the prior PDFs,
	sample $T_A$ and $\log h_p$ from such prior PDF and derive the corresponding prior-based PDF of $\varepsilon_{xx}$.
       In \Cref{fig:ToTUQworkflow} we show: i) the most probable $x$-profiles of $\varepsilon_{xx}$
          obtained by the two forward UQ analyses, 
          i.e., the modes of the two PDFs of $\varepsilon_{xx}$ at each of the $L=120$ locations (dotted black line for the prior-based PDF,
          continuous red line for the data-informed PDF), 
          and ii) the associated uncertainty bands, i.e., the 5\% - 95\% quantile bands of the two PDFs (gray area for the prior-based PDF,
          pink area for the data-informed PDF).
          The figure also reports (continuous black line) the $x$-profile of the residual strains obtained from the part-scale thermomechanical
          analysis at the target value
          $\bar{\boldsymbol{\vv}} = (T_A; \log h_p)=(1339. 8$ $^\circ{\text{C}}; -3.75$):
          this profile is overlapping with the mode of the data-informed PDF, which means that the most likely residual strains profile identified
          by such PDF closely resambles the true profile.
          Moreover, the fact that the prior-based quantile band is much larger than the data-informed one suggests 
          that using the posterior PDF for the parameters greatly reduces
          the uncertainties in the prediction of the residual strains.
          To provide further insight in this uncertainty reduction,
          we select $6$ locations of the beam (marked by vertical dotted lines in \Cref{fig:ToTUQworkflow}),
          for which we plot the prior and data-informed PDF of the residual strain in \Cref{fig:pdf40}.
          The residual strain profile for the target value $\bar{\boldsymbol{\vv}}$
          is also shown in \Cref{fig:TargetAnsys}, where we overlap the profile with the geometry of the beam to provide more geometrical context to the
          information discussed in \Cref{fig:ToTUQworkflow,fig:pdf40}.


	\section{Conclusions}
	\label{sec:Conclusion}
	In the present work, we quantify and reduce the uncertainties involved in the PBF process of a part-scale thermomechanical model of an Inconel 625 beam using synthetic data. We first perform a GSA, by calculating the principal Sobol and total Sobol indices, to study the sensitivity of the AM model to the activation temperature and the gas and powder convection coefficients. This analysis allows us to set the gas convection coefficient to an arbitrary but fixed value since the AM model is essentially insensitive to this parameter. Then, applying an inverse UQ approach, we quantify the uncertainties associated to the activation temperature and the powder convection coefficient. In particular, we do not only provide the point estimate of the uncertain parameters, but also estimate the residual uncertainties of such parameters.
	A data-informed forward UQ is subsequently performed to predict residual strains and their associated PDFs. We employ different sparse-grid surrogate models to reduce the computational cost of the numerous analyses required by the UQ methodology (the whole procedure, GSA + inverse UQ + forward UQ, only requires 177 part-scale thermomechanical analyses, of which 100 are only used for validation purposes). The results show the ability of the proposed approach to reduce the uncertainties of the powder convection coefficient and the activation temperature, as well as how the prediction of residual strains based on posterior uncertainties is significantly more accurate than the prediction of residual strains based on prior uncertainties.
	
	 \rev{In summary, we are able to substantially improve calibration of a part-scale PBF model
          as well as the reliability of the corresponding residual strains prediction.
          This is obtained thanks to a structured UQ workflow which is
          much easier and more straight-forward than the usual trial-and-error calibration. It is also
          computationally lighter than standard calibration techniques,
          since it is based on surrogate models that allow us to considerably reduce the number of full-model
          simulations, and furthermore it provides
          richer results since it delivers not only an estimate of the strains but also of their uncertainty.
          From a practical point of view, an additional advantage is that this methodology is performed on beam deflection after support removal,
          a quantity much easier to measure than residual strains.}
	\rev{As a further outlook of this work, we aim at extending the present results adopting a multi-fidelity approach to build the surrogate models to be used in the various steps of the UQ procedure. In particular, we will consider the so-called multi-index stochastic collocation \cite{hajiali.eal:MISC1,piazzola.eal:ferry-paper,jakeman2019adaptive},
		which is the multi-fidelity extension of the sparse-grids method used here, but other methods such as multi-fidelity Gaussian processes \cite{yang2022uncertainty,kennedy2000predicting}
		or multi-fidelity radial basis functions \cite{piazzola.eal:ferry-paper} can be used as well. The goal of multi-fidelity methods is to build a surrogate model
		calling many times the PBF solver on coarse meshes (that has limited accuracy but is also cheap to run) and to correct the resulting surrogate model with only a handful of calls to the solver on the finest (hence most expensive) mesh. In this way, we can further
		reduce the cost of performing UQ analyses, by keeping to a minimum the number of samples on the highest fidelity.} Moreover, since the accuracy of the proposed approach has been verified, experimental measurements can be employed, instead of synthetic ones, in forthcoming works.
	\\
	\section*{Acknowledgments}
	\label{}
	This work was partially supported by the Italian Minister of University and Research through the MIUR-PRIN projects
        \lq\lq A BRIDGE TO THE FUTURE: Computational methods, innovative applications, experimental validations of new materials and technologies\rq\rq~(No. 2017L7X3CS) and
        \lq\lq XFAST-SIMS\rq\rq~(no. 20173C478N).
        Lorenzo Tamellini and Chiara Piazzola have been supported by the PRIN 2017 project 201752HKH8 \lq\lq Numerical Analysis for Full and Reduced Order Methods for the efficient and accurate solution of complex systems governed by Partial Differential Equations (NA-FROM-PDEs)\rq\rq. 
        \rev{Chiara Piazzola acknowledges the support of the Alexander von Humboldt Foundation. 
        Mihaela Chiappetta, Massimo Carraturo and Ferdinando Auricchio acknowledge partial financial support from the Ministry of Enterprise and Made in Italy and Lombardy Region through the project \lq\lq PRotesi innOvaTivE per applicazioni vaScolari ed ortopedIChe e mediante Additive Manufacturing\rq\rq - CUP : B19J22002460005, on the finance  Asse I, Azione 1.1.3 PON Businesses and Competitiveness 2014 - 2020.
     Lorenzo Tamellini, Ferdinando Auricchio and Alessandro Reali acknowledge the Research program
        CN00000013 \lq\lq National Centre for HPC, Big Data and Quantum Computing -- Spoke 6 - Multiscale Modelling \& Engineering Applications\rq\rq.}
	
	\appendix
	\section{Sparse-grid surrogate modeling}
	\label{Surrogate Model}
	In the present section we present the sparse-grid surrogate modeling approach adopted for our study. Following the notation introduced in \Cref{sec:UQ analysis},
	we consider the problem of approximating an $N$-variate scalar function $\qoiscal(\vv): \Gamma \rightarrow \mathbb{R}$,
	where $\vv \in \Gamma \subset \mathbb{R}^N$ (extension to $P$-valued
	functions $\qoi: \Gamma \rightarrow \mathbb{R}^P$ is immediate; it is enough to apply the same procedure to each component of $\qoi$).
	We also recall that $v_n$ are independent random variables with probability density function
	$\rho_n(v_n), n=1,\ldots,N$ and that therefore the joint probability density of $\vv$ over $\Gamma$ is the product $\rho(\vv) = \prod_{n=1}^N \rho_n(v_n)$.
	
	The first step in constructing the sparse-grid surrogate model is to define a set of collocation points for each parameter $v_n$.
	We denote the number of points  along $v_n$ by $K_n \in \mathbb{N}_+$,
	and define a discretization level for each parameter, i.e., a positive number $i_n \in \mathbb{N}_+$, $i_n\geq 1$, using a ``level-to-knots'' function $m$ that associates
	to each level a number of points:
	\begin{equation}\label{eq:lev2knots}
		m:\mathbb{\mathbb{N}}_+ \rightarrow \mathbb{N}_+ \text{ such that } m(i_n) = K_n.
	\end{equation}
	In this work, we have considered $m(i_n)=2i_n-1$ (i.e, at each level $i_n$ two more points with respect to the previous level are considered;
	cf. \Cref{eq:how-many-points-in-a-tensor}), but other choices are possible.      
	The set of collocation points at level $i_n$ along parameter $v_n$ is denoted by: 
	\begin{equation}
		\label{eq:1d-nodes}
		\mathcal{T}_{i_n} = \left\{y_{n,m(i_n)}^{(j_n)}: j_n=1, \ldots, m(i_n) \right\} \quad \text{ for } n=1,\ldots,N.  
	\end{equation}
	The positions of these points over $\Gamma_n$ is usually chosen on the basis of the PDF $\rho_n$ of the random variables $v_n$.
	As reported in \Cref{tab:grids}, in this work we have used symmetric Leja points whenever $v_n$ is a uniform random variable
	and symmetric Gaussian Leja points whenever $v_n$ is a Gaussian random variable (see \ref{sec:Leja_points} for details),
	but other choices are possible, see e.g. \cite{piazzola2022sparse}.
	Our choices have the advantage that Leja points are nested, i.e., $\mathcal{T}_{i_n} \subset \mathcal{T}_{l_n}$ if $l_n  > i_n$.

	The second step is the definition of tensor grids of $N$ dimensions, derived as the Cartesian product
	of the previously introduced univariate sets $\mathcal{T}_{i_n}$, and of their associated Lagrangian interpolants.
	In particular, by collecting the discretization levels $i_n$ in a multi-index $\ii \in \mathbb{N}^N_+$,
	considering the corresponding tensor grid $\mathcal{T}_{\ii} = \bigotimes_{n=1}^{N} \mathcal{T}_{i_n}$,
	with number of nodes $M_{\ii} = \prod_{n=1}^{N} m(i_n)$ we can write:
	\begin{equation*} 
		\mathcal{T}_{\ii} = \left\{\vv_{m(\ii)}^{(\jj)}\right\}_{\jj \leq m(\ii)},
		\quad  \text{ with } \quad
		\vv_{m(\ii)}^{(\jj)} = \left[v_{1,m(i_1)}^{(j_1)}, \ldots, v_{N,m(i_N)}^{(j_N)}\right]
		\text{ and } \jj \in \mathbb{N}^N_+,
	\end{equation*}
	where $m(\ii) = \left[m(i_1),\,m(i_2),\ldots,m(i_N) \right]$ and $\jj \leq m(\ii)$ means that $j_n \leq m(i_n)$ for every $n = 1,\ldots,N$. 
	The tensor-interpolant approximation (also called tensor-interpolant \emph{surrogate model}) of $\qoiscal(\vv)$, that we denote by $\mathcal{U}_{\ii}(\vv)$,
	is then an $N$-variate Lagrangian interpolant collocated at the grid nodes of $\mathcal{T}_{\ii}$
	and can be written as:
	\begin{equation}
		\label{eq:interp_tensor}
		\qoiscal(\vv) \approx \mathcal{U}_{\ii}(\vv) := \sum_{\jj \leq m(\ii)} f\left(v_{m(\ii)}^{(\jj)}\right) \mathcal{L}_{m(\ii)}^{(\jj)}(\vv),
	\end{equation}
	where $\left\{ \mathcal{L}_{m(\ii)}^{(\jj)}(\mathbf v) \right\}_{\jj \leq m(\ii)}$ are $N$-variate Lagrange polynomials,
	defined as tensor products of univariate Lagrange polynomials, i.e.  
	\begin{equation*}
		\mathcal{L}_{m(\ii)}^{(\jj)}(\mathbf v) = \prod_{n=1}^{N} \ell_{n,m(i_n)}^{(j_n)}(v_n)
		\quad \text{ with } \quad
		\ell_{n,m(i_n)}^{(j_n)}(v_n) = \prod_{k=1, k\neq j_n}^{m(i_n)} \frac{v_n-v_{n,m(i_n)}^{(k)}}{v_{n,m(i_n)}^{(k)}-v_{n,m(i_n)}^{(j_n)}}. 
	\end{equation*}
	
	The accuracy of the approximation $\qoiscal(\vv) \approx \mathcal{U}_{\ii}(\vv)$ increases as the number of collocation points in each $v_n$ grows,
	i.e., for $i_n \gg 1$, $n = 1,\ldots,N$.
	At the same time, the cost of constructing $\mathcal{U}_{\ii}(\vv)$ grows exponentially in $N$, since it requires evaluating $\qoiscal$ at $M_{\ii} = \prod_{n=1}^N m(i_n)$ points;
	this implies that even moderate choices of $i_n$, $n=1,\ldots N$ could be unfeasible for $N>2$ if evaluating $\qoiscal$ is an expensive operation.  
	To mitigate this problem, the sparse-grid surrogate model consists of an approximation of $\qoiscal(\vv)$ formed by a linear combination of several coarse $\mathcal{U}_{\ii}(\vv)$
	rather than by a single $\mathcal{U}_{\ii}(\vv)$ with $i_n \gg 1$, $n = 1, \ldots, N$.
	
	For this purpose, as a third step towards sparse-grid surrogate models
	we introduce the so-called univariate and multivariate detail operators: 
	\begin{alignat}{2}
		\label{eq:univariate detail}
		&\Delta_n[\mathcal{U}_{\ii}(\vv)]=\mathcal{U}_{\ii}(\vv)-\mathcal{U}_{\ii- \ee_n}(\vv) \text{ with } 1 \leq n \leq N; \\
		\label{eq:multivariate detail}
		&{\Delta}[\mathcal{U}_{\ii}(\vv)] = \bigotimes_{n=1}^N \Delta_n[\mathcal{U}_{\ii}(\vv)] = \Delta_1\left[ \, \cdots \left[ \Delta_N\left[ \mathcal{U}_{\ii}(\vv) \right] \, \right] \, \right],
	\end{alignat}
	where $\mathcal{U}_{\ii}(\vv) = 0$ when at least one component of $\ii$ is zero and $\ee_n$ the $n$-th canonical multi-index, i.e., $(\ee_n)_k = 1$ if $n=k$ and 0 otherwise.
	Multivariate detail operators can be evaluated as suitable linear combinations of certain approximations of the complete tensor approximations $\mathcal{U}_{\ii}$:
	\begin{align}
		\label{eq:combitec-delta}
		{\Delta}[\mathcal{U}_{\ii}](\vv)
		& = \Delta_1\left[ \, \cdots \left[ \Delta_N\left[ \mathcal{U}_{\ii} \right] \, \right] \, \right] 
		= \sum_{\jj \in \{0,1\}^N} (-1)^{\lVert \jj\rVert_1} \mathcal{U}_{\ii-\jj}(\vv).
	\end{align}
	Moreover, note that a hierarchical decomposition of $\mathcal{U}_{\ii}(\vv)$ holds:
	\begin{equation}
		\label{eq:sum}
		\mathcal{U}_{\ii}(\vv) = \sum_{\jj \leq \ii} {\Delta}[\mathcal{U}_{\jj}(\vv)].
	\end{equation}
	
	The fourth and final step to construct a sparse-grid surrogate model is to tweak such hierarchical decomposition. In detail, instead of summing over $\jj \leq \ii$ 
	we sum over a different collection of multi-indices $\mathcal{I}$ (from here on, multi-index set), 
	chosen according to criteria that will be made clearer in a moment:
	\[
	\qoiscal(\vv)  \approx \mathcal{S}_{\mathcal{I}} \qoiscal(\vv)
	= \sum_{\ii \in \mathcal{I}} {\Delta}[\mathcal{U}_{\ii}(\vv)].
	\]
	Furthermore, applying \Cref{eq:combitec-delta} we obtain a more practical expression, i.e., the so-called
	``combination technique'' \cite{wasilkowski1995explicit}, which is the form actually implemented in the Sparse-Grids Matlab-Kit:
	\begin{align}
		\qoiscal(\vv)  \approx \mathcal{S}_{\mathcal{I}} f(\vv)
		= \sum_{\ii \in \mathcal{I}} c_{\ii} \mathcal{U}_{\ii}(\mathbf{v}), \quad c_{\ii}: = \sum_{\substack{\jj \in \{0,1\}^N \\ \ii+\jj \in \mathcal{I}}} (-1)^{\lVert \jj\rVert_1}. \label{eq:sg_interp} 
	\end{align}
	This re-writing is valid only if $\mathcal{I}$ is \emph{downward-closed}, i.e., if $\mathcal{I}$ is such that if a certain multi-index $\ii$ is in $\mathcal{I}$
	all its ``previous'' multi-indices $\jj \leq \ii$ are also in the set. In formulae, we require that:
	\begin{equation}
		\label{eq:downward-closedness}
		\forall \ii \in \mathcal{I}, \	\ii - \mathbf{e}_n \in \mathcal{I}, \quad \forall n = 1,\ldots,N \mbox{ s.t. } i_n>0.  
	\end{equation}
	Coming back to the issue of choosing the multi-index set $\mathcal{I}$, the idea is to discard from the hierarchical decomposition in
	\Cref{eq:sum} the contributions that have a large cost and contribute little to the approximation (in a sense, dropping
	the high-order corrections). Under mild regularity assumptions of $\qoiscal(\vv)$, a simple yet effective choice to this end is 
	\begin{equation}\label{eq:I_sparse}
		\mathcal{I}_{sum} = \{ \ii \in \mathbb{N}^N_+ : \sum_{n=1}^N (i_n-1) \leq w \}  
	\end{equation}
	for some integer value $w$ (the larger $w$, the more accurate is the sparse-grid surrogate model). Note that conversely, choosing
	\begin{equation}
		\label{eq:I_tensor}
		\mathcal{I}_{max} = \{ \ii \in \mathbb{N}^N_+ : \max_{n=1,\ldots, N} (i_n-1) \leq w \}  
	\end{equation}
	one would obtain a tensor grid with $m(w+1)$ points per direction, i.e., $\mathcal{S}_{\mathcal{I}_{max}(w)}(\vv)  = \mathcal{U}_{\ii_w}(\vv)$ with $\ii_w = [w{+}1,w{+}1,\cdots]$;
	this is an immediate consequence of the decomposition in (\Cref{eq:sum}). More advanced options to
	tailor the set $\mcI$ to the function $\qoiscal$ are available in literature, and in particular it would
	be possible to use an adaptive algorithm, that adds multi-indices $\ii$ to $\mcI$ one by one given the values 
	of $\qoiscal$; see again \cite{piazzola2022sparse} for details.
	Finally, we call \emph{sparse grid} the collection of points needed to build the \emph{sparse-grid surrogate model} $\mathcal{S}_\mathcal{I}$, i.e. 
	\begin{equation}\label{eq:sg_grid} 
		\mathcal{G}_\mathcal{I} =  \bigcup_{\substack{\ii \in \mathcal{I} \\ c_{\ii} \neq 0 }} \mathcal{T}_{\ii}.
	\end{equation}
	
	\section{Leja points}
	\label{sec:Leja_points}
	
	Leja knots have been introduced for unweighted interpolation on intervals $[a,b]$, see \cite{narayan:Leja,piazzola2022sparse,nobile2014mathicse} and references therein,
	and are therefore a suitable choice when $v_n$ are uniform random variables.
	They are built recursively as:
	\begin{equation}\label{eq:leja}
		v_n^{(1)}=b, \quad v_n^{(2)}=a, \quad v_n^{(3)}=\frac{a+b}{2}, \quad \quad v_n^{(j)}= \argmax_{v_n \in [a\,b]} \prod_{k=1}^{j-1} \lvert v_n-v_n^{(k)}\rvert.
	\end{equation}
	Observe that by construction Leja knots are nested but not symmetric with respect to the mid-point $\frac{a+b}{2}$, which is also a desirable property.
	To fix this issue, the construction above can be changed by generating only the even elements of the sequence with the standard formula in (\Cref{eq:leja})
	and then symmetrizing them to obtain the odd elements, i.e.\
	\begin{align}  \label{eq:sym-leja}
		& v_n^{(1)}=b, \; v_n^{(2)}=a, \;v_n^{(3)}=\frac{a+b}{2}, \; \nonumber \\
		& v_n^{(2j)}= \argmax_{v_n \in [a\,b]} \prod_{k=1}^{2j-1} \lvert v_n-v_n^{(k)}\rvert, \; \nonumber \\
		& v_n^{(2j+1)}= \frac{a+b}{2} - \left(v_n^{(2j)} - \frac{a+b}{2}\right).  
	\end{align}

	It is furthermore possible to extend the construction of Leja knots to the case when $v_n \in \Gamma_n$ are
	Gaussian random variables (or more generally, random variables with a probability distribution other than uniform), see again \cite{narayan:Leja,piazzola2022sparse}.
	The knots thus obtained are the so-called Gaussian Leja knots (or in general, weighted Leja knots) and can be computed again recursively,
	by suitably introducing a weight in \Cref{eq:leja}, i.e., solving 
	\[
	v_n^{(j)}= \argmax_{v_n \in \Gamma_n} \sqrt{\rho_n(v_n)} \prod_{k=1}^{j-1} \lvert v_n-v_n^{(k)} \rvert,
	\] 
	where $\rho_n$ is the PDF of the random variable. Symmetric versions of Gaussian (weighted) Leja points can then be generated
	following the procedure that leads to \Cref{eq:sym-leja}.
	
	\bibliographystyle{unsrtnat}
	\bibliography{Reference}

\end{document}